\documentclass[11pt,titlepage,a4paper]{article}
\usepackage{bozhomac}  
\usepackage{bozhlogo}  
\usepackage{cite}	
\usepackage{varioref}	

\title{\bfseries	\vspace*{-1.7in}
{\huge Lagrangian quantum field theory\\[1ex] in momentum picture}
 \\[1.3ex]
{\LARGE II.\ Free spinor fields}
}

\vspace{1.7ex}

\author{
Bozhidar Z.\ Iliev
\thanks{Laboratory of Mathematical Modeling in Physics,
Institute for Nuclear Research and \mbox{Nuclear} Energy,
Bulgarian Academy of Sciences,
Boul.\ Tzarigradsko chauss\'ee~72, 1784 Sofia, Bulgaria}
\thanks{E-mail address: bozho@inrne.bas.bg}
\thanks{URL: http://theo.inrne.bas.bg/$\sim$bozho/}
}

\date{	
 \vspace{2.27ex}\ShortTitle{QFT in momentum picture: II}\\[0.27ex]
 \vspace{3.27ex}
\small
	\begin{tabular}{r@{$\colon\to~$}l}
 \vspace{0.09ex} Basic ideas	& June 4--9, 2001\\[0.09ex]
 \vspace{0.09ex} Began		& August 16, 2001	\\[0.09ex]
 \vspace{0.09ex} Ended		& September 14, 2001	\\[0.09ex]
 \vspace{0.09ex} Initial typeset& August 24 -- September 15, 2001
							\\[0.09ex]
 \vspace{0.09ex} Revised	& October, 2001		\\[0.09ex]
 \vspace{0.09ex} Last update	& April 27, 2004	\\[0.09ex]
 \vspace{0.27ex} Produced	& \fbox{\today}	\\[0.27ex]
	\end{tabular} \\[1.27ex]
\normalsize
	\begin{tabular}{r@{$\colon~$}l}
 \vspace{0.27ex} http://www.arXiv.org e-Print archive No. & hep-th/0405008
	\end{tabular} \\[-0.27ex]
 \vspace{4.27ex}{\Huge\BOZHO}	\\[4.27ex]
 \vspace{0.27ex}\Subject{Quantum field theory}
								\\[2.27ex]
	\begin{tabular}{r@{\hspace{0.512em}}|@{\hspace{0.512em}}l}
 \vspace{0.27ex}\MSC[2000]{81Q99, 81T99\\\hspace{0pt}}
&
 \vspace{0.27ex}
	\end{tabular} \\[1.27ex]
 \vspace{0.27ex}\\[0.27ex]
}

\newcommand{\opsi}{\overline{\psi}}	
\newcommand{\bpsi}{\Breve{\psi}}	
\newcommand{\bopsi}{\Breve{\opsi}}	
\newcommand{\bk}{\boldsymbol{k}}  	

 \newcommand{\Hil}{\mathcal{F}}		
	\newcommand{\base}{\mathit{M}}	

\newcommand{\ope}[2][{}]{\lindex[\mathcal{#2}]{}{#1}} 
\newcommand{\tope}[2][{}]{\ope[#1]{\Tilde{#2}}} 
\begin{document}		

\renewcommand{\thepage}{\roman{page}}

\renewcommand{\thefootnote}{\fnsymbol{footnote}} 
\maketitle				
\renewcommand{\thefootnote}{\arabic{footnote}}   

\tableofcontents		


\begin{abstract}

	Free spinor fields, with spin 1/2, are explored in details in
the momentum picture of motion in Lagrangian quantum field theory. The
field equations are equivalently written in terms of creation and
annihilation operators and on their base the anticommutation relations are
derived. Some problems concerning the vacuum and state vectors of free spinor
field are discussed. Several Lagrangians, describing free spinor fields, are
considered and the basic consequences of them are investigated.

\end{abstract}

\renewcommand{\thepage}{\arabic{page}}


\section {Introduction}
\label{Introduction}

	The paper is devoted to a detailed investigation, in
Lagrangian quantum field theory%
\footnote{~%
In this paper we considered only the Lagrangian (canonical) quantum field
theory in which the quantum fields are represented as operators, called field
operators, acting on some Hilbert space, which in general is unknown if
interacting fields are studied. These operators are supposed to satisfy some
equations of motion, from them are constructed conserved quantities
satisfying conservation laws, etc. From the view\ndash point of present\ndash
day quantum field theory, this approach is only a preliminary stage for more
or less rigorous formulation of the theory in which the fields are
represented via operator\ndash valued distributions, a fact required even for
description of free fields. Moreover, in non\ndash perturbative directions,
like constructive and conformal field theories, the main objects are the
vacuum mean (expectation) values of the fields and from these are
reconstructed the Hilbert space of states and the acting on it fields.
Regardless of these facts, the Lagrangian (canonical) quantum field theory is
an inherent component of the most of the ways of presentation of quantum
field theory adopted explicitly or implicitly in books
like~\cite{Bogolyubov&Shirkov,Bjorken&Drell,Roman-QFT,Ryder-QFT,
Akhiezer&Berestetskii,Ramond-FT,Bogolyubov&et_al.-AxQFT,Bogolyubov&et_al.-QFT}.
Besides, the Lagrangian approach is a source of many ideas for other
directions of research, like the axiomatic quantum field
theory~\cite{Roman-QFT,Bogolyubov&et_al.-AxQFT,Bogolyubov&et_al.-QFT}.%
},
of a free spinor (spin~$\frac{1}{2}$) field in momentum picture, introduced
in~\cite{bp-QFT-pictures} and developed in~\cite{bp-QFT-MP}. It
is a direct continuation of~\cite{bp-QFTinMP-scalars}, where similar
exploration of free scalar fields was carried out, and, respectively, most of
the methods of \emph{loc.~cit.}\ are \emph{mutatis mutandis} applied to
problems of free spinor fields. Most of the known fundamental results are
derived in a new way (and in a slightly modified form), but the work contains
and new ones; \eg the field equations in terms of creation and annihilation
operators and a uniform consideration of the massive and massless cases.

	The basic moments of the method, we will follow in this work, are the
following ones:
\\\indent
	(i) In Heisenberg picture is fixed a (second) non\ndash quantized and
non\ndash normally ordered  operator\ndash valued Lagrangian, which is
supposed to be polynomial (or convergent power series) in the field operators
and their first partial derivatives;
\\\indent
	(ii) As conditions additional to the Lagrangian formalism are
postulated the commutativity between the components of the momentum operator
(see~\eref{2.1} below) and the Heisenberg relations between the field
operators and momentum operator (see~\eref{2.28} below);
\\\indent
	(iii) Following the Lagrangian formalism in momentum picture, the
creation and annihilation operators are introduced and the dynamical
variables and field equations are written in their terms;
\\\indent
	(iv) From the last equations, by imposing some additional restrictions
on the creation and annihilation operators, the (anti)commutation relations
for these operators are derived;
\\\indent
	(v) At last, the vacuum and normal ordering procedure are defined, by
means of which the theory can be developed to a more or less complete form.

	The main difference of the above scheme from the standard one is that
we postulate the below\ndash written relations~\eref{2.1} and~\eref{2.28}
and, then, we look for compatible with them and the field equations
(anti)commutation relations. (Recall, ordinary the (anti)commutation
relations are postulated at first and the validity of the
equations~\eref{2.1} and~\eref{2.28} is explored after
that~\cite{Bjorken&Drell-2}.)

	The layout of the work is as follows.

	Section~\ref{Sect2} contains a review of the momentum picture of
motion in quantum field theory. As a new material, a consideration of the
angular momentum in it is added. On that base, in Sect.~\ref{Sect3}, a free
spinor field is described in momentum picture. In particular, the Dirac
equation in momentum picture is derived and special attention is paid to the
operators of momentum, charge and angular momentum.


	In Sect.~\ref{Sect5}, the (system of) Dirac equation(s) describing a
free spinor field is analyzed in terms of operators which, possibly, up to
normalization and pure phase factor, are identical with the Fourier
coefficients of the field in Heisenberg picture. From these operators, in
Sect.~\ref{Sect6}, is constructed a set of operators which admit
interpretation as creation and annihilation operators. Then, the last
operators are expressed via a new set of operators with the same physical
interpretation, which operators, up to a phase factor, coincide with the
creation/annihilation operators known from the investigation of free spinor
field in Heisenberg picture. The field's dynamical variables, \ie the
momentum, charge and angular momentum operators, are expressed via the
creation and annihilation operators in Sect.~\ref{Sect7}, which results in
expressions similar to ones in the momentum representation in Heisenberg
picture.

	In Sect.~\ref{Sect8}, the equations of motion for a free spinor field
are equivalently written as a system of operator equations in terms of
creation and annihilation operators. These equations are trilinear ones and
their form is similar to the one of the parafermi relations. The obtained
system of equations is analyzed in Sect.~\ref{Sect9}, where from it, under
some explicitly presented additional conditions, the anticommutation
relations for the creation and annihilation operators are derived.

	The concept of a vacuum for a free spinor field is introduced in
Sect.~\ref{Sect10}. Some problems in the theory are pointed and their
solution is described via the introduction of normal ordering of products
(compositions) of creation and annihilation operators. In Sect.~\ref{Sect11}
are discussed some general aspects regarding state vectors of free spinor
field.

	In Sect.~\ref{Sect12} are investigated different Lagrangians, which
do not differ by a full 4\ndash divergence, from which the quantum theory of
free spinor fields can be derived. They and their consequences are compared
from different positions and the `best' one of them is pointed out. It is the
one which is charge symmetric; so that in it is encoded the spin\ndash
statistics theorem for free spinor fields.

	Section~\ref{Conclusion} closes the work.

\vspace{1ex}

	The books~\cite{Bogolyubov&Shirkov,Roman-QFT,Bjorken&Drell} will be
used as standard reference works on quantum field theory. Of course, this is
more or less a random selection between the great number of (text)books and
papers on the theme to which the reader is referred for more details or other
points of view. For this end, e.g.,~\cite{Itzykson&Zuber,Ryder-QFT,Schweber} or
the literature cited
in~\cite{Bogolyubov&Shirkov,Roman-QFT,Bjorken&Drell,Itzykson&Zuber,
Ryder-QFT,Schweber} may be helpful.

	Throughout this paper $\hbar$ denotes the Planck's constant (divided
by $2\pi$), $c$ is the velocity of light in vacuum, and $\iu$ stands for the
imaginary unit. The superscripts $\dag$ and $\top$ means respectively
Hermitian conjugation and transposition of operators or matrices, the
superscript $\ast$ denotes complex conjugation, and the symbol $\circ$
denotes compositions of mappings/operators.

	By $\delta_{fg}$, or $\delta_f^g$ or $\delta^{fg}$  ($:=1$ for $f=g$,
$:=0$ for $f=g$) is denoted the Kronecker $\delta$\ndash symbol, depending on
arguments $f$ and $g$, and $\delta^n(y)$, $y\in\field[R]^n$, stands for the
$n$\ndash dimensional Dirac $\delta$\ndash function; $\delta(y):=\delta^1(y)$
for $y\in\field[R]$.

	The Minkowski spacetime is denoted by $\base$. The Greek indices run
from 0 to $\dim\base-1=3$. All Greek indices will be raised and lowered by
means of the standard 4\ndash dimensional Lorentz metric tensor
$\eta^{\mu\nu}$ and its inverse $\eta_{\mu\nu}$ with signature
$(+\,-\,-\,-)$. The Latin indices $a,b,\dots$ run from 1 to $\dim\base-1=3$
and, usually, label the spacial components of some object. The Einstein's
summation convention over indices repeated on different levels is assumed
over the whole range of their values.

	At the end, a technical remark is in order. The derivatives with
respect to operator\ndash valued (non\ndash commuting) arguments will be
calculated according to the rules of the classical analysis of commuting
variables, which is an everywhere silently accepted
practice~\cite{Bogolyubov&Shirkov,Bjorken&Drell-2}. As it is demonstrated
in~\cite{bp-QFT-action-principle}, this is not quite correct. We shall pay
attention on that item at the corresponding places in the text. With an
exception of the consequences of a `charge symmetric' Lagrangian, considered
and explained in Sect.~\ref{Sect12}, this method for calculation of
derivatives with respect to operators does not lead to incorrect results
when free spinor fields are involved.


\section{The momentum picture}
	\label{Sect2}

	In this section, we present a summary of the momentum picture in
quantum field theory, introduce in~\cite{bp-QFT-pictures} and developed
in~\cite{bp-QFT-MP}.

	Let us consider a system of quantum fields, represented in Heisenberg
picture of motion by field operators $\tope{\varphi}_i(x)\colon\Hil\to\Hil$,
$i=1,\dots,n\in\field[N]$, acting on the system's Hilbert space $\Hil$ of
states and depending on a point $x$ in Minkowski spacetime $\base$. Here and
henceforth, all quantities in Heisenberg picture will be marked by a tilde
(wave) ``$\tope{\mspace{6mu}}\mspace{3mu}$'' over their kernel symbols. Let
$\tope{P}_\mu$ denotes the system's (canonical) momentum vectorial operator,
defined via the energy\ndash momentum tensorial operator $\tope{T}^{\mu\nu}$
of the system, viz.
	\begin{equation}
			\label{2.0}
\tope{P}_\mu
:=
\frac{1}{c}\int\limits_{x^0=\const} \tope{T}_{0\mu}(x) \Id^3\bs x .
	\end{equation}
Since this operator is Hermitian, $\tope{P}_\mu^\dag=\tope{P}_\mu$, the
operator
	\begin{equation}	\label{12.112}
\ope{U}(x,x_0)
 =
\exp\Bigl( \iih \sum_\mu (x^\mu-x_0^\mu)\tope{P}_{\mu}  \Bigr) ,
	\end{equation}
where $x_0\in\base$ is arbitrarily fixed and $x\in\base$,%
\footnote{~%
The notation $x_0$, for a fixed point in $\base$, should not be confused with
the zeroth covariant coordinate $\eta_{0\mu}x^\mu$ of $x$ which, following
the convention $x_\nu:=\eta_{\nu\mu}x^\mu$, is denoted by the same symbol
$x_0$. From the context, it will always be clear whether $x_0$ refers to a
point in $\base$ or to the zeroth covariant coordinate of a point
$x\in\base$.%
}
is unitary, \ie
\(
\ope{U}^\dag(x_0,x)
:= (\ope{U}(x,x_0))^\dag
 = \ope{U}^{-1}(x,x_0)
:= (\ope{U}(x,x_0))^{-1}
\)
and, via the formulae
	\begin{align}	\label{12.113}
\tope{X}\mapsto \ope{X}(x)
	&= \ope{U}(x,x_0) (\tope{X})
\\			\label{12.114}
\tope{A}(x)\mapsto \ope{A}(x)
	&= \ope{U}(x,x_0)\circ (\tope{A}(x)) \circ \ope{U}^{-1}(x,x_0) ,
	\end{align}
realizes the transition to the \emph{momentum picture}. Here $\tope{X}$ is a
state vector in system's Hilbert space of states $\Hil$ and
$\tope{A}(x)\colon\Hil\to\Hil$ is (observable or not) operator\ndash valued
function of $x\in\base$ which, in particular, can be polynomial or convergent
power series in the field operators $\tope{\varphi}_i(x)$; respectively
$\ope{X}(x)$ and $\ope{A}(x)$ are the corresponding quantities in momentum
picture.
	In particular, the field operators transform as
	\begin{align}	\label{12.115}
\tope{\varphi}_i(x)\mapsto \ope{\varphi}_i(x)
     = \ope{U}(x,x_0)\circ \tope{\varphi}_i(x) \circ \ope{U}^{-1}(x,x_0) .
	\end{align}
	Notice, in~\eref{12.112} the multiplier $(x^\mu-x_0^\mu)$ is regarded
as a real parameter (in which $\tope{P}_\mu$ is linear). Generally,
$\ope{X}(x)$ and $\ope{A}(x)$ depend also on the point $x_0$ and, to be
quite correct, one should write $\ope{X}(x,x_0)$ and $\ope{A}(x,x_0)$ for
$\ope{X}(x)$ and $\ope{A}(x)$, respectively. However, in the most situations
in the present work, this dependence is not essential or, in fact, is not
presented at all. For that reason, we shall \emph{not} indicate it explicitly.

	As it was said above, we consider quantum field theories in which the
components $\tope{P}_\mu$ of the momentum operator commute between themselves
and satisfy the Heisenberg relations/equations with the field operators, \ie
we suppose that $\tope{P}_\mu$ and $\tope{\varphi}_i(x)$ satisfy the
relations:
	\begin{align}	\label{2.1}
& [\tope{P}_\mu, \tope{P}_\nu ]_{\_} = 0
\\			\label{2.28}
& [\tope{\varphi}_i(x), \tope{P}_\mu]_{\_} = \ih\pd_\mu \tope{\varphi}_i(x).
	\end{align}
Here $[A,B]_{\pm}:=A\circ B \pm B\circ A$, $\circ$ being the composition of
mappings sign, is the commutator/anticommutator of operators (or matrices)
$A$ and $B$. The momentum operator $\tope{P}_\mu$ commutes with the
`evolution' operator $\ope{U}(x,x_0)$ (see below~\eref{12.118}) and its
inverse,
	\begin{equation}	\label{2.2}
	[ \tope{P}_\mu, \ope{U}(x,x_0) ]_{\_} =0
\qquad
	[ \tope{P}_\mu, \ope{U}^{-1}(x,x_0) ]_{\_} =0 ,
	\end{equation}
due to ~\eref{2.1} and~\eref{12.112}. So, the momentum operator remains
unchanged in momentum picture, \viz we have (see~\eref{12.114}
and~\eref{2.2})
	\begin{equation}	\label{2.3}
	 \ope{P}_\mu = \tope{P}_\mu.
	\end{equation}

	Since from~\eref{12.112} and~\eref{2.1} follows
	\begin{equation}	\label{12.116}
\ih \frac{\pd\ope{U}(x,x_0)}{\pd x^\mu}
=
\ope{P}_{\mu} \circ \ope{U}(x,x_0)
\qquad
\ope{U}(x_0,x_0) = \id_\Hil ,
	\end{equation}
we see that, due to~\eref{12.113}, a state vector $\ope{X}(x)$ in
momentum picture is a solution of the initial\ndash value problem
	\begin{equation}	\label{12.117}
\ih \frac{\pd\ope{X}(x)}{\pd x^\mu}
=
\ope{P}_{\mu}  (\ope{X}(x))
\qquad
\ope{X}(x)|_{x=x_0}=\ope{X}(x_0) = \tope{X}
	\end{equation}
which is a 4-dimensional analogue of a similar problem for the Schr{\"o}dinger
equation in quantum mechanics~\cite{Messiah-QM,Dirac-PQM,Prugovecki-QMinHS}.

	By virtue of~\eref{12.112}, or in view of the independence of
$\ope{P}_{\mu} $ of $x$, the solution of~\eref{12.117} is
	\begin{equation}	\label{12.118}
\ope{X}(x)
= \ope{U}(x,x_0) (\ope{X}(x_0))
= \e^{\iih(x^\mu-x_0^\mu)\ope{P}_{\mu} } (\ope{X}(x_0)).
	\end{equation}
Thus, if $\ope{X}(x_0)=\tope{X}$  is an eigenvector of
$\ope{P}_{\mu} $ ($=\tope{P}_{\mu} $)
with eigenvalues $p_\mu$,
	\begin{equation}	\label{12.119}
\ope{P}_{\mu}  (\ope{X}(x_0)) = p_\mu \ope{X}(x_0)
\quad
( =p_\mu \tope{X} = \tope{P}_{\mu}  (\tope{X}) ) ,
	\end{equation}
we have the following \emph{explicit} form of the state vectors
	\begin{equation}	\label{12.120}
\ope{X}(x)
=
\e^{ \iih(x^\mu-x_0^\mu)p_\mu } (\ope{X}(x_0)).
	\end{equation}
It should clearly be understood, \emph{this is the general form of all state
vectors} as they are eigenvectors of all (commuting)
observables~\cite[p.~59]{Roman-QFT}, in particular, of the momentum operator.

	In momentum picture, all of the field operators happen to be
constant in spacetime, \ie
	\begin{equation}	\label{2.4}
\varphi_i(x)
= \ope{U}(x,x_0)\circ \tope{\varphi}_i(x) \circ \ope{U}^{-1}(x,x_0)
= \varphi_i(x_0)
= \tope{\varphi}_i(x_0)
=: \varphi_{(0)\, i} .
	\end{equation}
Evidently, a similar result is valid for any (observable or not such)
function of the field operators which is polynomial or convergent power series
in them and/or their first partial derivatives. However, if $\tope{A}(x)$ is an
arbitrary operator or depends on the field operators in a different way, then
the corresponding to it operator $\ope{A}(x)$ according to~\eref{12.114} is,
generally, not spacetime\ndash constant and depends on the both points $x$
and $x_0$. As a rules, if $\ope{A}(x)=\ope{A}(x,x_0)$ is independent of $x$,
we, usually, write $\ope{A}$ for $\ope{A}(x,x_0)$, omitting all arguments.

	It should be noted, the Heisenberg relations~\eref{2.28} in momentum
picture transform into the identities $\pd_\mu\varphi_i=0$ meaning that the
field operators $\varphi_i$ in momentum picture are spacetime constant
operators (see~\eref{2.4}). So, in momentum picture, the Heisenberg
relations~\eref{2.28} are incorporated in the constancy of the field
operators.

	Let $\tope{L}$ be the system's Lagrangian (in Heisenberg picture). It
is supposed to be polynomial or convergent power series in the field
operators and their first partial derivatives, \ie
$\tope{L}=\tope{L}(\varphi_i(x),\pd_\nu\varphi_i(x))$ with $\pd_\nu$ denoting
the partial derivative operator relative to the $\nu^{\mathrm{th}}$
coordinate $x^\nu$. In momentum picture it transforms into
	\begin{equation}	\label{2.5}
\ope{L} =\tope{L}(\varphi_i(x), y_{j\nu})
\qquad
y_{j\nu}=\iih[\varphi_j,\ope{P}_{\nu} ]_{\_}  ,
	\end{equation}
\ie in momentum picture one has simply to replace the field operators in
Heisenberg picture with their values at a fixed point $x_0$ and the partial
derivatives  $\pd_\nu\tope{\varphi}_j(x)$ in Heisenberg picture with the
above\ndash defined quantities $y_{j\nu}$.
	The (constant) field operators $\varphi_i$ satisfy the following
\emph{algebraic Euler\ndash Lagrange equations in momentum picture}:%
\footnote{~%
In~\eref{12.129} and similar expressions appearing further, the derivatives
of functions of operators with respect to operator arguments are calculated
in the same way as if the operators were ordinary (classical)
fields/functions, only the order of the arguments should not be changed.
This is a silently accepted practice in the
literature~\cite{Roman-QFT,Bjorken&Drell}. In the most cases such a procedure
is harmless, but it leads to the problem of non\ndash unique definitions of
the quantum analogues of the classical conserved quantities, like the
energy\ndash momentum and charge operators. For some details on this range of
problems in quantum field theory, see~\cite{bp-QFT-action-principle}. In
Sect.~\ref{Sect12}, we shall met a Lagrangian whose field equations are
\emph{not} the Euler\ndash Lagrange equations~\eref{12.129} obtained as just
described.%
}
	\begin{equation}	\label{12.129}
\Bigl\{
\frac{\pd\tope{L}(\varphi_j,y_{l\nu})} {\pd \varphi_i}
-
\iih
\Bigl[
\frac{\pd\tope{L}(\varphi_j,y_{l\nu})} {y_{i\mu}}
,
\ope{P}_{\mu}
\Bigr]_{\_}
\Bigr\}
\Big|_{ y_{j\nu}=\iih[\varphi_j,\ope{P}_{\nu} ]_{\_} }
= 0 .
	\end{equation}
Since $\ope{L}$ is supposed to be polynomial or convergent power series  in
its arguments, the equations~\eref{12.129}  are \emph{algebraic}, not
differential, ones. This result is a natural one in view of~\eref{2.4}.

	Suppose a quantum system under consideration possesses a charge (\eg
electric one) and angular momentum, described by respectively the current
operator $\tope{J}_\mu(x)$ and (total) angular momentum tensorial density
operator
	\begin{equation}	\label{2.9}
\tope{M}_{\mu\nu}^{\lambda} (x)
=
-\tope{M}_{\nu\mu}^{\lambda} (x)
=
  x_\mu \Sprindex[\tope{T}]{\nu}{\lambda}
- x_\nu \Sprindex[\tope{T}]{\mu}{\lambda}
+ \tope{S}_{\mu\nu}^{\lambda} (x)
	\end{equation}
with $x_\nu:=\eta_{\nu\mu}x^\mu$ and
 $\tope{S}_{\mu\nu}^{\lambda}(x)= - \tope{S}_{\nu\mu}^{\lambda}(x)$
being the spin angular momentum (density) operator. The (constant, time\ndash
independent) conserved quantities corresponding to them, the charge operator
$\tope{Q}$ and total angular momentum operator $\tope{M}_{\mu\nu}$,
respectively are
	\begin{gather}
			\label{2.10}
\tope{Q} := \frac{1}{c} \int\limits_{x^0=\const} \tope{J}_0(x) \Id^3\bs x
\\			\label{2.11}
\tope{M}_{\mu\nu} = \tope{L}_{\mu\nu}(x) + \tope{S}_{\mu\nu}(x) ,
	\end{gather}
where
	\begin{subequations}	\label{2.12}
      \begin{align}
      		\label{2.12a}
\tope{L}_{\mu\nu}(x)
& :=
\frac{1}{c} \int\limits_{x^0=\const}
\{
  x_\mu \Sprindex[\tope{T}]{\nu}{0}(x) - x_\nu \Sprindex[\tope{T}]{\mu}{0}(x)
\} \Id^3\bs x
\\			\label{2.12b}
\tope{S}_{\mu\nu}(x)
& := \frac{1}{c} \int\limits_{x^0=\const} \tope{S}_{\mu\nu}^0(x) \Id^3\bs x
      \end{align}
	\end{subequations}
are the orbital and spin, respectively, angular momentum operators (in
Heisenberg picture). Notice, we write $\tope{L}_{\mu\nu}(x)$ and
$\tope{S}_{\mu\nu}(x)$, but, as a result of~\eref{2.12}, these operators may
depend only on the zeroth (time) coordinate of $x\in\base$. When working in
momentum picture, in view of~\eref{12.114}, the following representations
turn to be useful:
      \begin{gather}
      		\label{2.6}
\ope{P}_\mu = \tope{P}_\mu
=
\frac{1}{c} \int\limits_{x^0=\const}
	\ope{U}^{-1}(x,x_0)\circ \ope{T}_{0\mu} \circ \ope{U}(x,x_0)
	\Id^3\bs{x}
\displaybreak[1]\\      		\label{2.13}
\tope{Q}
=
\frac{1}{c} \int\limits_{x^0=\const}
	\ope{U}^{-1}(x,x_0)\circ \ope{J}_{0} \circ \ope{U}(x,x_0)
	\Id^3\bs{x}
\displaybreak[1]\\      		\label{2.14}
\tope{L}_{\mu\nu} (x)
=
\frac{1}{c} \int\limits_{x^0=\const}
	\ope{U}^{-1}(x,x_0)\circ \{
  x_\mu \Sprindex[\ope{T}]{\nu}{0} - x_\nu \Sprindex[\ope{T}]{\mu}{0}
\} \circ \ope{U}(x,x_0)
	\Id^3\bs{x}
\displaybreak[1]\\      		\label{2.15}
\tope{S}_{\mu\nu} (x)
=
\frac{1}{c} \int\limits_{x^0=\const}
	\ope{U}^{-1}(x,x_0)\circ \ope{S}_{\mu\nu}^0 \circ \ope{U}(x,x_0)
	\Id^3\bs{x} .
      \end{gather}
These expressions will be employed essentially in the present paper.

	The conservation laws $\frac{\od\tope{Q}}{\od x^0}=0$ and
$\frac{\od\tope{M}_{\mu\nu}}{\od x^0}=0$ (or, equivalently,
 $\pd^\mu\tope{J}_\mu=0$ and $\pd_\lambda\tope{M}_{\mu\nu}^{\lambda}=0$),
can be rewritten as
	\begin{equation}	\label{2.16}
\pd_\mu\tope{Q} = 0
\qquad
\pd_\lambda\tope{M}_{\mu\nu} = 0
	\end{equation}
since~\eref{2.10}--\eref{2.12} imply
$\pd_a\tope{Q} = 0$ and $\pd_a\tope{M}_{\mu\nu} = 0$ for $a=1,2,3$.

	As a result of the skewsymmetry of the operators~\eref{2.11}
and~\eref{2.12} in the subscripts $\mu$ and $\nu$, their spacial components
form a (pseudo\ndash)vectorial operators. If $e^{abc}$, $a,b,c=1,2,3$,
denotes the 3\ndash dimensional Levi\ndash Civita (totally) antisymmetric
symbol, we put $\tope{\bs M}:=(\tope{\bs M}^1,\tope{\bs M}^2,\tope{\bs M}^3)$
with $\tope{\bs M}^a:=e^{abc}\tope{M}_{bc}$ and similarly for the orbital and
spin angular momentum operators. Then~\eref{2.11} and the below written
equation~\eref{2.25} imply
	\begin{align}	\label{2.31}
& \tope{\bs M} = \tope{\bs L}(x) + \tope{\bs S}(x)
\\			\label{2.32}
& \ope{\bs M}(x,x_0)
= \tope{\bs L}(x) + (\bs x -\bs x_0)\times \ope{\bs P} + \tope{\bs S}(x),
	\end{align}
where $\bs x:=(x^1,x^2,x^3)=-(x_1,x_2,x_3)$, $\times$ denotes the Euclidean
cross product, and
\(
\ope{\bs P}
:= (\ope{P}^1,\ope{P}^2,\ope{P}^3)=-(\ope{P}_1,\ope{P}_2,\ope{P}_3) .
\)
Obviously, the correction in~\eref{2.32} to $\tope{\bs M}$ can be interpreted
as a one due to an additional orbital angular momentum when the origin, with
respect to which it is determined, is change from $x$ to $x_0$.

	The consideration of $\tope{Q}$ and$\tope{M}_{\mu\nu}$ as generators
of constant phase transformations and 4\ndash rotations, respectively, leads
to the following
relations~\cite{Bjorken&Drell,Bogolyubov&Shirkov,Itzykson&Zuber}
	\begin{align}	\label{2.17}
& [\tope{\varphi}_i(x), \tope{Q}]_{\_}
	= \varepsilon(\tope{\varphi}_i) q_i \tope{\varphi}_i(x)
\\			\label{2.18}
& [\tope{\varphi}_i(x), \tope{M}_{\mu\nu}]_{\_}
=
\ih\{
x_\mu\pd_\nu\tope{\varphi}_i(x) - x_\nu\pd_\mu\tope{\varphi}_i(x)
+ I_{i\mu\nu}^{j} \tope{\varphi}_j(x)
\} .
	\end{align}
Here: $q_i=\const$ is the charge of the $i^\text{th}$ field,
 $q_j=q_i$ if $\tope{\varphi}_j=\tope{\varphi}_i^\dag$,
$\varepsilon(\tope{\varphi}_i) = 0$ if
		$\tope{\varphi}_i^\dag = \tope{\varphi}_i$,
$\varepsilon(\tope{\varphi}_i) = \pm 1$ if
		$\tope{\varphi}_i^\dag \not= \tope{\varphi}_i$
with
$\varepsilon(\tope{\varphi}_i) + \varepsilon(\tope{\varphi}_i^\dag) = 0$,
and the constants $I_{i\mu\nu}^{j} = -I_{i\nu\mu}^{j}$ characterize the
transformation properties of the field operators under 4\ndash rotations.
(If $\varepsilon(\tope{\varphi}_i)\not=0$, it is a convention whether to put
$\varepsilon(\tope{\varphi}_i)=+1$ or $\varepsilon(\tope{\varphi}_i)=-1$ for a
fixed $i$.)
	Besides, the operators~\eref{2.10}--\eref{2.12} are Hermitian,
	\begin{equation}	\label{2.19}
\tope{Q}^\dag = \tope{Q}, \quad
\tope{M}_{\mu\nu}^\dag = \tope{M}_{\mu\nu}, \quad
\tope{L}_{\mu\nu}^\dag = \tope{L}_{\mu\nu}, \quad
\tope{S}_{\mu\nu}^\dag = \tope{S}_{\mu\nu}, \quad
	\end{equation}
and satisfy the relations%
\footnote{~%
The author is completely aware of the fact that in the literature, for
instance in~\cite[p.~77, eq.~(2-87)]{Roman-QFT} or
in~\cite[eq.~(2.187)]{Ryder-QFT}, the relation~\eref{2.21} is written with an
opposite sign, \ie with $+\ih$ instead of $-\ih$ on its r.h.s. (In this
case~\eref{2.21} is part of the commutation relations characterizing the Lie
algebra of the Poincar\'e group --- see,
e.g.,~\cite[pp.~143--147]{Bogolyubov&et_al.-AxQFT}
or~\cite[sect.~7.1]{Bogolyubov&et_al.-QFT}.) However, such a choice of the
sign in~\eref{2.21} contradicts to the
explicit form of $\tope{P}_\mu$ and $\tope{L}_{\mu\nu}$ in terms of creation
and annihilation operators (see sections~\ref{Sect7} and~\ref{Sect8}) in the
framework of Lagrangian formalism. For this reason and since the
relation~\eref{2.21} is external to the Lagrangian formalism, we
accept~\eref{2.21} as it is written below. In connection with~\eref{2.21} ---
see below equation~\eref{8.15} and the paragraph containing it.%
}

	\begin{align}	\label{2.20}
& [\tope{Q}, \tope{P}_\mu]_{\_} = 0
\\			\label{2.21}
& [\tope{M}_{\mu\nu}, \tope{P}_\lambda]_{\_}
= 
- \ih\{ \eta_{\lambda\mu}\tope{P}_\nu  - \eta_{\lambda\nu}\tope{P}_\mu \} .
	\end{align}
Combining the last two equalities with~\eref{12.112} and~\eref{2.1}, we,
after a simple algebraic calculations, obtain%
\footnote{~\label{CommutativityWithMomentum}%
To derive equation~\eref{2.23}, notice that~\eref{2.21} implies
\(
[\tope{M}_{\mu\nu}, \tope{P}_{\mu_1}\circ\dots\circ\tope{P}_{\mu_n} ]_{\_}
=
- \sum_{i=1}^{n}
\bigl( \eta_{\mu\mu_i} \tope{P}_{\nu} - \eta_{\nu\mu_i} \tope{P}_{\mu} \bigr)
\tope{P}_{\mu_1}\circ\dots\circ
\tope{P}_{\mu_{i-1}}\circ\tope{P}_{\mu_{i+1}} \circ \dots \circ
\tope{P}_{\mu_n},
\)
due to $[A,B\circ C]_{\_} = [A,B]_{\_}\circ C+ B\circ [A,C]_{\_}$,
and expand the exponent in~\eref{12.112} into a power series.
More generally, if $[A(x),\tope{P}_\mu]_{\_}=B_\mu(x)$ with
$[B_\mu(x),\tope{P}_\nu]_{\_}=0$, then
\(
[A(x),\ope{U}(x,x_0)]_{\_}
= \iih (x^\mu-x_0^\mu)B_\mu(x) \circ \ope{U}(x,x_0);
\)
in particular, $[A(x),\tope{P}_\mu]_{\_}=0$ implies
 $[A(x),\ope{U}(x,x_0)]_{\_}=0$. Notice, we consider $(x^\mu-x_0^\mu)$ as a
real parameter by which the corresponding operators are multiplied and which
operators are supposed to be linear in it.%
}
	\begin{align}	\label{2.22}
& [\tope{Q}, \ope{U}(x,x_0)]_{\_} = 0
\\			\label{2.23}
& [\tope{M}_{\mu\nu}, \ope{U}(x,x_0)]_{\_}
=
- \{ (x_\mu-x_{0\,\mu}) \tope{P}_\nu
   - (x_\nu-x_{0\,\nu}) \tope{P}_\mu \} \circ \ope{U}(x,x_0) .
	\end{align}
Consequently, in accord with~\eref{12.114}, in momentum picture the charge and
angular momentum operators respectively are
	\begin{align}	\label{2.24}
\ope{Q}(x)
& = \tope{Q} := \ope{Q}
\\ \notag
\ope{M}_{\mu\nu}
& = \ope{U}(x,x_0) \circ \tope{M}_{\mu\nu} \circ \ope{U}^{-1}(x,x_0)
  = \tope{M}_{\mu\nu} + [\ope{U}(x,x_0),\tope{M}_{\mu\nu}]_{\_} \circ
  						\ope{U}^{-1}(x,x_0)
\\ \notag
& =
\tope{M}_{\mu\nu}
+ (x_\mu-x_{0\,\mu}) \ope{P}_\nu - (x_\nu-x_{0\,\nu}) \ope{P}_\mu
\\			\label{2.25}
& =
\tope{L}_{\mu\nu}
+ (x_\mu-x_{0\,\mu}) \ope{P}_\nu - (x_\nu-x_{0\,\nu}) \ope{P}_\mu
+ \tope{S}_{\mu\nu}
= \ope{L}_{\mu\nu} + \ope{S}_{\mu\nu} ,
	\end{align}
where
	\begin{equation}	\label{2.25new}
	\begin{split}
\tope{L}_{\mu\nu}(x)
  := \ope{U}(x,x_0) \circ\ope{L}_{\mu\nu}(x)\circ \ope{U}^{-1}(x,x_0)
\\
\tope{S}_{\mu\nu}(x)
  := \ope{U}(x,x_0) \circ\ope{S}_{\mu\nu}(x)\circ \ope{U}^{-1}(x,x_0)
	\end{split}
	\end{equation}
and~\eref{2.3} was taken into account. Notice, the correction to
$\tope{M}_{\mu\nu}$ on the r.h.s.\ of~\eref{2.25} is typical for the one of
classical orbital angular momentum when the origin, with respect to which it
is determined, is changed from $x$ to $x_0$.%
\footnote{~
In Sections~\ref{Sect7} and~\ref{Sect8}, for a free spinor field, it will be
proved that
 \(
\tope{S}_{\mu\nu}
= \lindex[\mspace{-6mu}{\tope{S}_{\mu\nu}}]{}{0}
+ \lindex[\mspace{-6mu}{\tope{S}_{\mu\nu}}]{}{1}
 \),
 \(
\tope{L}_{\mu\nu}
= \lindex[\mspace{-6mu}{\tope{L}_{\mu\nu}}]{}{0}
+ \lindex[\mspace{-6mu}{\tope{L}_{\mu\nu}}]{}{1}
 \),
where $\lindex[\mspace{-6mu}{\tope{S}_{\mu\nu}}]{}{0}$ and
$\lindex[\mspace{-6mu}{\tope{L}_{\mu\nu}}]{}{0}$ are such that
 \(
\pd_\lambda\lindex[\mspace{-6mu}{\tope{S}_{\mu\nu}}]{}{0}
=
\pd_\lambda\lindex[\mspace{-6mu}{\tope{L}_{\mu\nu}}]{}{0}
= 0
 \),
\(
\tope{M}_{\mu\nu}
=
\lindex[\mspace{-6mu}{\tope{S}_{\mu\nu}}]{}{0}
+
\lindex[\mspace{-6mu}{\tope{L}_{\mu\nu}}]{}{0}
\)
and
	\begin{align}	\label{2.25-1}
& [\lindex[\mspace{-6mu}{\tope{S}_{\mu\nu}}]{}{0} , \ope{P}_\lambda]_{\_} = 0,
\intertext{which implies}	\label{2.25-2}
& [\lindex[\mspace{-6mu}{\tope{S}_{\mu\nu}}]{}{0} , \ope{U}(x,s_0)]_{\_} = 0.
\intertext{Amongst other things, from here follow the equations}
		\label{2.25-3}
& \lindex[\mspace{-6mu}{\ope{S}_{\mu\nu}}]{}{0}
	= \lindex[\mspace{-6mu}{\tope{S}_{\mu\nu}}]{}{0}
\\					\label{2.25-4}
& \lindex[\mspace{-6mu}{\ope{L}_{\mu\nu}}]{}{0}
= \lindex[\mspace{-6mu}{\tope{L}_{\mu\nu}}]{}{0}
   + (x_\mu-x_{0\,\mu}) \ope{P}_\nu - (x_\nu-x_{0\,\nu}) \ope{P}_\mu .
	\end{align}%
}

	In momentum picture, by virtue of~\eref{12.114}, the
relations~\eref{2.17} and~\eref{2.18} respectively read
($ \varepsilon(\varphi) := \varepsilon(\tope{\varphi}) $)
	\begin{align}	\label{2.26}
& [\ope{\varphi}_i, \ope{Q}]_{\_}
	= \varepsilon(\ope{\varphi}_i) q \ope{\varphi}_i
\\			\label{2.27}
& [\ope{\varphi}_i, \ope{M}_{\mu\nu}(x,x_0)]_{\_}
=
x_\mu [\varphi_i ,\ope{P}_\nu]_{\_} - x_\nu [\varphi_i ,\ope{P}_\mu]_{\_}
+ \ih I_{i\mu\nu}^{j} \varphi_j .
	\end{align}
The first of these equation is evident. To derive the second one, we notice
that, by virtue of the Heisenberg relations/equations~\eref{2.28},
the equality~\eref{2.18} is equivalent to
	\begin{align}
			\label{2.29}
& [\tope{\varphi}_i(x), \tope{M}_{\mu\nu}]_{\_}
=
  x_\mu [\tope{\varphi}_i(x) ,\tope{P}_\nu]_{\_}
- x_\nu [\tope{\varphi}_i(x) ,\tope{P}_\mu]_{\_}
+ \ih I_{i\mu\nu}^{j} \tope{\varphi}_j(x)
	\end{align}
from where~\eref{2.27} follows.

	It should be emphasized, the Heisenberg relations~\eref{2.17}
and~\eref{2.18}, as well as the commutation relations~\eref{2.20}
and~\eref{2.21}, are external to the Lagrangian formalism. For this reason,
one should be quite careful when applying them unless they are explicitly
proved in the framework of Lagrangian scheme.


\section
[Description of free spinor field in momentum picture]
{Description of free spinor field in momentum picture}
\label{Sect3}

	A spinor field of spin $\frac{1}{2}$ is described by four operators
$\tope{\psi}_\mu$ which are collected in a matrix operator $\tope{\psi}$
considered as a 4\ndash component column, \ie
\(
\tope{\psi}(x)
:= \bigl(
\tope{\psi}_0(x),\tope{\psi}_1(x),\tope{\psi}_2(x),\tope{\psi}_3(x)
\bigr)^\top,
\)
where $\top$ is the sign of matrix transposition. In Heisenberg picture, the
theory of free spinor field is derived from the
Lagrangian~\cite{Bogolyubov&Shirkov}%
\footnote{~%
The theory of free spinor field can be derived also from the Lagrangian
\(
L =
\ih c \tope{\opsi}\gamma^\mu\circ\pd_\mu\tope{\psi}
- mc^2\tope{\opsi} \circ\tope{\psi} ,
\)
as it is done in~\cite{Roman-QFT,Bjorken&Drell-2}. The problem of a selection
of Lagrangian for free spinor field will be considered in Sect.~\ref{Sect12}.%
}
	\begin{equation}	\label{3.1}
\tope{L}
=
\frac{1}{2}\ih c\{
  \tope{\opsi}(x)\gamma^\mu\circ(\pd_\mu\tope{\psi}(x))
- (\pd_\mu\tope{\opsi}(x))\gamma^\mu\circ \tope{\psi}(x)
\}
- mc^2 \tope{\opsi}(x)\circ \tope{\psi}(x) .
	\end{equation}
Here: $\gamma^\mu$ are the Dirac's matrices (see equations~\eref{3.1-1},
\eref{3.15}, and~\eref{3.24} below), $\circ$ is the composition of
mappings/operators sign, and $\tope{\opsi}:=\tope{\psi}^\dag\gamma^0$, with
\( \tope{\psi}^\dag := \bigl(
\tope{\psi}_0^\dag,\tope{\psi}_1^\dag,\tope{\psi}_2^\dag,\tope{\psi}_3^\dag
\bigr) ,
\)
is the Dirac conjugate spinor to $\tope{\psi}$. Besides, in expressions like
$\tope{\opsi}\circ\tope{\psi}$ and  $\tope{\psi}^\dag\gamma^0$  a matrix
multiplication is understood, i.e., in these examples, we have
$\tope{\opsi}\circ\tope{\psi}:=\sum_{\mu}\tope{\opsi}_\mu\circ\tope{\psi}_\mu$
and
\(
\tope{\psi}^\dag\gamma^0
:=\Bigl(
\sum_{\mu} \tope{\psi}^\dag_\mu\gamma^0_{\mu0},\dots,
\sum_{\mu} \tope{\psi}^\dag_\mu\gamma^0_{\mu3}
\Bigr).
\)
	As a particular realization of the Dirac's gamma matrices, we shall
use the following one~\cite{Bogolyubov&Shirkov,Bjorken&Drell}:
	\begin{equation}	\label{3.1-1}
	\begin{split}
\gamma^0 & =
	\begin{pmatrix}
1 & 0 & 0 & 0 \\
0 & 1 & 0 & 0 \\
1 & 0 &-1 & 0 \\
1 & 0 & 0 &-1
	\end{pmatrix}
\quad
\gamma^1 =
	\begin{pmatrix}
0 & 0 & 0 & 1 \\
0 & 0 & 1 & 0 \\
0 &-1 & 0 & 0 \\
-1& 0 & 0 & 0
	\end{pmatrix}
\\
\gamma^2 &=
	\begin{pmatrix}
0 & 0 & 0 & -\iu \\
0 & 0 &\iu& 0 \\
0 &\iu& 0 & 0 \\
-\iu& 0 & 0 & 0
	\end{pmatrix}
\quad\
\gamma^3 =
	\begin{pmatrix}
0 & 0 & 1 & 0 \\
0 & 0 & 0 &-1 \\
-1& 0 & 0 & 0 \\
 0& 1 & 0 & 0
	\end{pmatrix} \ .
	\end{split}
	\end{equation}

	In momentum picture, in view of~\eref{12.115}, the spinor's
components and their Dirac conjugate transform into
	\begin{equation}	\label{3.2}
	\begin{split}
\psi_\mu(x) = \ope{U}(x,x_0)\circ\tope{\psi}_\mu(x)\circ \ope{U}^{-1}(x,x_0)
=
\tope{\psi}_\mu(x_0) = \ope{\psi}_\mu(x_0) =:\psi_\mu
\\
\opsi_\mu(x) = \ope{U}(x,x_0)\circ\tope{\opsi}_\mu(x)\circ \ope{U}^{-1}(x,x_0)
=
\tope{\opsi}_\mu(x_0) = \ope{\opsi}_\mu(x_0) =:\opsi_\mu .
	\end{split}
	\end{equation}
If we identify $\ope{U}(x,x_0)$ with $\ope{U}(x,x_0)\openone_4$, where
$\openone_4:=\diag(1,1,1,1,)$ is the unit (identity) $4\times4$ matrix, the
last equalities can be rewritten as
	\begin{equation}\tag{\ref{3.2}$^\prime$}	\label{3.2'}
	\begin{split}
\psi(x) = \ope{U}(x,x_0)\circ\tope{\psi}(x)\circ \ope{U}^{-1}(x,x_0)
=
\tope{\psi}(x_0) = \ope{\psi}(x_0) =:\psi
\\
\opsi(x) = \ope{U}(x,x_0)\circ\tope{\opsi}(x)\circ \ope{U}^{-1}(x,x_0)
=
\tope{\opsi}(x_0) = \ope{\opsi}(x_0) =:\opsi .
	\end{split}
	\end{equation}

	Regardless of the explicit dependence of $\psi$ and $\opsi$ on the
point $x_0\in\base$, further it will not be indicated as insignificant for
our present work. By virtue of~\eref{12.129}, the Lagrangian~\eref{3.1} in
momentum picture reads
	\begin{gather}	\label{3.3}
\ope{L} = \tope{L}(\psi,\opsi,y_\mu,\overline{y}_\mu)
=
\frac{1}{2} c \{
\opsi\gamma^\mu\circ[\psi,\ope{P}_\mu]_{\_}
- [\opsi,\ope{P}_\mu]_{\_}\gamma^\mu\circ \psi
\}
-m c^2\opsi\circ\psi.
\intertext{where}
			\notag
	\begin{split}
y_\mu:= \iih  [\psi,\ope{P}_\mu]_{\_}
:= \iih
\bigl( [\psi_0,\ope{P}_\mu]_{\_}, \dots, [\psi_3,\ope{P}_\mu]_{\_}\bigr)^\top
\\
\overline{y}_\mu:= \iih  [\opsi,\ope{P}_\mu]_{\_}
:= \iih
\bigl( [\opsi_0,\ope{P}_\mu]_{\_}, \dots, [\opsi_3,\ope{P}_\mu]_{\_}\bigr) .
	\end{split}
	 \end{gather}
(Notice, to simplify the notation, here, as we did above with
$\ope{U}(x,x_0)$, we identify $\ope{P}_\mu$ with the matrix operator
$\ope{P}_\mu\openone_4$. For the same reason, we identify $\gamma^\mu$ with
the matrix operator $\gamma^\mu\id_\Hil$, $\Hil$ being the system's Hilbert
space of states.) Therefore, we have:~%
\footnote{~%
As pointed at the end of the Introduction, the calculation of the derivatives
in~\eref{3.4} is not quite correct mathematically. However, the field
equations~\eref{3.5} and the formulae~\eref{3.10}--\eref{3.13} below are
correct. For their rigorous derivation, see~\cite{bp-QFT-action-principle}.%
}
	\begin{equation}	\label{3.4}
	\begin{split}
\frac{\pd\ope{L}}{\pd\psi} = \frac{\pd\tope{L}}{\pd\psi}
& =
-\frac{1}{2}c [\opsi,\ope{P}_\mu]_{\_} \gamma^\mu - m c^2\opsi
\qquad
\pi^\mu = \frac{\pd\ope{L}}{\pd y_\mu} = +\frac{1}{2} \ih c \opsi \gamma^\mu
\\
\frac{\pd\ope{L}}{\pd\opsi} = \frac{\pd\tope{L}}{\pd\opsi}
& =
+\frac{1}{2}c \gamma^\mu [\psi,\ope{P}_\mu]_{\_} - m c^2\psi
\qquad
\overline{\pi}^\mu
=
\frac{\pd\ope{L}}{\pd \overline{y}_\mu}= - \frac{1}{2} \ih c \gamma^\mu \psi.
	\end{split}
	\end{equation}
Consequently, the equations of motion~\eref{12.129} now read:
	\begin{gather}	\label{3.5}
\gamma^\mu [\psi,\ope{P}_\mu]_{\_} - m c\psi = 0
\qquad
[\opsi,\ope{P}_\mu]_{\_} \gamma^\mu + m c\opsi = 0
\intertext{where we have applied the equality}
			\label{3.6}
[\gamma^\mu,\ope{P}_\mu]_{\_} = 0
	\end{gather}
as
\(
  [\gamma^\mu,\ope{P}_\nu]_{\_}
= [\gamma^\mu\id_\Hil,\ope{P}_\nu\openone_4]_{\_}
= [ [\gamma_{\lambda\kappa}^\mu\id_\Hil,\ope{P}_\nu]_{\_}
  ]_{\lambda,\kappa=0}^{3}
\equiv  0
\)
since $\gamma_{\lambda\kappa}^\mu$ are constant complex numbers.
The first of the equations~\eref{3.5} is the \emph{Dirac equation in momentum
picture} and the second one is its Dirac conjugate. They correspond to the
famous equations
	\begin{equation}	\label{3.7}
\ih  \gamma^\mu \pd_\mu\tope{\psi} - mc \tope{\psi} = 0
\qquad
\ih \pd_\mu\tope{\opsi} \gamma^\mu  + mc \tope{\opsi} = 0
	\end{equation}
in Heisenberg picture. It is worth noting, \emph{the equations~\eref{3.5} are
valid in any picture of motion} which cannot be said with respect
to~\eref{3.7}. Indeed, since the transitions to an arbitrary picture of
motion is achieved via a unitary operator
$\ope{U}\colon\Hil\to\Hil$~\cite{bp-QFT-pictures} and
	\begin{equation}	\label{3.8}
[\gamma^\mu, \ope{U}]_{\_} = 0
	\end{equation}
(see the proof of~\eref{3.6} above), this statement is a consequence of the
algebraic structure of~\eref{3.5}.

	We shall comment on the choice of the Lagrangians~\eref{3.1}
and~\eref{3.3} in Sect.~\ref{Sect12}.

	We emphasize, the fields $\psi$ and $\opsi$ will be treated as
independent field variables in this work. However, because of the connection
$\opsi=\psi^\dag\gamma^0$, most of the relations regarding $\opsi$ can be
obtained from the ones concerning $\psi$ by Dirac conjugation, \ie via the
change $(\cdots)\mapsto\overline{(\cdots)}:=(\cdots)^\dag\gamma^0$, where the
dots stand for any spinor-matrix operator expression.

	It is well known, a spinor filed possesses energy-momentum, electric
charge and angular momentum. In Heisenberg picture, the operators of these
conserved operator quantities respectively are:%
\footnote{~%
See, for instance,~\cite{Bogolyubov&Shirkov} or~\cite{Itzykson&Zuber}.
However, the order of the operators in the compositions below is \emph{not
proved} in these books, there are only some arguments justifying the made
choice. For a rigorous proof --- see~\cite{bp-QFT-action-principle}.%
}
	\begin{align}
			\label{3.10}
& \tope{T}_{\mu\nu}
=
  \tope{\pi}_{\mu}\circ(\pd_\nu\tope{\psi})
+ (\pd_\nu\tope{\opsi})\circ\tope{\overline{\pi}}_{\mu}
=
\frac{1}{2}\ih c\{
  \tope{\opsi}\gamma_\mu\circ (\pd_\nu\tope{\psi})
- (\pd_\nu\tope{\opsi})\gamma_\mu\circ \tope{\psi} \}
\\			\label{3.11}
& \tope{J}_\mu
=
\frac{q}{\ih}\circ \{ \tope{\pi}_\mu\circ \tope{\psi}
- \tope{\opsi}\circ \tope{\overline{\pi}}_\mu \}
=
q c \tope{\opsi}\gamma_\mu \tope{\psi}
\\			\label{3.12}
& \tope{M}_{\mu\nu}^{\lambda}
=
  x_\mu \Sprindex[\tope{T}]{\nu}{\lambda}
- x_\nu \Sprindex[\tope{T}]{\mu}{\lambda}
+ \tope{S}_{\mu\nu}^{\lambda}
	\end{align}
where the spin angular momentum operator density
$\tope{S}_{\mu\nu}^{\lambda}$ is%
\footnote{~%
We adopt the definition of spin angular momentum operator density
from~\cite[\S~78, eq.~(13.47)]{Bjorken&Drell}
or~\cite[sect.~3.3.1, eq.~(3.115)]{Itzykson&Zuber}. It agrees with the
commutation relation~\eref{2.21}.
In~\cite[eqs.~(2.16) and~(7.31)]{Bogolyubov&Shirkov} is used a definition
with an opposite sign.%
}
       \begin{align}
\tope{S}_{\mu\nu}^{\lambda}
 = + \sum_{j=1,2} \{
\tope{\pi}^{\lambda}\circ ( I_{\psi\mu\nu} \tope{\psi} )
+
( \tope{\opsi} I_{\opsi\mu\nu}) \circ \tope{\overline{\pi}}^{\lambda}
\}
			\label{3.13}
 =
 \frac{1}{4} \hbar c
\tope{\opsi}
\{ \gamma^\lambda \sigma_{\mu\nu} + \sigma_{\mu\nu} \gamma^\lambda \}
\circ \tope{\psi} ,
	\end{align}
with
	\begin{equation}	\label{3.13-1}
\sigma^{\mu\nu}
:=
\frac{\iu}{2}( \gamma^\mu\gamma^\nu - \gamma^\nu\gamma^\mu),
	\end{equation}
and the coefficients $I_{\psi\mu\nu}$ and $I_{\opsi\mu\nu}$,
characterizing the transformation properties of $\tope{\psi}$ and
$\tope{\opsi}$ under 4\ndash rotations,
are~\cite{Bogolyubov&Shirkov,Roman-QFT}.
	\begin{equation}	\label{3.13new}
I_{\psi\mu\nu}
=-\frac{1}{2}\iu \sigma_{\mu\nu}
\qquad
I_{\opsi\mu\nu}
=+\frac{1}{2}\iu  \sigma_{\mu\nu} .
	\end{equation}
Often the spacial components of $\tope{S}_{\mu\nu}^{0}$, defining the spin
angular momentum via~\eref{2.12b}, are combined into the (pseudo\ndash)vector
$\frac{1}{2}e^{abc}\tope{S}_{bc}^{0}$ with $a,b,c=1,2,3$. Equation~\eref{3.13}
implies
	\begin{gather}
			\label{3.14}
\tope{S}^a
=
\frac{1}{2} e^{abc}\tope{S}_{bc}^{0} (x)
=
+ \frac{1}{2}\hbar c \tope{\psi}^\dag(x) \sigma^a \circ\tope{\psi}(x)
=
- \frac{1}{2}\hbar c \tope{\psi}^\dag(x) \sigma_a \circ\tope{\psi}(x)
\intertext{where $\sigma^a:=e^{abc}\sigma_{bc}$ and the relation}
			\label{3.15}
\gamma^\mu\gamma^\nu + \gamma^\nu\gamma^\mu = 2\eta^{\mu\nu} \openone_4
\intertext{was applied. So, the vectorial spin operator
$\tope{\bs S}=(\tope{S}^1,\tope{S}^2,\tope{S}^3)$, appearing in~\eref{2.31}
and~\eref{2.32}, is}
			\label{3.16}
\tope{\bs S}(x)
=
\frac{1}{2}\hbar \int\limits_{x^0=\const}
\tope{\psi}^\dag(x)\bs{\sigma}\circ\tope{\psi}(x) \Id^3\bs x
\qquad
\bs{\sigma}:=(\sigma^1,\sigma^2,\sigma^3) .
	\end{gather}

	In accord with~\eref{12.114} and~\eref{2.4}, the dynamical
characteristics~\eref{3.10}--\eref{3.13} in momentum picture are
	\begin{align}
			\label{3.17}
& \ope{T}_{\mu\nu}
=
\frac{1}{2} c\{
  \ope{\opsi}\gamma_\mu\circ [\psi,\ope{P}_\nu]_{\_}
- [\opsi,\ope{P}_\nu]_{\_}\gamma_\mu  \circ \ope{\psi}   \}
\\			\label{3.18}
& \ope{J}_\mu
=
q c \ope{\opsi}\gamma_\mu \ope{\psi}
\\			\label{3.19}
&\ope{M}_{\mu\nu}^{\lambda}
=
  x_\mu \Sprindex[\ope{T}]{\nu}{\lambda}
- x_\nu \Sprindex[\ope{T}]{\mu}{\lambda}
+ \ope{S}_{\mu\nu}^{\lambda}
\\
			\label{3.20}
& \ope{S}_{\mu\nu}^{\lambda}
=
\frac{1}{4} \hbar c
\ope{\opsi}
\{ \gamma^\lambda \sigma_{\mu\nu} + \sigma_{\mu\nu} \gamma^\lambda \}
\circ \ope{\psi}
\\\intertext{In particular, the vector~\eref{3.14} takes the form}
			\label{3.21}
&
\ope{S}^a
=
e^{abc}\ope{S}_{bc}^{0}
=
  \frac{1}{2}\hbar c \ope{\psi}^\dag \sigma^a \circ\ope{\psi}
=
- \frac{1}{2}\hbar c \ope{\psi}^\dag \sigma_a \circ\ope{\psi}
	\end{align}
in momentum picture.

	To specify the relations~\eref{2.26} and~\eref{2.27} for a spinor
field, we, by convention, put $\varepsilon(\psi)=+1$ and
$\varepsilon(\opsi)=-1$ and get
	\begin{equation}	\label{3.22}
[\psi, \ope{Q}]_{\_} =  q \psi
\quad
[\opsi, \ope{Q}]_{\_} = - q \opsi
	\end{equation}
\vspace{-4.4ex}
	\begin{subequations}	\label{3.23}
	\begin{align}
			\label{3.23a}
& [\psi, \ope{M}_{\mu\nu}(x,x_0)]_{\_}
=
x_\mu [\psi ,\ope{P}_\nu]_{\_} - x_\nu [\psi ,\ope{P}_\mu]_{\_}
+ \frac{1}{2}\hbar \sigma_{\mu\nu}\psi
\\			\label{3.23b}
& [\opsi, \ope{M}_{\mu\nu}(x,x_0)]_{\_}
=
x_\mu [\opsi ,\ope{P}_\nu]_{\_} - x_\nu [\opsi ,\ope{P}_\mu]_{\_}
- \frac{1}{2}\hbar \opsi \sigma_{\mu\nu}
	\end{align}
	\end{subequations}
due to~\eref{3.13new}. The last terms in~\eref{3.23} have their origin in the
spin angular momentum and the remaining ones are due to the orbital angular
momentum.

	Besides~\eref{3.15}, the gamma matrices are suppose to satisfy the
conditions $(\gamma^0)^\dag=\gamma^0$ and $(\gamma^a)^\dag=-\gamma^a$ for
$a=1,2,3$, \ie (do not sum over $\mu$!)
	\begin{equation}	\label{3.24}
\gamma^{\mu\,\dag}
:= (\gamma^\mu)^\dag
 = \eta^{\mu\mu} \gamma^\mu
 = \gamma^0 \gamma^\mu \gamma^0 .
	\end{equation}
As a consequence of~\eref{3.13-1} and~\eref{3.24}, we have
(do not sum over $\mu$ and $\nu$!)
	\begin{equation}	\label{3.25}
(\sigma^{\mu\nu})^\dag
 = \eta^{\mu\mu} \eta^{\nu\nu} \sigma^{\mu\nu} .
	\end{equation}
Important  corollaries from~\eref{3.13-1} and~\eref{3.15} are the equalities
(do not sum over $\mu$ and $\nu$!)
	\begin{align}	\label{3.26}
[\sigma^{\mu\nu} , \gamma^\lambda ]_{\_}
& =
-2\iu ( \eta^{\lambda\mu}\gamma^\nu - \eta^{\lambda\nu}\gamma^\mu )
\\			\label{3.27}
\sigma^{\mu\nu} \gamma^0
& =
\eta^{\mu\mu} \eta^{\nu\nu} \gamma^0 \sigma^{\mu\nu}  ,
	\end{align}
the last of which implies%
\footnote{~%
The equations~\eref{3.27} and~\eref{3.28} are also corollaries from~\eref{3.26}
with $\lambda=0$ and
$\sigma^{0a}=\iu\gamma^0\gamma^a=-\iu\gamma^a\gamma^0$.%
}
	\begin{equation}	\label{3.28}
\gamma^0 \sigma_{0a} + \sigma_{0a} \gamma^0 = 0
\qquad
\gamma^0 \sigma_{ab} = \sigma_{ab} \gamma^0 .
	\end{equation}
Combining~\eref{3.25} and~\eref{3.27}, we get
	\begin{equation}	\label{3.28new}
(\sigma^{\mu\nu})^\dag = \gamma^0 \sigma^{\mu\nu} \gamma^0 .
	\end{equation}

	As result of~\eref{3.20} and~\eref{3.28}, we have
	\begin{equation}	\label{3.29}
\ope{S}_{0a}^{0} = 0
\qquad
\ope{S}_{ab}^{0}
=
\frac{1}{2} \hbar c
\ope{\opsi} \gamma^0 \sigma_{ab} \circ \ope{\psi} .
	\end{equation}
From here, we see that the vector $\ope{S}^a$ and the components
$\ope{S}_{\mu\nu}^{0}$ carry one and the same information. Notice, the
equations~\eref{3.29} are specific for the Lagrangian~\eref{3.1}; for
instance, the first of them does not hold for the Lagrangian~\eref{12.3} ---
see~\cite[eq.~(13.47)]{Bjorken&Drell-2}.

%

	According to the Dirac equation~\eref{3.5}, the operators
	\begin{equation}	\label{4.1}
\ope{M}\colon \psi\mapsto
		+ \frac{1}{c} \gamma^\mu [\psi,\ope{P}_\mu]_{\_}
\qquad
\overline{\ope{M}}\colon \opsi\mapsto
		- \frac{1}{c} [\opsi,\ope{P}_\mu]_{\_} \gamma^\mu,
	\end{equation}
when acting on the solutions of~\eref{3.5}, have the mass parameter $m$ of
the spinor field as eigenvalue. Applying the relations~\eref{3.15}, the
Jacobi identity and~\eref{2.1}, we see that
	\begin{equation}	\label{4.2}
\ope{M}^2 := \ope{M}\circ \ope{M}
=
\overline{\ope{M}}^2 := \overline{\ope{M}}\circ \overline{\ope{M}}
\colon \varphi\mapsto
		\frac{1}{c^2} [[\varphi,\ope{P}_\mu]_{\_}, \ope{P}_\mu]_{\_}
\qquad
\varphi = \psi,\opsi
	\end{equation}
has a meaning of a square-of-mass operator as its eigenvalues on the
solutions of~\eref{3.5} are equal to $m^2$. Therefore the operator
 $\frac{1}{c^2} [[(\cdot),\ope{P}_\mu]_{\_}, \ope{P}_\mu]_{\_}$,
rather than $\ope{P}_\mu\circ\ope{P}^\mu$, determines the square of the mass
(parameter) of a free spinor field. Obviously, this conclusion is valid in any
picture of motion. This situation is similar to the one with free scalar
fields~\cite{bp-QFTinMP-scalars}. It is well known, the operator
$\ope{P}_\mu\circ\ope{P}^\mu$ has a sense of a square-of-mass operator for
the field's states (on which it acts).


\section {Analysis of the Dirac equation(s)}
\label{Sect5}

	The Dirac equations~\eref{3.5}, together with the equality~\eref{2.6}
and the explicit expression~\eref{3.10} for the energy-momentum operator,
form a complete algebraic\ndash functional system of equations for
determination of the spinor field $\psi$ and its Dirac conjugate
$\opsi=\psi^\dag\gamma^0$ (and its momentum operator $\ope{P}_\mu$ too). This
situation is similar to the one concerning a free charged scalar field,
investigated in~\cite{bp-QFTinMP-scalars}; respectively similar results hold
for the structure of spinor field operators. Moreover, there is now a
simplification due to the \emph{linearity} of the Dirac equation relative to
the momentum operator, contrary to the scalar field case, when the
corresponding Klein\ndash Gordon equations are \emph{quadratic} in
momentum operator.%
\footnote{~%
In Heisenberg picture, the last statement is equivalent to the one that
Dirac and Klein\ndash Gordon equations are respectively first and second
order partial differential equations.%
}

	To begin with, we single out the `degenerate' solutions
	\begin{gather}
			\label{5.1}
[\psi,\ope{P}_\mu]_{\_} = 0
\quad
[\opsi,\ope{P}_\mu]_{\_} = 0
\qquad\text{for $m=0$}
\\\intertext{of the Dirac equation~\eref{3.5}, which solutions, in view
of~\eref{2.28}, in Heisenberg picture read}
			\label{5.2}
\tope{\psi}(x) = \tope{\psi}(x_0) = \psi \ (=\const)
\quad
\tope{\opsi}(x) = \tope{\opsi}(x_0) = \opsi \ (=\const)
\quad\text{for $m=0$} .
	\end{gather}

	According to equations~\eref{3.17}--\eref{3.20} (see
also~\eref{3.28}), the energy\ndash mo\-men\-tum, charge and (spin) angular
momentum operators for the solutions~\eref{5.1} respectively are:
	\begin{align}
				\label{5.3}
& \ope{T}_{\mu\nu} = 0
\\				\label{5.4}
& \ope{J}_\mu = q c \opsi \gamma^\mu\circ\psi
\\				\label{5.5}
& \ope{M}_{\mu\nu}^{\lambda}
= \ope{S}_{\mu\nu}^{\lambda}
=
\frac{1}{4}\hbar c \opsi
( \gamma^\lambda\sigma_{\mu\nu} + \sigma_{\mu\nu}\gamma^\lambda ) \circ\psi .
\\\intertext{Since~\eref{5.3} and~\eref{2.6} imply}
				\label{5.6}
& \ope{P}_\mu = 0,
	\end{align}
the equalities~\eref{5.4} and~\eref{5.5} hold in momentum and
Heisenberg pictures for any constant spinors $\psi$ and $\opsi$. Thus, the
solutions~\eref{5.1} (or~\eref{5.2} in Heisenberg picture) describe a
massless spinor field with vanishing energy\ndash momentum characteristics
which, however, carries, generally, a non\ndash (if $\psi\not=0$) zero charge
and spin angular momentum. It seems, such a spinor fields has not been observed
until now.%

\footnote{~%
This situation is completely different from a similar one, when free scalar
fields are concerned; in the last case, solutions like~\eref{5.1} are
unobservable in principle as all their dynamical characteristics vanish ---
see~\cite{bp-QFTinMP-scalars}.%
}

	Now, we are ready to describe the general structure of the solutions
of the Dirac equations~\eref{3.5}.
	\begin{Prop}	\label{Prop5.1}
The solutions of~\eref{2.28} and the Dirac equations~\eref{3.5} can be
written as
	\begin{subequations}	\label{5.7}
	\begin{gather}
			\label{5.7a}
\psi
=
\int\Id^3\bk
	\bigl\{
	  f_+(\bk) \psi(k)\big|_{k_0=+\sqrt{m^2c^2+{\bk}^2} }
	+ f_-(\bk) \psi(k)\big|_{k_0=-\sqrt{m^2c^2+{\bk}^2} }
	\bigr\}
\\			\label{5.7b}
\opsi
=
\int\Id^3\bk
	\bigl\{
  \overline{f}_+(\bk) \opsi(k)\big|_{k_0=+\sqrt{m^2c^2+{\bk}^2} }
+ \overline{f}_-(\bk) \opsi(k)\big|_{k_0=-\sqrt{m^2c^2+{\bk}^2} }
	\bigr\}
	\end{gather}
	\end{subequations}
or, equivalently as
	\begin{gather}
			\label{5.8}
\psi = \int\Id^4k \delta(k^2-m^2c^2) f(k) \psi(k)
\quad
\opsi = \int\Id^4k \delta(k^2-m^2c^2) \overline{f}(k) \opsi(k) .
	\end{gather}
Here: $k=(k^0,k^1,k^2,k^3)$ is a 4-vector with dimension of 4-momentum,
$k^2=k_\mu k^\mu=k_0^2 -k_1^2 -k_2 -k_3^2=k_0^2 -{\bk}^2$ with $k_\mu$ being
the components of $k$ and $\bk:=(k^1,k^2,k^3)=-(k_1,k_2,k_3)$ being the
3\ndash dimensional part of $k$,
$\delta(\cdot)$ is the (1\ndash dimensional) Dirac delta function,
the spinor operators $\psi(k),\opsi(k)\colon\Hil\to\Hil$ are solutions of
the equations
	\begin{subequations}	\label{5.9}
	\begin{align}	\label{5.9a}
& [ \psi(k), \ope{P}_\mu ]_{\_} = - k_\mu \psi(k)
&&
[ \opsi(k), \ope{P}_\mu ]_{\_} = - k_\mu \opsi(k)
\\	   		\label{5.9b}
& \{ (\gamma^\mu k_\mu + mc\openone_4)\psi(k) \} \big|_{k^2=m^2c^2} = 0
&&
\{ \opsi(k) ( -\gamma^\mu k_\mu + mc\openone_4) \} \big|_{k^2=m^2c^2} = 0 ,
	\end{align}
	\end{subequations}
$f_{\pm}(\bk)$ and $\overline{f}_{\pm}(\bk)$ are complex\ndash valued functions
(resp.\ distributions (generalized functions)) of $\bk$ for solutions
different from~\eref{5.1} (resp.\ for the solutions~\eref{5.1}), and
$f$ and $\overline{f}$ are complex\ndash valued functions (resp.\
distributions) of $k$ for solutions different from~\eref{5.1} (resp.\ for the
solutions~\eref{5.1}). Besides, we have the relations
\(
f(k)|_{k_0=\pm\sqrt{m^2c^2+{\bs k}^2} }
	= 2\sqrt{m^2c^2+{\bs k}^2} f_{\pm}(\bs k)
\)
and
\(
\overline{f}(k)|_{k_0=\pm\sqrt{m^2c^2+{\bs k}^2} }
	= 2\sqrt{m^2c^2+{\bs k}^2} \overline{f}_{\pm}(\bs k)
\)
for solutions different from~\eref{5.1}.
	\end{Prop}

	\begin{Rem}	\label{Rem5.1}
	Evidently, in~\eref{5.7} and~\eref{5.8} enter only the solutions
of~\eref{5.9} for which
	\begin{align}	\label{5.10}
k^2 := k_\mu k^\mu =k_0^2 - {\bk}^2 = m^2c^2 .
	\end{align}
This circumstance is a consequence of the fact that the solutions
of~\eref{3.5} are also solutions of the Klein\ndash Gordon
equation, viz.~\cite{bp-QFTinMP-scalars}
	\begin{equation}	\label{5.11}
m^2c^2 \psi - [[\psi,\ope{P}_\mu]_{\_} , \ope{P}^\mu]_{\_} = 0
\qquad
m^2c^2 \opsi - [[\opsi,\ope{P}_\mu]_{\_} , \ope{P}^\mu]_{\_} = 0 .
	\end{equation}
To prove this one should apply to the first (resp.\ second) equation
in~\eref{3.5} the operator
$\gamma^\nu[(\cdot) , \ope{P}_\nu]_{\_} + mc\openone_4\id_\Hil(\cdot) $
(resp.\
$[(\cdot) , \ope{P}_\nu]_{\_}\gamma^\nu - mc\openone_4\id_\Hil(\cdot) $)
and, then, to apply the Jacobi identity, \eref{2.1} and~\eref{3.15}.
	\end{Rem}

	\begin{Rem}	\label{Rem5.2}
	Obviously, the solutions~\eref{5.1} correspond to~\eref{5.9a} with
$\ope{P}_\mu=0$. Hence
	\begin{gather}	\label{5.12}
\tope{\psi}(x,0) = \psi(0) = \const
\quad
\tope{\opsi}(x,0) = \opsi(0) = \const
\quad
\ope{P}_\mu = \tope{P}_\mu = 0
\intertext{with (see~\eref{12.114})}
			\label{5.13}
	\begin{split}
\tope{\psi}(x,k) := \ope{U}^{-1}(x,x_0)\circ \psi(k)\circ \ope{U}(x,x_0)
\quad
\tope{\opsi}(x,k):= \ope{U}^{-1}(x,x_0)\circ \opsi(k)\circ \ope{U}(x,x_0) .
	\end{split}
	\end{gather}
These solutions, in terms of~\eref{5.7} or~\eref{5.8}, are described by  $m=0$
and, for example,
$f_\pm(\bk)=\overline{f}_\pm(\bk)=(\frac{1}{2}\pm a)\delta^3(\bk)$ for some
$a\in\field[C]$ or $f(k)=\overline{f}(k)$ such that
$ f(k)|_{k_0=\pm|\bs k|} = (1\pm 2a) |\bs k| \delta^3(\bs k)$, respectively.
(Here $\delta^3(\bk):=\delta(k^1)\delta(k^2)\delta(k^3)$ is the 3\ndash
dimensional Dirac delta\ndash function.)
To prove that, use the equality
$\delta(y^2-b^2)=\frac{1}{b}(\delta(y+b)+\delta(y-b))$ for $b>0$. In that
case, the equations~\eref{5.9a} reduce to
	\begin{equation}	\label{5.13new}
\gamma^\mu k_\mu \psi(k)  = 0
\qquad
\opsi(k) \gamma^\mu k_\mu = 0
	\end{equation}
as in it $m=0$. Since the rank of the matrix $\gamma^\mu k_\mu$ is equal to
one for $k^2=m^2c^2=0$ but $\bk\not=0$ (see below), the
equations~\eref{5.13new} have non\ndash vanishing solutions in this case only
if distributions are taken into account.
	\end{Rem}

	\begin{Rem}	\label{Rem5.3}
	Since $\opsi=\psi^\dag\gamma^0$, from~\eref{5.7} (resp.~\eref{5.8})
is clear that there should exist some connection between $f_\pm(\bk) \psi(k)$
 and $\overline{f}_\pm(\bk) \opsi(k)$ with $k_0=+\sqrt{m^2c^2+{\bk}^2}$
(resp.\ between $f(k) \psi(k)$ and  $\overline{f}(k) \opsi(k)$).
A simple examination of ~\eref{5.7} (resp.~\eref{5.8}) reveals that the Dirac
conjugation can either transforms these expressions into each other or
`changes' the signs plus and minus in them according to:
	\begin{subequations}	\label{5.14}
	\begin{align}
			\label{5.14a}
& \overline{
\bigl( f_\pm(\bk)\psi(k) \big|_{k_0=\pm\sqrt{m^2c^2+{\bs k}^2} } \bigr)
	}
=
 - \overline{f}_\mp(-\bk)\opsi(-k) \big|_{k_0=\mp\sqrt{m^2c^2+{\bs k}^2} } .
\\			\label{5.14b}
& \overline{
\bigl( \overline{f}_\pm(\bk)\opsi(k) \big|_{k_0=\pm\sqrt{m^2c^2+{\bs k}^2} }
\bigr)
	}
=
 - f_\mp(-\bk)\psi(-k) \big|_{k_0=\mp\sqrt{m^2c^2+{\bs k}^2} }
	\end{align}
	\end{subequations}
\vspace{-4ex}
	\begin{subequations}	\label{5.15}
	\begin{align}
			\label{5.15a}
& \overline{ \bigl( f(k)\psi(k) \bigr)	}
=
  \overline{f}(-k)\opsi(-k)
\\			\label{5.15b}
& \overline{ \bigl( \overline{f}(k)\opsi(k) \bigr)	}
=
  {f}(-k)\psi(-k) .
	\end{align}
	\end{subequations}
 From the below presented proof of proposition~\ref{Prop5.1} and the comments
after it, it will be clear that ~\eref{5.14} and~\eref{5.15} should be
accepted. Notice, the above equations mean that $\opsi(k)$ is \emph{not} the
Dirac conjugate of $\psi(k)$.
	\end{Rem}

	\begin{Proof}
	The proposition was proved for the solutions~\eref{5.1} in
remark~\ref{Rem5.2}. So, below we suppose that $(k,m)\not=(0,0)$.

	The equivalence of~\eref{5.7} and~\eref{5.8} follows from
$\delta(y^2-b^2)=\frac{1}{b}(\delta(y+b)+\delta(y-b))$ for $b>0$.

	Since the solutions $\psi$ and $\opsi$ of the Dirac
equations~\eref{3.5} are also solutions of the Klein\ndash Gordon
equations~\eref{5.11}, the representations~\eref{5.7} and the
equalities~\eref{5.14} and~\eref{5.15}, with  $\psi(k)$ and $\opsi(k)$
satisfying~\eref{5.9a}, follow from the proved in~\cite{bp-QFTinMP-scalars}
similar proposition~4.1\typeout{!!!!!!!!!!!!!!!!! Bozho, check this citation
before a publication!!!!!!!!!!!!!!!!}
describing the structure of the solutions of the Klein\ndash Gordon equation
in momentum picture.%
\footnote{~%
One can prove the representations~\eref{5.7}, under the conditions~\eref{5.9},
by repeating \emph{mutatis mutandis} the proof
of~\cite[proposition~4.1]{bp-QFTinMP-scalars}\typeout{!!!!!!!!!!!!!!!!!
Bozho, check this citation before a publication!!!!!!!!!!!!!!!!}. From it the
equalities~\eref{5.14} and~\eref{5.15} rigorously follow too.%
}

	At the end, inserting~\eref{5.7} or~\eref{5.8} into~\eref{3.5}, we
obtain the equations~\eref{5.9b} due to~\eref{5.9a}.
	\end{Proof}

	From the proof of proposition~\ref{Prop5.1}, as well as from the one
of~\cite[proposition~4.1]{bp-QFTinMP-scalars}\typeout{!!!!!!!!!!!!!!!!!
Bozho, check this citation before a publication!!!!!!!!!!!!!!!!}
the next two conclusions can be made. On one hand, the conditions~\eref{5.9a}
ensure that~\eref{5.7} and~\eref{5.8} are solutions of~\eref{2.28} and the
Klein\ndash Gordon equations~\eref{5.11}, while~\eref{5.9b} single out
between them the ones satisfying the Dirac equations~\eref{3.5}. On the other
hand, since up to a phase factor and, possibly, normalization constant, the
expressions $f(k)\psi(k)$ and $\overline{f}(k)\opsi(k)$ coincide with the
Fourier images of respectively $\tope{\psi}(x)$ and $\tope{\opsi}(x)$ in
Heisenberg picture, we can write
	\begin{equation}	\label{5.16}
\psi = \int\delta(k^2-m^2c^2) \underline{\psi}(k) \Id^4k
\quad
\tope{\psi} (x)
=
\int\delta(k^2-m^2c^2) \underline{\psi}(k)
\e^{\iu\frac{1}{\hbar}(x^\mu-x_0^\mu)k_\mu} \Id^4k
	\end{equation}
and similarly for $\opsi$ (with
$\overline{\underline{\opsi}(k)}=\underline{\opsi}(-k)$), where
$\underline{\psi}(k)$ are suitably normalized solutions of~\eref{5.9}.
Therefore, up to normalization factor, the Fourier images of $\tope{\psi}(x)$
and $\tope{\opsi}(x)$ are
	\begin{equation}	\label{5.16-1}
\underline{\tope{\psi}} (k)
=
\e^{\iih x_0^\mu k_\mu} \underline{\psi} (k)
\qquad
\underline{\tope{\opsi}} (k)
=
\e^{\iih x_0^\mu k_\mu} \underline{\opsi} (k)
	\end{equation}
where $x_0$ is a fixed point (see Sect.~\ref{Sect2}). So, the momentum
representation of free spinor field in Heisenberg picture is an appropriately
chosen operator base for the solutions of the Dirac equation in momentum
picture. This conclusion allows us freely to apply in momentum picture the
existing results concerning that basis in Heisenberg picture.

	The equations~\eref{5.9b} are well known and explicitly solved in the
textbooks where the Dirac equation, in momentum representation of Heisenberg
picture, is explored~\cite{Bjorken&Drell,Bogolyubov&Shirkov} (for a summary,
see, e.g.,~\cite[asppendix~1]{Roman-QFT}
or~\cite[appendix~3A]{Bogolyubov&Shirkov}). Here are some facts about them,
which we shall need further in the present paper; for references --- see
\emph{loc.\ cit.}

	Working in a representation in which $\gamma^0$ is diagonal
(see~\eref{3.1-1}), by a direct calculation, one can prove that the rank of
the matrices $\pm\gamma^\mu k_\mu+mc\openone_4$, with
$k_0=\pm\sqrt{m^2c^2+\bk^2}$, is equal
to~0, if $(\bs k,m) = (\bs 0,0)$
and
to~2, 
if $(\bs k,m) \not= (\bs 0,0)$.%
\footnote{~%
For $(\bs k,m)=(\bs 0,0)$ the statement is evident.
Usually (see, e.g.,~\cite{Bogolyubov&Shirkov}),
when $m\not=0$ (but
$\bs k\not=0$), it is proved in a frame in which $\bk=\bs0$ and a subsequent
return to a general one;
when $\bs k\not=0$ and $m=0$, the same method can be
used with the only modification that a frame in which $k^1=k^2=0$ has to be
employed.
In the last case, the Dirac equation is replaced with a system of the
so\ndash called Weyl equations~\cite{Bjorken&Drell,Bogolyubov&Shirkov}.%
}
Since the $\gamma$\ndash matrices are defined up to a change
$\gamma^\mu\mapsto O\gamma^\mu O^{-1}$, $O$ being a non\ndash degenerate
matrix (usually taken to be unitary), from here follows that any one of the
equations~\eref{5.9b} has $r$ linearly independent solutions, where
$r=\infty$ for $(\bs k,m)=(\bs 0,0)$
and
$r=2$ for $(\bs k,m)\not=(\bs 0,0)$.
So, in these cases~\eref{5.9b} has respectively
\emph{infinitely many} linearly independent solutions
and \emph{two} linearly independent solutions.
	Since the case $(\bs k,m)=(\bs 0,0)$, corresponds to the
`degenerate' solutions~\eref{5.1}, which require different treatment, we
shall exclude it from our further considerations in this section.%
\footnote{~%
On the exploration of the consequences of~\eref{5.1}, see the paragraph
containing equation~\eref{6.15} in Sect.~\ref{Sect6}.%
}
	In the case $m\not=0$, we shall label these  $r=2$ linearly
independent solutions with an index $s$ taking the values~$1$ and~$2$,
$s=1,2$.%
\footnote{~%
Usually~\cite{Bjorken&Drell,Bogolyubov&Shirkov}, the index $s$ is referred as
the \emph{polarization} or \emph{spin} index (parameter, variable).%
}
	Define the operator spinors
 $\psi_{s,(\pm)}(k)$ and
$\opsi_{s,(\pm)}(k)$,
where the index $(\pm)$ indicates the sign of $k_0=\pm\sqrt{m^2c^2+\bk^2}$
($\not=0$) in~\eref{5.9b}, as linearly independent solutions of respectively
the equations
	\begin{gather}
			\label{5.17}
	\begin{split}
\bigl( \gamma^\mu k_\mu\big|_{k_0=\pm\sqrt{m^2c^2+\bk^2}} +mc\openone_4
\bigr)
\psi_{s,(\pm)}(k)
= 0
\\
\opsi_{s,(\pm)}(k)
\bigl(- \gamma^\mu k_\mu\big|_{k_0=\pm\sqrt{m^2c^2+\bk^2}} +mc\openone_4
\bigr)
= 0 .
	\end{split}
\\\intertext{If $m=0$, we set by definition $s=0$ and label the linearly
independent solutions of~\eref{5.9b} by the signs "$+$" and "$-$".
Respectively, for $m=0$, we define the operator spinors
 $\psi_{0,(\pm)}(k)$ and
$\opsi_{0,(\pm)}(k)$
as linearly independent solutions of~\eref{5.17} with $m=0$.
As a result of~\eref{5.9a}, the solutions of~\eref{5.17} satisfy also the
equations}
	\begin{split}
			\label{5.18}
[ \psi_{s,(\pm)}(k) ,\ope{P}_\mu]_{\_}
= - k_\mu\big|_{k_0=\pm\sqrt{m^2c^2+\bk^2}} \psi_{s,(\pm)}(k)
\\
[ \opsi_{s,(\pm)}(k) ,\ope{P}_\mu]_{\_}
= - k_\mu\big|_{k_0=\pm\sqrt{m^2c^2+\bk^2}} \opsi_{s,(\pm)}(k) .
	\end{split}
	\end{gather}

	Since any solution of the first (resp.\ second) equation
in~\eref{5.9b}  can be represented as a linear combination of
$\psi_{s,(\pm)}(k)$ (resp.\ $\opsi_{s,(\pm)}(k)$), we can
rewrite~\eref{5.7} as
	\begin{subequations} \label{5.19}
	\begin{gather}
			\label{5.19a}
\psi
= \sum_{s} \int\Id^3\bk \bigl\{
  f_{s,+}(\bk) \psi_{s,(+)}(k)
+ f_{s,-}(\bk) \psi_{s,(-)}(k)
\bigr\} \bigr|_{k_0=\sqrt{m^2c^2+\bk^2}}
\\			\label{5.19b}
\opsi
= \sum_{s} \int\Id^3\bk \bigl\{
  \overline{f}_{s,+}(\bk) \opsi_{s,(+)}(k)
+ \overline{f}_{s,-}(\bk) \opsi_{s,(-)}(k)
\bigr\} \bigr|_{k_0=\sqrt{m^2c^2+\bk^2}}
	\end{gather}
	\end{subequations}
where $f_{s,\pm}(\bk)$ and $\overline{f}_{s,\pm}(\bk)$ are some complex-valued
functions of $\bk$ for solutions different from~\eref{5.1}. Regarding the
solutions~\eref{5.1}, in view of remark~\ref{Rem5.2}, for them~\eref{5.19}
holds too for some distributions $f_{s,\pm}(\bk)$ and
$\overline{f}_{s,\pm}(\bk)$ and some operators $\psi_{s,(\pm)}(k)$ and
$\opsi_{s,(\pm)}(k)$, which can be chosen in different ways%
\footnote{~%
For details, see below ~\eref{6.15} and the paragraph containing it.
Recall, the solutions~\eref{5.1} can be describe by
$(k,m,\ope{P}_\mu)=(0,0,0)$ in which case~\eref{5.9} (and,
hence,~\eref{5.17} and~\eref{5.18}) reduce to the identity $0=0$.%
}

	Below we shall  need a system of \emph{classical}, not
operator-valued, suitably normalized solutions of the equations~\eref{5.17}.
The idea of their introduction is to be separated the invariant operator
properties of the spinor field from its particular `matrix' representation as
a collection of (operator) components which depend on some concrete reference
frame with respect to which it is studied. This will be done in
Sect.~\ref{Sect6} by expanding the operator\ndash valued quantum spinors as
linear combinations of the mentioned system, in fact a basis, of classical
spinors, the coefficients of which expansion are frame\ndash independent
invariant operators characterizing the spinor field.

	Let $(k,m)\not=(0,0)$. Consider \emph{classical} 4-spinors
$v^{s,\pm}(\bk)$ and their Dirac conjugate
\(
\overline{v}^{s,\pm}(\bk)
:= ({v}^{s,\mp}(\bk))^\dag\gamma^0
= \overline{({v}^{s,\pm}(\bk))}
=: {v}^{\dag\, s,\pm}(\bk) \gamma^0,
\)
where $s=1,2$ for $m\not=0$, $s=0$ for $m=0$, and ``$\dag$'' means Hermitian
conjugation (\ie matrix transposition combined with complex conjugation in
the classical case), which are linearly independent solutions of the
equations
	\begin{gather}
			\label{5.20}
	\begin{split}
\bigl(\pm \gamma^\mu k_\mu\big|_{k_0=+\sqrt{m^2c^2+\bk^2}} +mc\openone_4
\bigr)
{v}^{s,\pm}(\bk)
= 0
\\
\overline{v}^{s,\pm}(\bk)
\bigl(\mp \gamma^\mu k_\mu\big|_{k_0=+\sqrt{m^2c^2+\bk^2}} +mc\openone_4
\bigr)
= 0
	\end{split}
\intertext{and satisfy the conditions}
			\label{5.21}
\bigl({v}^{s,\pm}(\bk)\bigr)^\dag =: {v}^{\dag\,s,\mp}(\bk)
\qquad
\overline{ {v}^{s,\pm}(\bk) } =: \overline{v}^{s,\mp}(\bk)
=
{v}^{\dag\,s,\mp}(\bk) \gamma^0
\\			\label{5.22}
{v}^{\dag\,s,\pm}(\bk) {v}^{s',\mp}(\bk) = \delta^{ss'}
\qquad
\overline{v}^{s,\pm}(\bk) v^{s',\mp}(\bk)
=
\pm\frac{mc}{\sqrt{m^2c^2+\bk^2}}  \delta^{ss'} ,
	\end{gather}
where $s,s'=0$ for $m=0$ (and $k\not=0$) and $s,s'=1,2$ for  $m\not=0$.
From here the following relations can be
derived~\cite[subsect.~7.2]{Bogolyubov&Shirkov}:
	\begin{gather}
			\label{5.23}
{v}^{\dag\,s,\pm}(\bk) {v}^{s',\pm}(-\bk) = 0
\\			\label{5.24}
\left.	\begin{split}
{v}^{\dag\,s,\pm}(\bk)
\{ k^a\gamma^b - k^b\gamma^a - mc\gamma^a\gamma^b \}
{v}^{s,\pm}(-\bk)
= 0 \qquad a,b=1,2,3
\\
\sum_{a} k^a
\{  {v}^{\dag\,s,\pm}(\bk) (\gamma^a\gamma^b-\gamma^b\gamma^a)
    {v}^{s',\mp}(\bk)
\}
= 0 \qquad a,b=1,2,3
	\end{split}
\right\}
\\			\label{5.25}
\sum_{s} {v}^{s,\pm}_\mu(\bk) \overline{v}^{s,\mp}_\nu(\bk)
=
\frac{(\gamma^\lambda k_\lambda \mp mc\openone_4)_{\mu\nu}}
     {2 k_0}
\Big|_{k_0=\sqrt{m^2c^2+\bk^2}}
\qquad \text{ for } m\not=0 .
	\end{gather}
These formulae will be applied in different calculations in the next sections.

	The explicit form of the spinors $v^{s,\pm}(\bk)$ for $m\not=0$
can be found in~\cite[pp.~617--618]{Roman-QFT} or
in~\cite[sect.~2.2.1]{Itzykson&Zuber}, where the notation
$u(\bk;s):=v^{s,-}(\bk)$ and $v(\bk;s):=v^{s,+}(\bk)$ is used. Since
\(
(\gamma^\mu k_\mu + mc\openone_4) (\gamma^\nu k_\nu - mc\openone_4)
  = ( k^2-m^2c^2 ) \openone_4
\)
and $k^2=m^2c^2$, we can write
	\begin{equation}	\label{5.26}
  v^{s,\pm}(\bk)
= \bigl( v^{\dag\, s,\mp}(\bk) \bigr)^\dag
= A_\pm(\bk)
  ( \mp \gamma^\mu k_\mu\big|_{k_0=\sqrt{m^2c^2+\bk^2}} + mc\openone_4)
  v^{s,\pm}_0,
	\end{equation}
where $A_\pm(\bk)$ are some normalization constants, that can be found by
using~\eref{5.22},%
\footnote{~%
Explicitly, we have
\(
A_\pm(\bk) = \{2mc (mc +\sqrt{m^2c^2+\bk^2} ) \}^{-1/2}
\)
for $m\not=0$; see~\cite[Appendix~A, eq.~(A1-29)]{Roman-QFT}. For $m=0$, one
may set
\(
A_\pm(\bk) = \pm \{2 \bk^2\}^{-1/2} .
\)%
}
and $v^{s,\pm}_0$ are constant spinors given by:
	\begin{subequations}	\label{5.27}
	\begin{align}	\label{5.27a}
	\begin{split}
v^{1,-}_0 &= (1,0,0,0)^\top \quad
v^{2,-}_0 = (0,1,0,0)^\top \quad
\\
v^{1,+}_0 &= (0,0,1,0)^\top \quad
v^{2,+}_0 = (0,0,0,1)^\top \quad
	\end{split}
\Bigg\}\qquad\text{for $m\not=0$}
\\			\label{5.27b}
v^{0,-}_0 &= (0,0,1,0)^\top \quad
v^{0,+}_0 = (0,0,0,1)^\top
\qquad\text{for $m=0$} .
	\end{align}
	\end{subequations}


\section {Frequency decompositions}
\label{Sect6}

	As a consequence of the results of Sect.~\ref{Sect5}, one can expect
the existence of decompositions of the constant Dirac spinors $\psi$ and
$\opsi$ similar to the one in Heisenberg
picture~\cite{Bjorken&Drell,Bogolyubov&Shirkov}. Such expansions, in fact,
exist and can be introduced in almost the same way as it was done for a
free charged scalar field in~\cite{bp-QFTinMP-scalars}; the only differences
being the spinor (polarization) index $s$ in~\eref{5.19} and the multi\ndash
component character of the spinors.

	Let us set
	\begin{gather}
			\label{6.1}
	\begin{split}
\psi_s^\pm(k) & :=
	\begin{cases}
f_{s,\pm}(\pm \bk) \psi_{s,(\pm)}(\pm k)	& \text{for $k_0\ge0$} \\
0						& \text{for $k_0<0$}
	\end{cases}
\\
\opsi_s^\pm(k) & :=
	\begin{cases}
\overline{f}_{s,\pm}(\pm \bk) \opsi_{s,(\pm)}(\pm k)
					& \text{for $k_0\ge0$} \\
0					& \text{for $k_0<0$}
	\end{cases}
	\end{split}
\\			\label{6.1new}
 \psi^\pm(k) := \sum_{s}  \psi_s^\pm(k)
\qquad
\opsi^\pm(k) := \sum_{s} \opsi_s^\pm(k)
\\
	\begin{split}	\label{6.1new2}
\psi_s^\pm(\bk)   := \psi_s^\pm(k)\big|_{k_0=\sqrt{m^2c^2+\bk^2}}
\qquad
\opsi_s^\pm(\bk) := \opsi_s^\pm(k)\big|_{k_0=\sqrt{m^2c^2+\bk^2}}
\\
\psi^\pm(\bk)   := \psi^\pm(k)\big|_{k_0=\sqrt{m^2c^2+\bk^2}}
		 = \sum_{s} \psi_s^\pm(\bk)
\\
\opsi^\pm(\bk) := \opsi^\pm(k)\big|_{k_0=\sqrt{m^2c^2+\bk^2}}
		 = \sum_{s} \opsi_s^\pm(\bk) ,
	\end{split}
	\end{gather}
where  $k^2=m^2c^2$ and (the summation is over) $s=0$ for $m=0$ (and
$k\not=0$) or $s=1,2$ for $m\not=0$.%
\footnote{~%
On the case $(k,m)=(0,0)$, or, more generally, the case regarding
the solutions~\eref{5.1} --- \emph{vide infra} the paragraph containing
equation~\eref{6.15}.%
}

	The equalities~\eref{5.15} imply
	\begin{equation}	\label{6.2}
\overline{ \bigl(  \psi^{\pm}(\bk) \bigr) }  =  \opsi^{\mp}(\bk)
\qquad
\overline{ \bigl( \opsi^{\pm}(\bk) \bigr) }  =   \psi^{\mp}(\bk)
	\end{equation}
which mean that $\opsi^{\pm}(\bk) $ is \emph{not} the Dirac conjugate of
 $\psi^{\pm}(\bk) $.

	Combining~\eref{6.1},~\eref{5.18},~\eref{5.19} and
proposition~\ref{Prop5.1}, we get:
	\begin{gather}
			\label{6.3}
\psi = \psi^+ + \psi^-
\qquad
\opsi = \opsi^+ + \opsi^-
\\			\label{6.4}
	\begin{split}
\psi^\pm
:= \sum_{s}\int\Id^3\bs k \psi_s^\pm (\bk)
 = \int\Id^3\bs k \psi^\pm (\bk)
\quad
\opsi^\pm
:= \sum_{s}\int\Id^3\bs k \opsi_s^\pm (\bk)
 = \int\Id^3\bs k \opsi^\pm (\bk)
	\end{split}
	\end{gather}
\vspace{-3ex}
	\begin{subequations}	\label{6.5}
	\begin{align}
			\label{6.5a}
 [\psi^\pm(\bk),\ope{P}_\mu]_{\_}
	& = \mp k_\mu \psi^\pm(\bk)
\qquad
k_0=\sqrt{m^2c^2+{\bs k}^2}
\\			\label{6.5b}
[\opsi^\pm(\bk),\ope{P}_\mu]_{\_}
	& = \mp k_\mu \opsi^\pm(\bk)
\qquad
k_0=\sqrt{m^2c^2+{\bs k}^2} .
	\end{align}
	\end{subequations}
Notice, now equations~\eref{5.9b} are incorporated in the
definitions~\eref{6.1} via the equations~\eref{5.17}.

	To reveal the physical meaning of the operators introduced, we shall
rewrite~\eref{3.22} and~\eref{3.23} in their terms:%
\footnote{~%
To derive rigorously~\eref{6.6} and~\eref{6.7} from~\eref{3.22}
and~\eref{3.23}, respectively, one has to take into account that $\psi(k)$
and $\opsi(k)$ are, up to a phase factor and, possibly, normalization factor,
the Fourier images of $\tope{\psi}(x)$ and $\tope{\opsi}(x)$, respectively
(see~\eref{5.16}).%
}
	\begin{equation}	\label{6.6}
[\psi^\pm(\bk), \ope{Q}]_{\_} =  q \psi^\pm(\bk)
\qquad
[\opsi^\pm(\bk), \ope{Q}]_{\_} = - q \opsi^\pm(\bk)
	\end{equation}
\vspace{-3.4ex}
	\begin{subequations}	\label{6.7}
	\begin{align}
			\label{6.7a}
& [\psi^\pm(\bk), \ope{M}_{\mu\nu}(x)]_{\_}
=
\{
\mp (x_\mu k_\nu - x_\nu k_\mu) \big|_{k_0=\sqrt{m^2c^2+\bk^2}} \openone_4
	+ \frac{1}{2}\hbar \sigma_{\mu\nu}
\} \psi^\pm(\bk)
\\			\label{6.7b}
& [\opsi^\pm(\bk), \ope{M}_{\mu\nu}(x)]_{\_}
=
\opsi^\pm(\bk)\{
\mp (x_\mu k_\nu - x_\nu k_\mu) \big|_{k_0=\sqrt{m^2c^2+\bk^2}} \openone_4
	- \frac{1}{2}\hbar \sigma_{\mu\nu}
\}
	\end{align}
	\end{subequations}
where~\eref{6.5} was taken into account. Recall, here $\ope{Q}$ and
$\ope{M}_{\mu\nu}$ stand for the charge and (total) angular momentum
operators, respectively, and the spin matrices $\sigma_{\mu\nu}$ are defined
via~\eref{3.13-1}. Besides, the last terms in~\eref{6.7} are due to the spin
angular momentum while the remaining ones originate from the orbital angular
momentum. We should remind, the equations~\eref{6.6} and~\eref{6.7} originate
from~\eref{2.17} and~\eref{2.18}, which are external to the Lagrangian
formalism. Therefore the below\ndash presented results, in particular the
physical interpretation of the creation and annihilation operators, should be
accepted with some reserve. However, after the establishment of the particle
interpretation of the theory (see Sections~\ref{Sect10} and~\ref{Sect11}), the
results of this section will be confirmed (see also Sect.~\ref{Sect12}).

	Let  $\ope{X}_p$, $\ope{X}_e$ and $\ope{X}_m$ denote state vectors of
a spinor field with fixed respectively 4\ndash momentum $p_\mu$, (total)
charge $e$ and (total) angular momentum $m_{\mu\nu}(x)$, \ie
	\begin{subequations}	\label{6.8}
	\begin{align}
			\label{6.8a}
& \ope{P}_\mu(\ope{X}_p) = p_\mu \ope{X}_p
\\			\label{6.8b}
& \ope{Q}(\ope{X}_e) = e \ope{X}_e
\\			\label{6.8c}
& \ope{M}_{\mu\nu}(x)(\ope{X}_m) = m_{\mu\nu}(x) \ope{X}_m.
	\end{align}
	\end{subequations}
Combining these equations
with~\eref{6.5}--\eref{6.7}, we obtain%
\footnote{~%
Expressions like $\psi_s^\pm(\bk)(\ope{X}_p)$ should be understand as a
vector\ndash columns of vectors of the  form
\(
\bigl( \psi_{s,0}^\pm(\bk)(\ope{X}_p),\dots,
	\psi_{s,3}^\pm(\bk)(\ope{X}_p) \bigr)^\top.
\)
Similarly, the quantity
\(
\opsi_s^\pm(\bk)(\ope{X}_p)
:=
\bigl( \opsi_{s,0}^\pm(\bk)(\ope{X}_p),\dots,
	\opsi_{s,3}^\pm(\bk)(\ope{X}_p) \bigr)
\)
is a vector-row of vectors.%
}
	\begin{subequations}	\label{6.9}
	\begin{align}
			\label{6.9a}
& \ope{P}_\mu\bigl( \psi^\pm(\bk) (\ope{X}_p) \bigr)
  = (p_\mu \pm k_\mu) \psi^\pm(\bk) (\ope{X}_p)
	\qquad k_0=\sqrt{m^2c^2+\bk^2}
\\			\label{6.9b}
& \ope{P}_\mu\bigl( \opsi^\pm(\bk) (\ope{X}_p) \bigr)
  = (p_\mu \pm k_\mu) \opsi^\pm(\bk) (\ope{X}_p)
	\qquad k_0=\sqrt{m^2c^2+\bk^2}
	\end{align}
	\end{subequations}
\vspace{-4ex}
	\begin{subequations}	\label{6.10}
	\begin{align}
			\label{6.10a}
	\begin{split}
 \ope{Q}\bigl( \psi (\ope{X}_e) \bigr)  = (e-q) \psi (\ope{X}_e)
\quad
& \ope{Q}\bigl( \opsi (\ope{X}_e) \bigr) = (e+q) \opsi (\ope{X}_e)
	\end{split} 
\displaybreak[1]\\			\label{6.10b}
	\begin{split}
 \ope{Q}\bigl( \psi^\pm (\ope{X}_e) \bigr)
  = (e-q) \psi^\pm (\ope{X}_e)
\quad
& \ope{Q}\bigl( \opsi^\pm (\ope{X}_e) \bigr)
  = (e+q) \opsi^\pm (\ope{X}_e)
	\end{split} 
\displaybreak[1]\\			\label{6.10c}
	\begin{split}
 \ope{Q}\bigl( \psi^\pm(\bk) (\ope{X}_e) \bigr)
  = (e-q) \psi^\pm(\bk) (\ope{X}_e)
\quad
& \ope{Q}\bigl( \opsi^\pm(\bk) (\ope{X}_e) \bigr)
  = (e+q) \opsi^\pm(\bk) (\ope{X}_e)
	\end{split} 
	\end{align}
	\end{subequations}
\vspace{-4ex}
	\begin{subequations}	\label{6.11}
	\begin{align}
			\label{6.11a}
	\begin{split}
\ope{M}_{\mu\nu}(x)
&  \bigl( \psi^\pm(\bk) (\ope{X}_m) \bigr)
=
\{ m_{\mu\nu}(x) \openone_4
 \pm (x_\mu k_\nu - x_\nu k_\mu) \big|_{k_0=\sqrt{m^2c^2+\bk^2}} \openone_4
	- \frac{1}{2}\hbar \sigma_{\mu\nu}
\}  \psi^\pm(\bk) (\ope{X}_m)
	\end{split}
\\			\label{6.11b}
	\begin{split}
\ope{M}_{\mu\nu}(x)
&  \bigl( \opsi^\pm(\bk) (\ope{X}_m) \bigr)
=
\opsi^\pm(\bk)(\ope{X}_m)
\{ m_{\mu\nu}(x) \openone_4
 \pm (x_\mu k_\nu - x_\nu k_\mu) \big|_{k_0=\sqrt{m^2c^2+\bk^2}} \openone_4
	+ \frac{1}{2}\hbar \sigma_{\mu\nu}
\} .
	\end{split}
	\end{align}
	\end{subequations}

If the field configuration happens to be such that
	\begin{equation}	\label{6.12}
[\psi,\ope{P}_\mu]_{\_} = A_\mu\psi
\quad
[\opsi,\ope{P}_\mu]_{\_} = \opsi \, \overline{A}_\mu
\qquad
\overline{A}_\mu:=-\gamma^0 A^\dag_\mu\gamma^0
	\end{equation}
for some $4\times4$ matrices $A_\mu$, then to~\eref{6.11} can be added the
equations (cf.~\eref{6.10a} and~\eref{6.10b}):
	\begin{subequations}	\label{6.13}
	\begin{align}
			\label{6.13a}
	\begin{split}
\ope{M}_{\mu\nu}(x)
&  \bigl( \psi (\ope{X}_m) \bigr)
=
\{ m_{\mu\nu}(x) \openone_4
 - (x_\mu A_\nu - x_\nu A_\mu) \big|_{k_0=\sqrt{m^2c^2+\bk^2}} \openone_4
	- \frac{1}{2}\hbar \sigma_{\mu\nu}
\}  \psi (\ope{X}_m)
	\end{split}
\\			\label{6.13b}
	\begin{split}
\ope{M}_{\mu\nu}(x)
&  \bigl( \opsi (\ope{X}_m) \bigr)
=
\opsi (\ope{X}_m)
\{ m_{\mu\nu}(x) \openone_4
 - (x_\mu \overline{A}_\nu - x_\nu \overline{A}_\mu)
	\big|_{k_0=\sqrt{m^2c^2+\bk^2}} \openone_4
	+ \frac{1}{2}\hbar \sigma_{\mu\nu}
\} .
	\end{split}
	\end{align}
	\end{subequations}
In particular, for the `exotic' solutions~\eref{5.1}, we have
(cf.~\eref{6.10}):
	\begin{subequations}	\label{6.14}
	\begin{align}
			\label{6.14a}
	\begin{split}
& \ope{M}_{\mu\nu}(x)  \bigl( \psi (\ope{X}_m) \bigr)
=
\{ m_{\mu\nu}(x) \openone_4  - \frac{1}{2}\hbar \sigma_{\mu\nu}
\}  \psi (\ope{X}_m)
\\
& \ope{M}_{\mu\nu}(x) \bigl( \opsi (\ope{X}_m) \bigr)
=
\opsi (\ope{X}_m)
\{ m_{\mu\nu}(x) \openone_4  + \frac{1}{2}\hbar \sigma_{\mu\nu} \}
	\end{split} \Bigg\}
\displaybreak[1]\\			\label{6.14b}
	\begin{split}
& \ope{M}_{\mu\nu}(x)  \bigl( \psi^\pm (\ope{X}_m) \bigr)
=
\{ m_{\mu\nu}(x) \openone_4  - \frac{1}{2}\hbar \sigma_{\mu\nu}
\}  \psi^\pm (\ope{X}_m)
\\
& \ope{M}_{\mu\nu}(x) \bigl( \opsi^\pm (\ope{X}_m) \bigr)
=
\opsi^\pm(\ope{X}_m)
\{ m_{\mu\nu}(x) \openone_4  + \frac{1}{2}\hbar \sigma_{\mu\nu} \}
	\end{split} \Bigg\}
\displaybreak[1]\\			\label{6.14c}
	\begin{split}
& \ope{M}_{\mu\nu}(x)  \bigl( \psi^\pm(\bk) (\ope{X}_m) \bigr)
=
\{ m_{\mu\nu}(x) \openone_4  - \frac{1}{2}\hbar \sigma_{\mu\nu}
\}  \psi^\pm(\bk) (\ope{X}_m)
\\
& \ope{M}_{\mu\nu}(x) \bigl( \opsi^\pm(\bk) (\ope{X}_m) \bigr)
=
\opsi^\pm(\bk) (\ope{X}_m)
\{ m_{\mu\nu}(x) \openone_4  + \frac{1}{2}\hbar \sigma_{\mu\nu} \} .
	\end{split} \Bigg\}
	\end{align}
	\end{subequations}
Besides, the equations~\eref{6.10} remain the same for the
solutions~\eref{5.1} and ~\eref{6.9} reduce to the identity $0=0$ for them.

	The equations~\eref{6.9} (resp.~\eref{6.10}) show that the
eigenvectors of the momentum (resp.\ charge) operator are mapped into such
vectors by the operators $\psi^{\pm}(\bk)$ and $\opsi^{\pm}(\bk)$
(resp.\
$\psi$, $\psi^{\pm}$, $\psi^{\pm}(\bk)$,
$\opsi$, $\opsi^{\pm}$, and $\opsi^{\pm}(\bk)$).
However, by virtue of the equalities~\eref{6.11}--\eref{6.14}, no one of the
operators
$\psi$, $\psi^{\pm}$, $\psi^{\pm}(\bk)$,
$\opsi$, $\opsi^{\pm}$, and $\opsi^{\pm}(\bk)$
maps an eigenvector of the angular momentum operator into such a vector. The
cause for this fact are the matrices $\pm\frac{1}{2}\hbar\sigma_{\mu\nu}$,
appearing in~\eref{6.11}--\eref{6.14}, which generally are  non\ndash
diagonal~\cite{Bjorken&Drell-1,Bogolyubov&Shirkov} and, consequently mix the
components of the matrix vectors
$\psi(\ope{X}_m)$,   $\psi^{\pm}(\ope{X}_m)$,     $\psi^{\pm}(\bk)(\ope{X}_m)$,
$\opsi(\ope{X}_m)$, $\opsi^{\pm}(\ope{X}_m)$, and $\opsi^{\pm}(\bk)(\ope{X}_m)$
in~\eref{6.11}--\eref{6.14}. Since the matrices
$\pm\frac{1}{2}\hbar\sigma_{\mu\nu}$ have a dimension of angular momentum
and, obviously, originate from the `pure spinor' properties of spin
$\frac{1}{2}$ fields, we shall refer to them as
\emph{spin\ndash mixing angular momentum matrices} or simply as
\emph{spin\ndash mixing matrices}; by definition, the spin\ndash mixing
matrix of the field $\psi$ (resp.\ its Dirac conjugate $\opsi$) is
$-\frac{1}{2}\hbar\sigma_{\mu\nu}$ (resp.\
$+\frac{1}{2}\hbar\sigma_{\mu\nu}$). More generally, if $\ope{X}$  is a
state vector and
\(
 \ope{M}_{\mu\nu}(x)  \bigl( \psi^\pm(\bk) (\ope{X}) \bigr)
=
\{ l_{\mu\nu}(x) \openone_4  + s_{\mu\nu} \}  \psi^\pm(\bk) (\ope{X})
\)
or
\(
 \ope{M}_{\mu\nu}(x) \bigl( \opsi^\pm(\bk) (\ope{X}) \bigr)
=
\opsi^\pm(\bk) (\ope{X})
\{ \overline{l}_{\mu\nu}(x) \openone_4  + \overline{s}_{\mu\nu} \},
\)
where $l_{\mu\nu}$ and $\overline{l}_{\mu\nu}$ are some operators
and $s_{\mu\nu}$ and $\overline{s}_{\mu\nu}$ are matrices, not proportional
to the unit matrix $\openone_4$, with operator entries, then we shall say
that the operators $\psi^\pm(\bk)$ or $\opsi^\pm(\bk)$ have respectively
spin\ndash mixing (angular momentum) matrices $s_{\mu\nu}$ and
$\overline{s}_{\mu\nu}$ relative to the state vector $\ope{X}$; we shall
abbreviate this by saying that the states
$\psi^\pm(\bk) (\ope{X})$ and $\opsi^\pm(\bk) (\ope{X})$
have spin\ndash mixing matrices
$s_{\mu\nu}$ and $\overline{s}_{\mu\nu}$,
respectively.

	The other additional terms in, e.g., equation~\eref{6.11} are
$\pm (x_\mu k_\nu - x_\nu k_\mu) \big|_{k_0=\sqrt{m^2c^2+\bk^2}} \openone_4$.
They do not mix the components of
$\psi^{\pm}(\bk)(\ope{X}_m)$ and $\opsi^{\pm}(\bk)(\ope{X}_m)$. These
terms may be associated with the orbital angular momentum of the (matrix)
state vectors
$\psi^{\pm}(\bk)(\ope{X}_m)$ and $\opsi^{\pm}(\bk)(\ope{X}_m)$ .

	Thus, from~\eref{6.9}--\eref{6.11}, the following conclusions can be
made:
\renewcommand{\theenumi}{\roman{enumi}}
	\begin{enumerate}
\item
	The operators $\psi^+(\bk)$ and $\opsi^+(\bk)$
(respectively $\psi^-(\bk)$ and $\opsi^-(\bk)$) increase (respectively
 decrease) the state's 4\ndash momentum by the quantity
 $(\sqrt{m^2c^2+\bk^2},\bk)$.

\item
	The operators $\psi$, $\psi^\pm$ and $\psi^\pm(\bk)$
(respectively $\opsi$, $\opsi^\pm$ and $\opsi^\pm(\bk)$) decrease
 (respectively increase) the states' charge by $q$.

\item
	The operators $\psi^+(\bk)$ and $\opsi^+(\bk)$
(respectively $\psi^-(\bk)$ and $\opsi^-(\bk)$) increase (respectively
decrease) the state's orbital angular momentum by
$(x_\mu k_\nu -x_\nu k_\mu)\bigr|_{k_0=\sqrt{m^2c^2+\bk^2}}$.

\item
	The operators $\psi^\pm(\bk)$
(respectively $\opsi^\pm(\bk)$ ) possess spin\ndash mixing angular momentum
matrices $-\frac{1}{2}\hbar\sigma_{\mu\nu}$ (resp.\
$+\frac{1}{2}\hbar\sigma_{\mu\nu}$) relative to states with fixed total
angular momentum.
	\end{enumerate}

	In this way, the operators $\psi^\pm(\bk)$ and $\opsi^\pm(\bk)$
obtain an interpretation of creation and annihilation operators of particles
(quanta) of a spinor field, \viz
\\\indent
	(a) the operator $\psi^+(\bk)$ (respectively $\psi^-(\bk)$)
creates (respectively annihilates) a particle with 4\ndash momentum
$(\sqrt{m^2c^2+\bk^2},\bk)$, charge $(-q)$ (resp.\ $(+q)$), orbital angular
momentum $(x_\mu k_\nu -x_\nu k_\mu)\bigr|_{k_0=\sqrt{m^2c^2+\bk^2}}$, and
spin\ndash mixing angular momentum matrix
$\bigl(-\frac{1}{2}\hbar\sigma_{\mu\nu}\bigr)$ and
\\\indent
	(b) the operator $\opsi^+(\bk)$ (respectively $\opsi^-(\bk)$)
creates (respectively annihilates) a particle with 4\ndash momentum
$(\sqrt{m^2c^2+\bk^2},\bk)$, charge $(+q)$ (resp.\ $(-q)$), orbital angular
momentum $(x_\mu k_\nu -x_\nu k_\mu)\bigr|_{k_0=\sqrt{m^2c^2+\bk^2}}$, and
spin\ndash mixing angular momentum matrix
$\bigl(+\frac{1}{2}\hbar\sigma_{\mu\nu}\bigr)$.

	Let us say a few words on the solutions~\eref{5.1} of the Dirac
equations~\eref{3.5} in momentum picture. If we try to describe them in terms
of ordinary operators, not operator\ndash valued distributions, in the scheme
developed we should put $(k,m)=(0,0)$. But the operators $\psi^\pm(\bs0)$
and $\opsi^\pm(\bs0)$ do not change a state's 4\ndash momentum
(see~\eref{6.9} with $p_\mu=k_\mu=0$, $m=0$ and $\ope{P}_\mu=0$) and orbital
angular momentum and produce states with spin\ndash mixing angular momentum
matrix $-\frac{1}{2}\hbar\sigma_{\mu\nu}$
(resp.\ $+\frac{1}{2}\hbar\sigma_{\mu\nu}$) (see~\eref{6.14c}).
	Since the interpretation of the operators $\psi^\pm(\bk)$ and
$\opsi^\pm(\bk)$ for $(k,m)\not=(0,0)$ is connected with a non\ndash
vanishing 4\ndash momentum $(\sqrt{m^2c^2+\bk^2},\bk)$ and/or non\ndash
vanishing orbital momentum
\(
(x_\mu k_\nu - x_\nu k_\mu) \big|_{k_0=\sqrt{m^2c^2+\bk^2}} \openone_4 ,
\)
the interpretation of $\psi^\pm(\bk)$ and $\opsi^\pm(\bk)$ in the limit
$(k,m)\to(0,0)$ as creation/annihilation operators is lost. This is exactly
the case if we make the limit $(k,m)\to(0,0)$ in~\eref{5.17}, \eref{5.18},
and~\eref{6.3}--\eref{6.5} (see remark~\ref{Rem5.2}). But, as we said already,
 $\psi^\pm(\bk)$ and $\opsi^\pm(\bk)$ should be operator\ndash valued
spinor distributions (generalized functions) for the solutions~\eref{5.1}.
For them~\eref{6.9} transform into
	\begin{equation}	\label{6.15}
k_\mu \psi^\pm(\bk) (\ope{X}_0) = 0
\quad
k_\mu \opsi^\pm(\bk) (\ope{X}_0) = 0
\qquad
k_0=\sqrt{\bk^2} ,
	\end{equation}
as a result of $m=0$, $\ope{P}_\mu=0$ and~\eref{6.5}. So, if we suppose these
equalities to hold for any $\ope{X}_0\not=0$, we get
	\begin{equation}	\label{6.15new}
k_\mu \psi^\pm(\bk)  = 0
\quad
k_\mu \opsi^\pm(\bk)  = 0
\qquad
k_0=\sqrt{\bk^2}
	\end{equation}
which convert~\eref{5.17} and~\eref{5.18} into identities, due to $m=0$ for
the solutions~\eref{5.1}. So, for example, we can set
	\begin{align}	\label{6.16}
&\psi^\pm(\bk)
= (2\pi\hbar)^{-3/2} \sum_{s} a_s^\pm(\bk) v^{s,\pm}(\bk)
&&
\opsi^\pm(\bk)
= (2\pi\hbar)^{-3/2} \sum_{s} a_s^{\dag\,\pm}(\bk) \overline{v}^{s,\pm}(\bk)
\\			\label{6.17}
& a_s^{\pm}(\bk) = \delta^3(\bk) \alpha_s^{\pm}(\bk)
&&
a_s^{\dag\,\pm}(\bk) = \delta^3(\bk) \alpha_s^{\dag\,\pm}(\bk)
	\end{align}
where $\alpha_s^{\pm}(\bk)$ and $\alpha_s^{\dag\,\pm}(\bk)$ are some
operators, ${v}^{s,\pm}(\bk)$ and $\overline{v}^{s,\pm}(\bk)$ are the
classical spinors defined via the equations~\eref{5.20}--\eref{5.22}, and
the constant factor $(2\pi\hbar)^{-3/2}$ is introduced for future
convenience. (The fact that the solutions~\eref{6.16}--\eref{6.17}
of~\eref{6.15} are not the only ones is inessential for the following.)
Consequently, when the solutions~\eref{5.1} are concerned, the operators
(operator\ndash valued distributions) $\psi^\pm(\bk)$ (resp.\
$\opsi^\pm(\bk)$) decrease (resp.\ increase) state's charge by $q$  and
spin\ndash mixing angular momentum matrix by
$\frac{1}{2}\hbar\sigma_{\mu\nu}$; they preserve the vanishing values of the
4\ndash momentum and orbital angular momentum of the states. However, if we
consider the solutions~\eref{5.1} as a limiting case when
$([\psi,\ope{P}_\mu]_{\_},m)\to(0,0)$ and
$([\opsi,\ope{P}_\mu]_{\_},m)\to(0,0)$, we, by convention, can say that:
	(a) the distributions $\psi^+(\bk)$ (resp.\ $\psi^-(\bk)$) create
(resp.\ annihilate) particles with vanishing 4\ndash momentum and angular
momentum, charge $(-q)$, and spin\ndash mixing angular momentum matrix
$-\frac{1}{2}\hbar\sigma_{\mu\nu}$, and
	(b) the distributions $\opsi^+(\bk)$ (resp.\ $\opsi^-(\bk)$)
create (resp.\ annihilate) particles with vanishing 4\ndash momentum and
angular momentum, charge $(+q)$, and spin\ndash mixing angular momentum
matrix $+\frac{1}{2}\hbar\sigma_{\mu\nu}$.

	Until now the operators $\psi_{s,(\pm)}(k)$ and
$\opsi_{s,(\pm)}(k)$, $s=0$ for $m=0$ and $s=1,2$ for $m\not=0$, entering
in definition~\eref{6.1}, were completely arbitrary linearly independent
solutions of~\eref{5.17} and~\eref{5.18}. As a result, the operators (or
operator\ndash valued spinor distributions) $\psi_s^\pm(\bk)$ and
$\opsi_s^\pm(\bk)$ are arbitrary linearly independent solutions of the
operator equations
	\begin{equation}	\label{6.19}
	\begin{split}
\bigl(\pm\gamma^\mu k_\mu\big|_{k_0=\sqrt{m^2c^2+\bk^2}}
	+ mc\openone_4\bigr) \psi_s^\pm(\bk)
 = 0
\quad
\opsi_s^\pm(\bk) \bigl(
\mp\gamma^\mu k_\mu\big|_{k_0=\sqrt{m^2c^2+\bk^2}} + mc\openone_4\bigr)
 = 0
	\end{split}
	\end{equation}
which, in form, coincide with the \emph{classical} equations~\eref{5.20}. This
fact makes it possible to be separated the `pure operator' part form the
`pure matrix\ndash spinor' part in the spinor operators $\psi_s^\pm(\bk)$ and
$\opsi_s^\pm(\bk)$. The existence of such decompositions is intuitively clear
as one can expect the dynamical variables to be expressible through some
invariant characteristics of the field (the `operator part') while the
transformation properties of the field should be expressible via some
frame\ndash dependent objects of the field (the `matrix part'). This
separation is most conveniently done by expanding $\psi_s^\pm(\bk)$ and
$\opsi_s^\pm(\bk)$ as follows (do not sum over $s$!)
	\begin{equation}	\label{6.20}
\psi_s^\pm(\bk) = (2\pi\hbar)^{-3/2} a_s^\pm(\bk) v^{s,\pm}(\bk)
\qquad
\opsi_s^\pm(\bk)
	= (2\pi\hbar)^{-3/2} a_s^{\dag\,\pm}(\bk) \overline{v}^{s,\pm}(\bk)
	\end{equation}
or, equivalently,
	\begin{equation}	\label{6.20new}
	\begin{split}
\psi^\pm(\bk)
= (2\pi\hbar)^{-3/2} \sum_{s} a_s^\pm(\bk) v^{s,\pm}(\bk)
\quad
\opsi^\pm(\bk)
=
(2\pi\hbar)^{-3/2} \sum_{s} a_s^{\dag\,\pm}(\bk) \overline{v}^{s,\pm}(\bk) .
	\end{split}
	\end{equation}
Here $v^{s,\pm}(\bk)$ and $\overline{v}^{s,\pm}(\bk)$ are defined
via~\eref{5.20}--\eref{5.22},
$a_s^{\pm}(\bk)$ and $a_s^{\dag\,\pm}(\bk)$ are some ($1\times1$
matrix) operators, for solutions different from~\eref{5.1}, or ($1\times1$
matrix) operator\ndash valued distributions, for the solutions~\eref{5.1},
acting on the system's (spinor field's) Hilbert space $\Hil$ of states,
$a_s^{\pm}(\bk),a_s^{\dag\,\pm}(\bk)\colon\Hil\to\Hil$,
and $a_s^{\pm}(\bk)$ and $a_s^{\dag\,\pm}(\bk)$ are such that
	\begin{equation}	\label{6.21}
\bigl(a_s^\pm(\bk)\bigr)^\dag  = a_s^{\dag\,\mp}(\bk)
\qquad
\bigl(a_s^{\dag\,\pm}(\bk)\bigr)^\dag  = a_s^{\mp}(\bk),
	\end{equation}
due to~\eref{6.2} and~\eref{5.21}. We have met the settings~\eref{6.20} above
(see~\eref{6.16}) when the special solutions~\eref{5.1} were considered,
in which case the distributions $a_s^{\pm}(\bk)$  and $a_s^{\dag\,\pm}(\bk)$
can be represented in the form~\eref{6.17}. The operators $a_s^{+}(\bk)$ and
$a_s^{\dag\,+}(\bk)$ (resp.\ $a_s^{-}(\bk)$ and $a_s^{\dag\,-}(\bk)$) will be
referred as the \emph{creation} (resp.\ \emph{annihilation}) operators (of
the field).

	From~\eref{6.9}--\eref{6.11} and~\eref{6.20new}, we,
applying~\eref{5.20}--\eref{5.22}, derive the relations
(with $k_0:=\sqrt{m^2c^2+\bk^2}$):%
\footnote{~%
The relations involving $a_s^\pm(\bk)$ are obtained from the similar ones
involving $\psi^\pm(\bk)$  by multiplying the latter from the left by
$v^{\dag\,s,\mp}(\bk)$. Analogously, the relations involving
$a_s^{\dag\,\pm}(\bk)$ are obtained from the similar ones involving
$\opsi^{\dag\,\pm}(\bk)$ by multiplying the latter from the right by
$\gamma^0 v^{s,\mp}(\bk)$.%

}
	\begin{subequations}	\label{6.22}
	\begin{gather}
			\label{6.22a}
	\begin{split}
 \ope{P}_\mu\bigl( a_s^\pm(\bk) (\ope{X}_p) \bigr)
  = (p_\mu \pm k_\mu) a_s^\pm(\bk) (\ope{X}_p)
\quad
 \ope{P}_\mu\bigl( a_s^{\dag\,\pm}(\bk) (\ope{X}_p) \bigr)
  = (p_\mu \pm k_\mu) a_s^{\dag\,\pm}(\bk) (\ope{X}_p)
	\end{split} 
\displaybreak[1]\\			\label{6.22b}
	\begin{split}
 \ope{Q}\bigl( a_s^\pm(\bk) (\ope{X}_e) \bigr)
  = (e-q) a_s^\pm(\bk) (\ope{X}_e)
\quad
 \ope{Q}\bigl( a_s^{\dag\,\pm}(\bk) (\ope{X}_e) \bigr)
  = (e+q) a_s^{\dag\,\pm}(\bk) (\ope{X}_e)
	\end{split} 
\displaybreak[1]\\			\label{6.22c}
\left.	\begin{split}
\ope{M}_{\mu\nu}(x)  \bigl( a_s^\pm(\bk) (\ope{X}_m) \bigr)
 =  \{ m_{\mu\nu}(x)
& \pm (x_\mu k_\nu - x_\nu k_\mu) \} a_s^\pm(\bk) (\ope{X}_m)
\\
&	- \frac{1}{2}\hbar \sum_{t} \Hat{\sigma}_{\mu\nu}^{st,\pm}(\bk)
 	a_t^\pm(\bk) (\ope{X}_m)
\\
\ope{M}_{\mu\nu}(x)   \bigl( a_s^{\dag\,\pm}(\bk) (\ope{X}_m) \bigr)
=  \{ m_{\mu\nu}(x)
& \pm (x_\mu k_\nu - x_\nu k_\mu) \} a_s^{\dag\,\pm}(\bk) (\ope{X}_m)
\\
&	+ \frac{1}{2}\hbar \sum_{t} \Hat{\sigma}_{\mu\nu}^{ts,\mp}(\bk)
	a_t^{\dag\,\pm}(\bk) (\ope{X}_m) ,
	\end{split}\right\}
	\end{gather}
	\end{subequations}
where
	\begin{equation}	\label{6.23}
\Hat{\sigma}_{\mu\nu}^{st,\pm} (\bk)
: =
v^{\dag\,s,\mp}(\bk) \sigma_{\mu\nu} v^{t,\pm}(\bk)
=
- \Hat{\sigma}_{\nu\mu}^{st,\pm} (\bk)
=
\overline{v}^{s,\mp}(\bk) \gamma^0 \sigma_{\mu\nu} v^{t,\pm}(\bk)
	\end{equation}
and the indices $s$ and $t$ take the value $0$ for $m=0$ and the values~1
and~2 for $m\not=0$. The numbers $\sigma_{\mu\nu}^{st,\pm} (\bk)$, appearing
in~\eref{6.22c}, generally depend on all of the arguments indicated. We shall
comment on them in Sect.~\ref{Sect7}.

	For the solutions~\eref{5.1}, the equations~\eref{6.22} reduce to:
	\begin{subequations}	\label{6.24}
	\begin{gather}
			\label{6.24a}
k_\mu a_s^\pm(\bk) (\ope{X}_0) = 0
\qquad
k_\mu a_s^{\dag\,\pm}(\bk) (\ope{X}_0) = 0
\displaybreak[1]\\			\label{6.24b}
	\begin{split}
 \ope{Q}\bigl( a_s^\pm(\bk) (\ope{X}_e) \bigr)
  = (e-q) a_s^\pm(\bk) (\ope{X}_e)
\quad
 \ope{Q}\bigl( a_s^{\dag\,\pm}(\bk) (\ope{X}_e) \bigr)
  = (e+q) a_s^{\dag\,\pm}(\bk) (\ope{X}_e)
	\end{split} 
\displaybreak[1]\\			\label{6.24c}
	\begin{split}
\ope{M}_{\mu\nu}(x)  \bigl( a_s^\pm(\bk) (\ope{X}_m) \bigr)
 =   m_{\mu\nu}(x) a_s^\pm(\bk) (\ope{X}_m)
   - \frac{1}{2}\hbar \sum_{t} \Hat{\sigma}_{\mu\nu}^{st,\pm}(\bk)
     a_t^\pm(\bk) (\ope{X}_m)
\\
\ope{M}_{\mu\nu}(x)   \bigl( a_s^{\dag\,\pm}(\bk) (\ope{X}_m) \bigr)
 =   m_{\mu\nu}(x) a_s^{\dag\,\pm}(\bk) (\ope{X}_m)
   + \frac{1}{2}\hbar \sum_{t} \Hat{\sigma}_{\mu\nu}^{ts,\mp}(\bk)
     a_t^{\dag\,\pm}(\bk) (\ope{X}_m),
	\end{split}
	\end{gather}
	\end{subequations}
where, the quantities $m_{\mu\nu}(x)$ in~\eref{6.24c} contain only a spin
angular momentum.

	As a consequence of~\eref{6.22}, the interpretation of
$a_s^{\pm}(\bk)$ and $a_s^{\dag\,\pm}(\bk)$ is almost the same as the one of
$\psi^{\pm}(\bk)$ and $\opsi^{\pm}(\bk)$, respectively, with an only
change concerning the spin angular momentum.

	If $m=0$, then $s,t=0$ and the equations~\eref{6.24c} say that the
particles created/annihilated by $a_0^\pm(\bk)$ possess spin angular momentum
$-\frac{1}{2}\hbar\Hat{\sigma}_{\mu\nu}^{00,\pm}(\bk)$
and the ones created/annihilated by $a_0^{\dag\,\pm}(\bk)$  have
a spin angular momentum equal to
$+\frac{1}{2}\hbar\Hat{\sigma}_{\mu\nu}^{00,\mp}(\bk)$.
However, for $m\not=0$, the eigenstates of $\ope{M}_{\mu\nu}$ are, generally,
\emph{not} mapped into such states by $a_0^{\pm}(\bk)$ and
$a_0^{\dag\,\pm}(\bk)$ as they are mixed through the quantities
$\Hat{\sigma}_{\mu\nu}^{st,\pm}(\bk)$ via the polarization indices they
carry. Now the matrices
 $-\frac{1}{2}\hbar \Hat{\sigma}_{\mu\nu}^{\pm}(\bk)$ and
 $+\frac{1}{2}\hbar \Hat{\sigma}_{\mu\nu}^{\mp}(\bk)$, with
	\begin{equation}	\label{6.25}
\Hat{\sigma}_{\mu\nu}^{\pm}(\bk)
:=
\bigl[ \Hat{\sigma}_{\mu\nu}^{st,\pm}(\bk) \bigr]_{s,t=1}^{2}
	\end{equation}
play a role of \emph{polarization-mixing matrices} since we can
rewrite~\eref{6.24c} for $m\not=0$ as
	\begin{gather}
			\label{6.26}
	\begin{split}
\ope{M}_{\mu\nu}(x)  \bigl( a^\pm(\bk) (\ope{X}_m) \bigr)
& =
\{  m_{\mu\nu}(x) \openone_2
  - \frac{1}{2}\hbar \Hat{\sigma}_{\mu\nu}^{\pm}(\bk)
\}
  a^\pm(\bk) (\ope{X}_m)
\\
\ope{M}_{\mu\nu}(x)   \bigl( a^{\dag\,\pm}(\bk) (\ope{X}_m) \bigr)
& =
a_t^{\dag\,\pm}(\bk) (\ope{X}_m)
\{  m_{\mu\nu}(x) \openone_2
  + \frac{1}{2}\hbar \Hat{\sigma}_{\mu\nu}^{\mp}(\bk)
\} ,
	\end{split}
	\end{gather}
where
\(
\openone_2 :=
	     \bigl( \begin{smallmatrix}
	     1&0 \\ 0&1
	     \end{smallmatrix} \bigr)
\)
is the unit $2\times2$ matrix,
\(
a^\pm(\bk) :=
	     \Big( \begin{smallmatrix}
	     a_1^\pm(\bk)  \\ a_2^\pm(\bk)
	     \end{smallmatrix} \Big) ,
\)
and
\(
a^{\dag\,\pm}(\bk) :=
		     \Big( \begin{smallmatrix}
	     	     a_1^{\dag\,\pm}(\bk)  \\ a_2^{\dag\,\pm}(\bk)
	     	     \end{smallmatrix} \Big) .
\)
So, we can say that $a^\pm(\bk)$ (resp.\ $a^{\dag\,\pm}(\bk)$)
creates/annihilates particles (states) with polarization\ndash mixing
matrices $-\frac{1}{2}\hbar\Hat{\sigma}_{\mu\nu}^{\pm}(\bk)$ (resp.\
$+\frac{1}{2}\hbar\Hat{\sigma}_{\mu\nu}^{\mp}$). This interpretation holds
also for the massless case, $m=0$, if we set $a^{\pm}(\bk):=a_0^{\pm}(\bk)$,
 $a^{\dag\,\pm}(\bk):=a_0^{\dag\,\pm}(\bk)$, and
 $\Hat{\sigma}_{\mu\nu}^{\pm}(\bk):=\Hat{\sigma}_{\mu\nu}^{00,\pm}(\bk)$
in that case. It is worth mentioning, as a consequence of~\eref{6.24}, the
particles and antiparticles of a free spinor field are always different
regardless of their mass $m$ and charge $q$.

	Ending this section, we must note, the interpretation of
$\psi^\pm(\bk)$, $\opsi^\pm(\bk)$, $a_s^\pm(\bk)$, and $a_s^{\dag\,\pm}(\bk)$
as operators creating/annihilating particles with fixed charge and spin
(polarization) mixing matrices is entirely based on the equations~\eref{3.22}
and~\eref{3.23}, or, more generally, on~\eref{2.26} and~\eref{2.27}, which
are \emph{external} to the Lagrangian formalism and whose validity depends on
the particular Lagrangian employed. In particular, for the
Lagrangian~\eref{3.3} (or~\eref{3.1} in Heisenberg picture), the
equalities~\eref{3.23} do not hold for $\mu=0$ and $\nu=1,2,3$ --- see,
e.g.,~\eref{3.29} and~\eref{2.15}. The below\ndash written
Lagrangian~\eref{12.4} (or~\eref{12.3} in Heisenberg picture), used
in~\cite{Bjorken&Drell}, is an example of a one for which all of the
equalities~\eref{3.22} and~\eref{3.23} hold. With an exception of the present
sections, the equations~\eref{3.22} and~\eref{3.23} are not used in this
work.


\section
[The dynamical variables in terms of creation and annihilation operators]
{The dynamical variables in terms of\\ creation and annihilation operators}
\label{Sect7}

	The main purpose of this section is a technical one: to be derived
expressions for the momentum, charge and angular momentum operators in terms
of the creation and annihilation operators introduced in Sect.~\ref{Sect6}.
The results will promote the interpretation of these operators and will be
applied significantly in the subsequent sections.

	As~\eref{6.3}--\eref{6.5} imply
	\begin{equation}	\label{7.1}
	\begin{split}
[\psi,\ope{P}_\mu]_{\_}
& =
\sum_{s}\int \{ k_\mu \big|_{k_0=\sqrt{m^2c^2+\bk^2}}
	(- \psi_s^{+}(\bk) + \psi_s^{-}(\bk)) \} \Id^3\bk
\\
[\opsi,\ope{P}_\mu]_{\_}
& =
\sum_{s}\int \{ k_\mu \big|_{k_0=\sqrt{m^2c^2+\bk^2}}
	(- \opsi_s^{+}(\bk) + \opsi_s^{-}(\bk)) \} \Id^3\bk,
	\end{split}
	\end{equation}
the energy-momentum operator~\eref{3.17}, by virtue of~\eref{6.3}
and~\eref{6.4}, can be written as
	\begin{multline}	\label{7.2}
\ope{T}_{\mu\nu}
=
\frac{1}{2} c \sum_{s,s'} \int \Id^3\bk \Id\bk'
\bigl\{
k'_\nu \big|_{k'_0=\sqrt{m^2c^2+(\bk')^2}}
\bigl(  \opsi_s^{+}(\bk) + \opsi_s^{-}(\bk) \bigr)
\gamma_\mu
 \circ
\bigl( - \psi_{s'}^{+}(\bk') + \psi_{s'}^{-}(\bk') \bigr)
\\
- k_\nu \big|_{k_0=\sqrt{m^2c^2+\bk^2}}
 \bigl( - \opsi_s^{+}(\bk) + \opsi_s^{-}(\bk) \bigr)
 \gamma_\mu \circ
 \bigl(   \psi_{s'}^{+}(\bk') + \psi_{s'}^{-}(\bk') \bigr)
\bigl\} .
	\end{multline}

	Since the spinors $\psi$ and $\opsi$ satisfy the Klein-Gordon
equation~\eref{5.11}, this is true also for their components and,
consequently, according
to~\cite[eq.~(6.4)]{bp-QFTinMP-scalars}\typeout{!!! Bozho, check eq. 6.4
?????}, the equality
	\begin{gather}
			\label{7.3}
\varphi^\varepsilon(\bk)\circ \varphi^{\varepsilon'}(\bk')\circ\ope{U}(x,x_0)
=
\e^{-\iih(x^\mu-x_0^\mu)(\varepsilon k_\mu + \varepsilon' k'_\mu)}
 \ope{U}(x,x_0) \varphi^\varepsilon(\bk) \circ \varphi^{\varepsilon'}(\bk')
	\end{gather}
holds for $\varepsilon,\varepsilon'=+,-$,
$k_0=\sqrt{m^2c^2+\bk^2}$, $k'_0=\sqrt{m^2c^2+(\bk')^2}$,
\(
\varphi^\varepsilon(\bk)
= \psi_{\alpha,s}^{\varepsilon}(\bk),\opsi_{\alpha,s}^{\varepsilon}(\bk),
\)
with $\alpha$ being an index denoting spinor's components, and
$\ope{U}(x,x_0)$ being the operator~\eref{12.112} by means of which the
transition from Heisenberg to momentum picture is performed.

	Substituting~\eref{7.2} into~\eref{2.6}, commuting $\ope{U}(x,x_0)$
according to~\eref{7.3} until it meets $\ope{U}^{-1}(x,x_0)$, performing the
integration over $x$ (resulting in  $(2\pi\hbar)^3\delta^3(\bk\pm\bk')$ and
a phase factor), and, at last, integrating over $\bk'$, we get%
\footnote{~
	The integrals, appearing in the transition from~\eref{2.21}
to~\eref{7.4} and similar ones required for the derivation of~\eref{7.7}
and~\eref{7.10} below, are of the type
	\begin{multline}	\label{7.4-1}
\mspace{-6mu}
J
 = \sum_{\alpha} \sum_{\varepsilon,\varepsilon'=+,-} \sum_{s, s'}
\int\Id^3\bs x \int\Id^3\bk\Id^3\bk'
\ope{U}^{-1}(x,x_0) \circ
\{ \opsi_s^\varepsilon(\bk)
A_{\varepsilon\varepsilon'}^{\alpha} (\bk,\bk')
   \circ \psi_{s'}^{\varepsilon'}(\bk')
\}
\circ \ope{U}(x,x_0)
\\
 = \sum_{\alpha} \sum_{\varepsilon,\varepsilon'=+,-} \sum_{s, s'}
\int\Id^3\bs x \int\Id^3\bk\Id^3\bk'
\{ \opsi_s^\varepsilon(\bk)
A_{\varepsilon\varepsilon'}^{\alpha} (\bk,\bk')
   \circ \psi_{s'}^{\varepsilon'}(\bk')
\}
\e^{-\iih (x^\lambda-x_0^\lambda)
	(\varepsilon k_\lambda + \varepsilon' k'_\lambda)} ,
	\end{multline}
where $A_{\varepsilon\varepsilon'}^{\alpha} (\bk,\bk')$, $\alpha=1,2,\dots$,
are some matrices, $k_0=\sqrt{m^2c^2+\bk^2}$, $k'_0=\sqrt{m^2c^2+{\bk'}^2}$,
and~\eref{7.3} was applied. Representing the exponent in the integrand as
\[
\e^{-\iih (x^0-x_0^0) (\varepsilon k_0 + \varepsilon' k'_0) }
\e^{+\iih \sum_{a=1}^{3} x_0^a (\varepsilon k_a + \varepsilon' k'_a) }
\e^{ - \iih \sum_{a=1}^{3} x^a (\varepsilon k_a + \varepsilon' k'_a) }
\]
and taking into account that the integral over $\bs x$ results in
 $(2\pi\hbar)^3 \delta^3(\varepsilon \bk + \varepsilon' \bk')$,
we see that, after trivial integration over $\bk'$, the above integral takes
the form
	\begin{multline}	\label{7.4-2}
J =
(2\pi\hbar)^3
\sum_{\alpha} \sum_{\varepsilon,\varepsilon'=+,-} \sum_{s, s'} \int\Id^3\bk
\e^{ - \iih (x^0-x_0^0) k_0 (\varepsilon 1 + \varepsilon' 1) }
  \opsi_s^\varepsilon(\bk)
A_{\varepsilon\varepsilon'}^{\alpha} (\bk,-(\varepsilon1)(\varepsilon'1) \bk)
  \circ \psi_{s'}^{\varepsilon'}(-(\varepsilon1)(\varepsilon'1) \bk) .
	\end{multline}
The reader can easily write the concrete form of the matrices
$A_{\varepsilon\varepsilon'}^{\alpha} (\bk,\bk')$ in the particular cases we
consider in this paper.%
} 
	\begin{multline}	\label{7.4}
\ope{P}_\mu
=
(2\pi\hbar)^3 \sum_{s,s'} \int\Id^3\bk
k_\mu \big|_{k_0=\sqrt{m^2c^2+\bk^2}}
\bigl\{
\bigl(	\opsi_{s}^+(\bk) \gamma^0 \circ \psi_{s'}^-(\bk)
-
	\opsi_{s}^-(\bk) \gamma^0 \circ \psi_{s'}^+(\bk) \bigr)
\\\displaybreak[1]
+
( \delta_{1\mu} + \delta_{2\mu} + \delta_{3\mu} )
\e^{ -\frac{2}{\ih} (x^0-x_0^0) \sqrt{m^2c^2+\bk^2} }
\opsi_{s}^+(\bk) \gamma^0 \circ \psi_{s'}^+(-\bk)
\\
-
( \delta_{1\mu} + \delta_{2\mu} + \delta_{3\mu} )
\e^{ +\frac{2}{\ih}(x^0-x_0^0) \sqrt{m^2c^2+\bk^2} }
\opsi_{s}^-(\bk) \gamma^0 \circ \psi_{s'}^-(-\bk)
\bigr\} .
	\end{multline}
Inserting the expansions~\eref{6.20} in the last result and
applying~\eref{5.22} (respectively~\eref{5.23}) to the different (resp.\
equal) frequency terms in the obtained expression, we derive the familiar
result (cf.~\cite{Bogolyubov&Shirkov,Bjorken&Drell-2})
	\begin{equation}	\label{7.5}
\ope{P}_\mu
=
\sum_{s}\int
  k_\mu |_{ k_0=\sqrt{m^2c^2+{\bs k}^2} }
\{
a_s^{\dag\,+}(\bk)\circ a_s^-(\bk)
-
a_s^{\dag\,-}(\bk)\circ a_s^+(\bk)
\}
\Id^3\bk .
	\end{equation}

	We turn now our attention to the charge operator in Heisenberg
picture.%
\footnote{~%
In momentum picture, it will be found in Sect.~\ref{Sect8}; see~\eref{8.17}.%
}
In view of the decompositions~\eref{6.3}--\eref{6.4}, the
current operator~\eref{3.18} reads
	\begin{equation}	\label{7.6}
\ope{J}_{\mu}
=
q c \sum_{s,s'} \int \Id^3\bk \Id\bk'
\bigl\{
\bigl(  \opsi_s^{+}(\bk) + \opsi_s^{-}(\bk) \bigr)
\gamma_\mu \circ
\bigl(  \psi_{s'}^{+}(\bk') + \psi_{s'}^{-}(\bk') \bigr)
\bigl\} .
	\end{equation}
Substituting the last equation into~\eref{2.13}, commuting the
operator $\ope{U}(x,x_0)$ with $\psi_s^{+}(\bk)$ and $\opsi_s^{+}(\bk)$
according to~\eref{7.3}, integrating over $x$ (which gives
$(2\pi\hbar)^3\delta^3(\bk\pm\bk')$ and a phase factor), and integrating over
$\bk'$, we obtain
       \begin{multline}	\label{7.7}
\tope{Q}
=
q (2\pi\hbar)^3 \sum_{s,s'} \Id^3\bk
\bigl\{
\bigl(	\opsi_{s}^+(\bk) \gamma^0 \circ \psi_{s'}^-(\bk)
+
	\opsi_{s}^-(\bk) \gamma^0 \circ \psi_{s'}^+(\bk) \bigr)
\\ +
\e^{ -\frac{2}{\ih}(x^0-x_0^0) \sqrt{m^2c^2+\bk^2} }
\bigl(  \opsi_{s}^+(\bk) \gamma^0 \circ \psi_{s'}^+(-\bk)
 +
\e^{ +\frac{2}{\ih}(x^0-x_0^0) \sqrt{m^2c^2+\bk^2} }
	\opsi_{s}^-(\bk) \gamma^0 \circ \psi_{s'}^-(-\bk) \bigr)
\bigr\} .
	\end{multline}
At the end, the insertion of~\eref{6.20} here
entails (cf.~\cite{Bogolyubov&Shirkov,Bjorken&Drell-2})
	\begin{equation}	\label{7.8}
\tope{Q}
=
q \sum_{s}\int
\{
a_s^{\dag\,+}(\bk)\circ a_s^-(\bk) + a_s^{\dag\,-}(\bk)\circ a_s^+(\bk)
\} \Id^3\bk
	\end{equation}
where~\eref{5.22} and~\eref{5.23} were used.

	Now comes the order of the angular momentum to be expressed via the
creation and annihilation operators. Combining~\eref{3.19}, \eref{3.20}
and~\eref{6.3}--\eref{6.4}, we see that
      \begin{gather}
			\label{7.9}
\ope{M}_{\mu\nu}^\lambda
= \ope{L}_{\mu\nu}^\lambda + \ope{S}_{\mu\nu}^\lambda
=
  x_\mu \Sprindex[\ope{T}]{\nu}{\lambda}(x)
- x_\nu \Sprindex[\ope{T}]{\mu}{\lambda}(x)
+ \ope{S}_{\mu\nu}^\lambda
	\end{gather}
where $\ope{T}_{\mu\nu}$ is given by~\eref{7.2} and the spin angular momentum
density is
	\begin{equation}	\label{7.10}
{S}_{\mu\nu}^\lambda
=
\frac{1}{4} \hbar c \sum_{s,s'} \int\Id^3\bk \Id^3\bk'
\bigl\{
\bigl(  \opsi_s^{+}(\bk) + \opsi_s^{-}(\bk) \bigr)
( \gamma^\lambda \sigma_{\mu\nu} + \sigma_{\mu\nu}\gamma^\lambda )
 \circ
\bigl(  \psi_{s'}^{+}(\bk') + \psi_{s'}^{-}(\bk') \bigr)
\bigl\} .
	\end{equation}
Taking into account~\eref{3.28}, substituting~\eref{7.10} into~\eref{2.15}
and performing with the r.h.s.\ of the obtained equality manipulations
similar to the ones leading from~\eref{7.6} to~\eref{7.7}, we derive the
following representation for the spin angular momentum in Heisenberg picture:
	\begin{subequations}	\label{7.11}
	\begin{equation}	\label{7.11a}
\tope{S}_{0a} = 0
	\end{equation}
\vspace{-3.5ex}
	\begin{multline}	\label{7.11b}
\tope{S}_{ab}
=
\frac{1}{2}\hbar (2\pi\hbar)^3 \sum_{s,s'} \int \Id^3\bk
\bigl\{
\bigl( \opsi_{s}^+(\bk) \gamma^0 \sigma_{ab} \circ \psi_{s'}^-(\bk)
+
       \opsi_{s}^-(\bk) \gamma^0 \sigma_{ab} \circ \psi_{s'}^+(\bk) \bigr)
\\\displaybreak[1]
+
\e^{ - \frac{2}{\ih}(x^0-x_0^0) \sqrt{m^2c^2+\bk^2} }
\opsi_{s}^+(\bk) \gamma^0 \sigma_{ab} \circ \psi_{s'}^+(-\bk)
+
\e^{ +\frac{2}{\ih} (x^0-x_0^0) \sqrt{m^2c^2+\bk^2} }
\opsi_{s}^-(\bk) \gamma^0 \sigma_{ab} \circ \psi_{s'}^-(-\bk)
\bigr\}
	\end{multline}
	\end{subequations}
where  $a,b=1,2,3$.

	Similar transformations of the orbital angular momentum~\eref{2.14}
give its value in Heisenberg picture as%
\footnote{~
	To prove~\eref{7.12}, one has to substitute~\eref{7.2}
into~\eref{2.14}, then to apply~\eref{7.3}, to calculate the space integral
over $x$, and, at the end, to integrate over $\bk'$. The integrals, one has
to calculate, are of the type
	\begin{multline}	\label{7.12-1}
J
 = \sum_{\alpha} \sum_{\varepsilon,\varepsilon'=+,-} \sum_{s, s'}
\int\Id^3\bs x \int\Id^3\bk\Id^3\bk'
x_a
\ope{U}^{-1}(x,x_0)
 \circ
\{ \opsi_s^\varepsilon(\bk)
A_{\varepsilon\varepsilon'}^{\alpha} (\bk,\bk')
   \circ \psi_{s'}^{\varepsilon'}(\bk')
\}
\circ\ope{U}(x,x_0)
\\
 = \sum_{\alpha} \sum_{\varepsilon,\varepsilon'=+,-} \sum_{s, s'}
\int\!\!\Id^3\bs x \int\!\!\Id^3\bk\Id^3\bk'
\{ \opsi_s^\varepsilon(\bk)
A_{\varepsilon\varepsilon'}^{\alpha} (\bk,\bk')
   \circ \psi_{s'}^{\varepsilon'}(\bk')
\}
x_a
\e^{-\iih (x^\lambda-x_0^\lambda)
	(\varepsilon k_\lambda + \varepsilon' k'_\lambda)} ,
	\end{multline}
where $a=1,2,3$, $A_{\varepsilon\varepsilon'}^{\alpha} (\bk,\bk')$,
$\alpha=1,2,\dots$, are some matrices, $k_0=\sqrt{m^2c^2+\bk^2}$, and
$k'_0=\sqrt{m^2c^2+{\bk'}^2}$. The integration over $x$ results in
\(
(2\pi\hbar)^3 \Bigl( -\ih
\frac{\pd}
{ \pd( \varepsilon k^a + \varepsilon' k^{\prime\, a} ) }
\Bigr) \delta^3 ( \varepsilon \bk + \varepsilon' \bk' ) .
\)
Simple manipulation with the remaining terms, by invoking the equality
\(
f(y)\frac{\pd\delta(y)}{\pd y} = -\frac{\pd f(y)} {\pd y} \delta(y)
\)
in the form
	\begin{multline*}
\int \Id^3\bs y \Id^3\bs z
f(\bs y,\bs z) \frac{\pd\delta^3(\bs y-\bs z)}{\pd(y^a-z^a)}
=
\int \Id^3\bs y \Id^3\bs z
\frac{1}{2}\bigl( f(\bs y,\bs z) - f(\bs z,\bs y) \bigr)
\frac{\pd\delta^3(\bs y-\bs z)}{\pd(y^a-z^a)}
\\
=
- \frac{1}{2} \int \Id^3\bs y \Id^3\bs z  \delta^3(\bs y-\bs z)
\Big( \frac{\pd}{\pd y^a} - \frac{\pd}{\pd z^a} \Bigr)
 f(\bs y,\bs z) ,
	\end{multline*}
gives the following result:
	\begin{multline}	\label{7.12-2}
J =
(2\pi\hbar)^3
  \sum_{\alpha} \sum_{\varepsilon,\varepsilon'=+,-} \sum_{s, s'}
  \int\Id^3\bk\Id^3\bk'
\delta^3(\varepsilon \bk + \varepsilon'\bk')
\e^{ - \iih (x^0-x_0^0) k_0 (\varepsilon 1 + \varepsilon' 1) }
\\ \times
\Bigl\{
(x^0-x_0^0) \frac{1}{2}\Bigl( \frac{k_a}{k_0} + \frac{k'_a}{k_0} \Bigr)
+ x_{0a}
+ \frac{1}{2} \ih
\Bigl( \varepsilon \frac{\pd}{\pd k^a}
     + \varepsilon'\frac{\pd}{\pd {k'}^a} \Bigr)
\Bigr\}
\bigl\{  \opsi_s^\varepsilon(\bk)
A_{\varepsilon\varepsilon'}^{\alpha} (\bk,\bk')
  \circ \psi_{s'}^{\varepsilon'}(\bk') \bigr\} .
	\end{multline}
The particular form of $A_{\varepsilon\varepsilon'}^{\alpha} (\bk,\bk')$ is
clear from~\eref{2.14} and~\eref{7.2}. So, applying several
times~\eref{7.12-2}, calculating the appearing derivatives, and, at last,
performing the trivial integration over $\bk$ or $\bk'$ by means of
 $\delta^3(\varepsilon \bk + \varepsilon'\bk')$,
one can derive~\eref{7.12} after simple, but lengthy and tedious algebraic
manipulations.%
}
\vspace{-3ex}
	\begin{multline}	\label{7.12}
\tope{L}_{\mu\nu} ( x_{0\, 0} )
=
x_{0\,\mu} \ope{P}_\nu  - x_{0\,\nu} \ope{P}_\mu
+
\frac{1}{2} \ih (2\pi\hbar)^3 \sum_{s,s'} \int\Id^3\bk
\Bigl\{
\opsi_{s}^+(\bk) \gamma^0
\Bigl( \xlrarrow{ k_\mu \frac{\pd}{\pd k^\nu} }
     - \xlrarrow{ k_\nu \frac{\pd}{\pd k^\mu} } \Bigr)
\circ \psi_{s'}^-(\bk)
\\  +
\opsi_{s}^-(\bk) \gamma^0
\Bigl( \xlrarrow{ k_\mu \frac{\pd}{\pd k^\nu} }
     - \xlrarrow{ k_\nu \frac{\pd}{\pd k^\mu} } \Bigr)
\circ \psi_{s'}^+(\bk)
\displaybreak[3]\\
-
\e^{-\frac{2}{\ih}(x^0-x_0^0) k_0}
\Bigl[
\Bigl(-\delta_{0\mu}\frac{k_\nu}{k_0} + \delta_{0\nu}\frac{k_\mu}{k_0} \Bigr)
 +
   ( \delta_{1\mu} + \delta_{2\mu} + \delta_{3\mu} )
   ( \delta_{1\nu} + \delta_{2\nu} + \delta_{3\nu} )
\\ \times
\Bigl( k_\mu \Bigl( \frac{\pd}{\pd k^\nu} + \frac{\pd}{\pd k^{\prime\,\nu}}
	     \Bigr)
-
       k_\nu \Bigl( \frac{\pd}{\pd k^\mu} + \frac{\pd}{\pd k^{\prime\,\mu}}
	      \Bigr)
\Bigr)
\Bigr]
\bigl( \opsi_{s}^+(\bk) \gamma^0 \circ \psi_{s'}^+(\bk') \bigr)
\displaybreak[2]\\
-
\e^{+\frac{2}{\ih}(x^0-x_0^0) k_0}
\Bigl[
\Bigl(-\delta_{0\mu}\frac{k_\nu}{k_0} + \delta_{0\nu}\frac{k_\mu}{k_0} \Bigr)
-
	( \delta_{1\mu} + \delta_{2\mu} + \delta_{3\mu} )
	( \delta_{1\nu} + \delta_{2\nu} + \delta_{3\nu} )
\\ \times
\Bigl( k_\mu \Bigl( \frac{\pd}{\pd k^\nu} + \frac{\pd}{\pd k^{\prime\,\nu}}
	     \Bigr)
-
       k_\nu \Bigl( \frac{\pd}{\pd k^\mu} + \frac{\pd}{\pd k^{\prime\,\mu}}
	      \Bigr)
\Bigr)
\Bigr]
\bigl( \opsi_{s}^-(\bk) \gamma^0 \circ \psi_{s'}^-(\bk') \bigr)
\Bigr\}
\Big|_{\begin{subarray}{l}
	\bk'=-\bk\\
	k_0=\sqrt{m^2c^2+\bk^2}
	\end{subarray}
} \ \ ,
	\end{multline}
where~\eref{7.4} was taken into account, the derivatives with respect to
$k_0$, like $\frac{\pd}{\pd k^0}\psi_{s}^{\pm}(\bk)$, must be set equal to
zero, and%
\footnote{~%
More generally, if $\omega\colon\{\Hil\to\Hil\}\to\{\Hil\to\Hil\}$ is a
mapping on the operator space over the system's Hilbert space, we put
 $A\xlrarrow{\omega}\circ B := -\omega(A)\circ B + A\circ \omega(B)$
for any $A,B\colon\Hil\to\Hil$. Usually~\cite{Bjorken&Drell,Itzykson&Zuber},
this notation is used for $\omega=\pd_\mu$.%
}
	\begin{multline}	\label{7.12-3new}
A(\bk) \xlrarrow{ k_\mu\frac{\pd}{\pd k^\nu} } \circ B(\bk)
:=
-
\Bigl( k_\mu\frac{\pd A(\bk)}{\pd k^\nu} \Bigr) \circ B(\bk)
+
\Bigl( A(\bk) \circ k_\mu\frac{\pd B(\bk)}{\pd k^\nu} \Bigr)
\\ =
k_\mu \Bigl(  A(\bk) \xlrarrow{ \frac{\pd}{\pd k^\nu} } \circ B(\bk) \Bigr)
	\end{multline}
for (matrix) operators $A(\bk)$ and $B(\bk)$ having $C^1$ dependence on
$\bk$.
	If the operators $A(\bk)$ and $B(\bk)$ tend to zero sufficiently fast
at spacial infinity, then, by integration by parts, one can prove the
equality
	\begin{multline}	\label{7.12-3new1}
\int\Id^3\bk
\Bigl\{ A(\bk)
\Bigl(
  \xlrarrow{ k_\mu\frac{\pd}{\pd k^\nu} }
- \xlrarrow{ k_\nu\frac{\pd}{\pd k^\mu} }
\Bigr) \circ B(\bk)
\Bigr\}
\Big|_{	k_0=\sqrt{m^2c^2+\bk^2} }
\\
=
2 \int\Id^3\bk
\Bigl\{ A(\bk) \circ
\Bigl(
  { k_\mu\frac{\pd}{\pd k^\nu} }
- { k_\nu\frac{\pd}{\pd k^\mu} }
\Bigr) B(\bk)
\Bigr\}
\Big|_{	k_0=\sqrt{m^2c^2+\bk^2} }
\\
=
-2 \int\Id^3\bk
\Bigl\{
\Bigl( \Bigl(
  { k_\mu\frac{\pd}{\pd k^\nu} }
- { k_\nu\frac{\pd}{\pd k^\mu} }
\Bigr) A(\bk) \Bigr) \circ B(\bk)
\Bigr\}
\Big|_{	k_0=\sqrt{m^2c^2+\bk^2} } \ .
	\end{multline}
By means of these equations, one can reduce (two times) the number of terms
in~\eref{7.12}, but we prefer to retain the `more (anti)symmetric' form of
the results by invoking the operation introduced via~\eref{7.12-3new}.

	Let us represent the spin and orbital angular momentum
operators~\eref{7.11} and~\eref{7.12} as sums of time\ndash dependent and
time\ndash independent terms:
	\begin{equation}	\label{7.12-3}
\tope{S}_{\mu\nu}
= \lindex[\mspace{-6mu}{\tope{S}_{\mu\nu}}]{}{0}
	+ \lindex[\mspace{-6mu}{\tope{S}_{\mu\nu}}]{}{1} (x^0)
\qquad
\tope{L}_{\mu\nu}
= \lindex[\mspace{-6mu}{\tope{L}_{\mu\nu}}]{}{0}
	+ \lindex[\mspace{-6mu}{\tope{L}_{\mu\nu}}]{}{1} (x^0) ,
	\end{equation}
where
	\begin{subequations}	\label{7.12-4}
	\begin{multline}	\label{7.12-4a}
\lindex[\mspace{-6mu}{\tope{S}_{\mu\nu}}]{}{1} (x^0)
=
\frac{1}{2}\hbar (2\pi\hbar)^3
	( \delta_{1\mu} + \delta_{2\mu} + \delta_{3\mu} )
	( \delta_{1\nu} + \delta_{2\nu} + \delta_{3\nu} )
\\\times
 \sum_{s,s'} \int \Id^3\bk
\bigl\{
\e^{ - \frac{2}{\ih}(x^0-x_0^0) \sqrt{m^2c^2+\bk^2} }
\opsi_{s}^+(\bk) \gamma^0 \sigma_{\mu\nu} \circ \psi_{s'}^+(-\bk)
\\+
\e^{ +\frac{2}{\ih} (x^0-x_0^0) \sqrt{m^2c^2+\bk^2} }
\opsi_{s}^-(\bk) \gamma^0 \sigma_{\mu\nu} \circ \psi_{s'}^-(-\bk)
\bigr\}
	\end{multline}
\vspace{-3ex}
	\begin{multline}	\label{7.12-4b}
\lindex[\mspace{-6mu}{\tope{L}_{\mu\nu}}]{}{1} (x^0)
=
- \frac{1}{2} \ih (2\pi\hbar)^3 \sum_{s,s'} \int\Id^3\bk
\Bigl\{
\e^{-\frac{2}{\ih}(x^0-x_0^0) k_0}
\Bigl[
\Bigl(-\delta_{0\mu}\frac{k_\nu}{k_0} + \delta_{0\nu}\frac{k_\mu}{k_0} \Bigr)
\\
+
   ( \delta_{1\mu} + \delta_{2\mu} + \delta_{3\mu} )
   ( \delta_{1\nu} + \delta_{2\nu} + \delta_{3\nu} )
\\ \times
\Bigl( k_\mu \Bigl( \frac{\pd}{\pd k^\nu} + \frac{\pd}{\pd k^{\prime\,\nu}}
	     \Bigr)
-
       k_\nu \Bigl( \frac{\pd}{\pd k^\mu} + \frac{\pd}{\pd k^{\prime\,\mu}}
	      \Bigr)
\Bigr)
\Bigr]
\bigl( \opsi_{s}^+(\bk) \gamma^0 \circ \psi_{s'}^+(\bk') \bigr)
\displaybreak[2]\\
+
\e^{+\frac{2}{\ih}(x^0-x_0^0) k_0}
\Bigl[
\Bigl(-\delta_{0\mu}\frac{k_\nu}{k_0} + \delta_{0\nu}\frac{k_\mu}{k_0} \Bigr)
-
	( \delta_{1\mu} + \delta_{2\mu} + \delta_{3\mu} )
	( \delta_{1\nu} + \delta_{2\nu} + \delta_{3\nu} )
\\ \times
\Bigl( k_\mu \Bigl( \frac{\pd}{\pd k^\nu} + \frac{\pd}{\pd k^{\prime\,\nu}}
	     \Bigr)
-
       k_\nu \Bigl( \frac{\pd}{\pd k^\mu} + \frac{\pd}{\pd k^{\prime\,\mu}}
	      \Bigr)
\Bigr)
\Bigr]
\bigl( \opsi_{s}^-(\bk) \gamma^0 \circ \psi_{s'}^-(\bk') \bigr)
\Bigr\}
\Big|_{\begin{subarray}{l}
	\bk'=-\bk\\
	k_0=\sqrt{m^2c^2+\bk^2}
	\end{subarray}
} \ .
	\end{multline}
	\end{subequations}

	Now we shall prove that
	\begin{equation}	\label{7.12-5}
\lindex[\mspace{-6mu}{\tope{S}_{\mu\nu}}]{}{1} (x^0)
+
\lindex[\mspace{-6mu}{\tope{L}_{\mu\nu}}]{}{1} (x^0)
= 0 .
	\end{equation}
For the purpose, define the quantities
	\begin{align}	\label{7.12-6}
\Hat{\tau}_{\mu\nu}^{ss',\pm}(\bk)
& :=
v^{\dag\, s,\pm}(\bk) \sigma_{\mu\nu} v^{s',\pm}(-\bk)
=
- \Hat{\tau}_{\nu\mu}^{ss',\pm}(\bk)
\\			\label{7.12-7}
\tau_{\mu\nu}^{ss',\pm}(x^0,\bk)
&  :=
	\begin{cases}
0					& \text{for $\mu=0$ or $\nu=0$} \\
\e^{\mp\frac{2}{\ih}(x^0-x_0^0) \sqrt{m^2c^2+\bk^2}}
\Hat{\tau}_{\mu\nu}^{ss',\pm}(\bk)	& \text{for $\mu,\nu\not=0$}
	\end{cases}
~~,
	\end{align}
in terms of which~\eref{7.12-4a} reads
	\begin{equation}	\label{7.12-8}
\lindex[\mspace{-6mu}{\tope{S}_{\mu\nu}}]{}{1} (x^0)
=
\frac{1}{2} \hbar \sum_{s,s'} \int\Id^3\bk
\bigl\{
\tau_{\mu\nu}^{ss',+}(x^0,\bk) a_s^{\dag\,+}(\bk) \circ a_{s'}^{+}(-\bk)
 +
\tau_{\mu\nu}^{ss',-}(x^0,\bk) a_s^{\dag\,-}(\bk) \circ a_{s'}^{-}(-\bk)
\bigr\}
	\end{equation}
as a result of the substitution of~\eref{6.20} into~\eref{7.12-4a}.
Inserting~\eref{5.26} into equation~\eref{7.12-6} and using~\eref{3.26}
and~\eref{3.15}, we get
	\begin{multline}	\label{7.12-9}
\mspace{-9mu}
\Hat{\tau}_{\mu\nu}^{ss',\pm}(-\bk)
=
\pm 2\iu A_{\pm}^\ast(-\bk) A_{\pm}(+\bk)
 (v_0^{s,\mp})^\top (\pm \gamma^\lambda k_\lambda + mc\openone_4)
 (k_\mu \gamma_\nu - k_\nu \gamma_\mu)  v_0^{s',\pm}
\\
=
\pm 2\iu A_{\pm}^\ast(-\bk) A_{\pm}(+\bk)
 (v_0^{s,\mp})^\top (k_\mu \gamma_\nu - k_\nu \gamma_\mu)
 (\mp \gamma^\lambda k_\lambda + mc\openone_4)
  v_0^{s',\pm} .
	\end{multline}
Substituting~\eref{6.20} into~\eref{7.12-4b}, we see, on one hand, that
 $\lindex[\mspace{-6mu}{\tope{L}_{0a}}]{}{1} (x^0) =0$ for $a=1,2,3$ as a
consequence of~\eref{5.23}. On the other hand, performing the differentiations
in the obtained equation, with $\mu,\nu\not=0$, by means of~\eref{5.26} (see
equation~\eref{7.23new1} below) and applying~\eref{7.12-9}, we get
\(
\lindex[\mspace{-6mu}{\tope{L}_{ab}}]{}{1} (x^0)
=
- \lindex[\mspace{-6mu}{\tope{S}_{ab}}]{}{1} (x^0) ,
\)
 $a,b=1,2,3$. These results, together with
 $\lindex[\mspace{-6mu}{\tope{S}_{0a}}]{}{1} (x^0) = 0$ (see~\eref{7.11a}),
complete the proof of~\eref{7.12-5}.
\hfill Q.E.D.

	Inserting~\eref{6.20} into~\eref{7.11b}, introducing the quantities
	\begin{equation}	\label{7.12-10}
\sigma_{\mu\nu}^{ss',\pm}(\bk)
:=
	\begin{cases}
0					& \text{for $\mu=0$ or $\nu=0$} \\
\Hat{\sigma}_{\mu\nu}^{ss',\pm}(\bk)	& \text{for $\mu,\nu\not=0$}
	\end{cases}
	\end{equation}
with $\Hat{\sigma}_{\mu\nu}^{ss',\pm}(\bk)$ defined by~\eref{6.23}, and
using~\eref{7.11a} and~\eref{7.12-3}, we obtain the time\ndash independent
part of the spin angular momentum in Heisenberg picture as%
\footnote{~%
In momentum picture, the spin angular momentum will be found in
Sect.~\ref{Sect8} --- see equation~\eref{8.6} below.%
}
	\begin{gather}
			\label{7.11-3}
\lindex[\mspace{-6mu}{\tope{S}_{\mu\nu}}]{}{0}
=
\frac{1}{2} \hbar
\sum_{s,s'} \!\int\!\!\Id^3\bk
\bigl\{
\sigma_{\mu\nu}^{s s',-}(\bk) a_{s}^{\dag\,+}(\bk)\circ a_{s'}^{-}(\bk)
+
\sigma_{\mu\nu}^{s s',+}(\bk) a_{s}^{\dag\,-}(\bk)\circ a_{s'}^{+}(\bk)
\bigr\}.
	\end{gather}

	We should note the that the quantities~\eref{6.23} and
hence~\eref{7.12-10}, generally, depend on all of the arguments indicated
and, for $m\not=0$, are not diagonal in $s$ and $s'$, \ie they are not
proportional to $\delta^{s s'}$($=1$ for $s=s'$, $=0$ for $s\not=s'$).
However, if $m=0$, the indices $s$ and $s'$ take the single value $0$,
$s,s'=0$, so that~\eref{7.11-3} reduces to
	\begin{equation}	\label{7.11-5}
\lindex[\mspace{-6mu}{\tope{S}_{\mu\nu}}]{}{0}  \big|_{m=0}
=
\frac{1}{2} \hbar
\int\Id^3\bk
\bigl\{
\sigma_{\mu\nu}^{00,-}(\bk) a_{0}^{\dag\,+}(\bk)\circ a_{0'}^{-}(\bk)
+
\sigma_{\mu\nu}^{00,+}(\bk) a_{0}^{\dag\,-}(\bk)\circ a_{0}^{+}(\bk)
\bigr\}.
	\end{equation}

	The `diagonal' part of the quantities~\eref{6.23} is real due
to~\eref{5.21}, \viz
	\begin{gather}	\label{7.11-6}
\bigl( \sigma_{\mu\nu}^{ss,\pm}(\bk) \bigr)^*
=
\sigma_{\mu\nu}^{ss,\pm}(\bk)  ,
\intertext{where the asterisk $\ast$ means complex conjugation, but, because
of (see~\eref{3.23} and do not sum over $\mu$ and $\nu$)}
			\label{7.11-7}
\bigl( \Hat{\sigma}_{\mu\nu}^{s s',\pm}(\bk) \bigr)^*
=
\eta_{\mu\mu}\eta_{\nu\nu}
v^{\dag\,s',\mp}(\bk) \sigma_{\mu\nu} v^{s,\pm}(\bk)
=
\eta_{\mu\mu}\eta_{\nu\nu}
\Hat{\sigma}_{\mu\nu}^{s' s,\pm}(\bk)  ,
	\end{gather}
the `non-diagonal' part of the quantities~\eref{6.23} is, generally, complex
for \emph{some} $\mu$ and $\nu$ and if $s\not=s'$ (for, of course,
$m\not=0$). However, using~\eref{6.21} and~\eref{7.11-7}, one can prove that
the sums over $s$ and $s'$ of the first/second terms in the integrand
in~\eref{7.11-3} are Hermitian operators (with, as it is well known, real
eigenvalues).

	To get a concrete understanding of the quantities~\eref{6.23}
and~\eref{7.12-6}, we shall present below their particular values in some
special frames. As $\sigma_{\mu\nu}^{ss'}(\bk)=-\sigma_{\nu\mu}^{ss'}(\bk)$,
not all of the quantities $\sigma_{\mu\nu}^{ss'}(\bk)$ are independent. Below
we present only the independent ones, corresponding to
$(\mu,\nu)=(0,1),(0,2),(0,3),(1,2),(2,3),(3,1)$.
We also omit the argument $\bk$ to save some space.

	If $m\not=0$, there is a frame such that $\bk=\bs 0$.
Following~\cite{Bogolyubov&Shirkov,Bjorken&Drell}, in this frame, we choose
\big(see~\eref{5.26} with $\bk=\bs0$ and $A_\pm(\bs 0)=\frac{1}{2mc}$\big)
	\begin{equation}	\label{7.11-8}
	\begin{split}
v^{\dag\,1,-} & = \bigl( v^{1,+} \bigr)^\top = (0,0,0,1) \quad
v^{\dag\,2,-}   = \bigl( v^{2,+} \bigr)^\top = (0,0,1,0)
\\
v^{\dag\,1,+} & = \bigl( v^{1,-} \bigr)^\top = (1,0,0,0) \quad
v^{\dag\,2,+}   = \bigl( v^{2,-} \bigr)^\top = (0,1,0,0) .
	\end{split}
	\end{equation}
Then, working with the representation~\eref{3.1-1} of the
$\gamma$-matrices~\cite{Bjorken&Drell,Bogolyubov&Shirkov}, from
equations~\eref{6.23} and~\eref{7.12-6}, we, via a straightforward
calculation, get:
	\begin{equation}	\label{7.11-9}
	\begin{split}
\Hat{\sigma}_{0a}^{s s',\pm} = 0 \quad
\Hat{\sigma}_{12}^{s s',\pm} =  \pm (-1)^s \delta^{s s'}
\qquad a=1,2,3 \quad s,s'=1,2
\\
\Hat{\sigma}_{23}^{s s',\pm} =
	\begin{cases}
1	&\text{for $s\not=s'$} \\
0	&\text{for $s=s'$}
	\end{cases}
\quad
\Hat{\sigma}_{31}^{s s',\pm} =
	\begin{cases}
\mp(-1)^s\, \iu	&\text{for $s\not=s'$} \\
0		&\text{for $s=s'$}
	\end{cases}
	\end{split}
\quad\Bigg\}
	\end{equation}
\vspace{-2ex}
	\begin{equation}	\label{7.11-9new}
\Hat{\tau}_{01}^{ss',\pm} = - \delta^{ss'} \iu \quad \!
\Hat{\tau}_{02}^{ss',\pm} = \pm (-1)^s \delta^{ss'} \quad \!
\Hat{\tau}_{03}^{ss',\pm} = \pm (-1)^s (1-\delta^{ss'}) \iu \quad \!
\Hat{\tau}_{ab}^{ss',\pm} = 0 .
	\end{equation}
Similarly, for $m\not=0$, in a frame in which $k^1=k^2=0$, we
put~\cite[sec.~7.3]{Bogolyubov&Shirkov}:
	\begin{equation}	\label{7.11-10}
	\begin{split}
v^{\dag\,1,-} & = \bigl( v^{1,+} \bigr)^\top = N^{-1} (\rho,0,1,0) \quad
v^{\dag\,2,-}   = \bigl( v^{2,+} \bigr)^\top = N^{-1} (0,-\rho,0,1)
\\
v^{\dag\,1,+} & = \bigl( v^{1,-} \bigr)^\top = N^{-1} (1,0,\rho,0) \quad
v^{\dag\,2,+}   = \bigl( v^{2,-} \bigr)^\top = N^{-1} (0,1,0,-\rho)
	\end{split}
	\end{equation}
where $\rho:=\frac{k^3}{mc + \sqrt{m^2c^2+\bk^2} }$ and $N=\sqrt{1+\rho^2}$.
Then, in the representation~\eref{3.1-1} of the $\gamma$\ndash matrices,
from~\eref{6.23} and~\eref{7.12-6}, we obtain for $m\not=0$ and
$k_0=\sqrt{m^2c^2+\bk^2}$:
	\begin{equation}	\label{7.11-11}
	\begin{split}
\Hat{\sigma}_{0a}^{s s',\pm} = 0 \quad
\Hat{\sigma}_{12}^{s s',\pm} = (-1)^s \delta^{s s'}
\qquad a=1,2,3 \quad s,s'=1,2
\\
\Hat{\sigma}_{23}^{s s',\pm} =
	\begin{cases}
\frac{mc}{k_0}
	&\text{for $s\not=s'$} \\
0	&\text{for $s=s'$}
	\end{cases}
\quad
\Hat{\sigma}_{31}^{s s',\pm} =
	\begin{cases}
(-1)^s \frac{mc}{k_0} \, \iu
	&\text{for $s\not=s'$} \\
0	&\text{for $s=s'$}
	\end{cases}
	\end{split}
\quad\Bigg\}
	\end{equation}
\vspace{-2ex}
	\begin{equation}	\label{7.11-11new}
	\begin{split}
& \Hat{\tau}_{01}^{ss',\pm} = - (1-\delta^{ss'}) \iu \quad
\Hat{\tau}_{02}^{ss',\pm} = (-1)^s (1-\delta^{ss'}) \quad
\Hat{\tau}_{03}^{ss',\pm} = (-1)^s \delta^{ss'} \iu \frac{mc}{k_0} \quad
\\
& \Hat{\tau}_{12}^{ss',\pm} = 0 \quad
\Hat{\tau}_{23}^{ss',\pm} = - (-1)^s 2(1-\delta^{ss'}) \frac{\rho}{N^2} \quad
\Hat{\tau}_{31}^{ss',\pm} = -  2(1-\delta^{ss'}) \iu \frac{\rho}{N^2}.
	\end{split}
\Bigg\}
	\end{equation}
Notice the appearance of the imaginary unit $\iu$ in the last formulae
in~\eref{7.11-9} and~\eref{7.11-11}, which is in conformity
with~\eref{7.11-7}; the rest of the results agree with~\eref{7.11-6}.

	For $m=0$, in a frame in which $k^1=k^2=0$, we set
\big(see~\eref{5.26} with $A_\pm(\bk)=\pm\frac{1}{\sqrt{2}
\sqrt{m^2c^2+\bk^2}}$\big)
	\begin{equation}	\label{7.11-12}
	\begin{split}
v^{\dag\,0,-} & = \bigl( v^{0,+} \bigr)^\top
		= \frac{1}{\sqrt{2}} (-\rho,0,1,0)\quad
v^{\dag\,0,+}   = \bigl( v^{0,-} \bigr)^\top
		=-\frac{1}{\sqrt{2}} (0,\rho,0,1),
	\end{split}
	\end{equation}
where
\(
\rho
:=\frac{k^3}{mc + \sqrt{m^2c^2+\bk^2}}
 =\frac{k^3}{\sqrt{(k^3)^2}}
\in\{-1,+1\},
\)
and, correspondingly, we get:
	\begin{equation}	\label{7.11-13}
\Hat{\sigma}_{0a}^{00,\pm} =
\Hat{\sigma}_{23}^{00,\pm} =
\Hat{\sigma}_{31}^{00,\pm} = 0
\quad
\Hat{\sigma}_{12}^{00,\pm} = -1
\qquad
a=1,2,3
	\end{equation}
\vspace{-3ex}
	\begin{equation}	\label{7.11-13new}
	\begin{split}
& \Hat{\tau}_{01}^{00,\pm} =   \pm \rho \iu \quad
\Hat{\tau}_{02}^{00,\pm} = - \rho \quad \!
\Hat{\tau}_{03}^{00,+  } = - \rho \iu
\\
& \Hat{\tau}_{03}^{00,-  } =  \Hat{\tau}_{12}^{00,\pm} = 0 \quad
\Hat{\tau}_{23}^{00,\pm} = -1 \quad
\Hat{\tau}_{31}^{00,\pm} = \mp \iu .
	\end{split}
\quad\Bigg\}
	\end{equation}

	To calculate the quantities~\eref{6.23} and~\eref{7.12-6} in an
arbitrary frame, one should use~\eref{5.26} or an equivalent to it explicit
form of the spinors $v^{s,\pm}(\bk)$  with $s=1,2$ for $m\not=0$ or $s=0$ for
$m=0$.

	If we write the spin vector~\eref{3.16} as
	\begin{gather}
			\label{7.11-14}
\bs{\tope{S}} = \int \bs{\tope{S}}(\bk) \Id^3\bk,
	\end{gather}
the results~\eref{7.11-5},~\eref{7.11-11},~\eref{7.11-13},
and~\eref{7.11-13new} imply
	\begin{align}	\label{7.13}
	\begin{split}
\lindex[\mspace{-6mu}\tope{S}]{}{0}^3(\bk) \big|_{m\not=0}
=
\tope{S}^3(\bk) \big|_{m\not=0}
& =
\frac{1}{2} \hbar \bigl\{
- a_{1}^{\dag\,+}(\bk) \circ a_{1}^{-}(\bk)
+ a_{2}^{\dag\,+}(\bk) \circ a_{2}^{-}(\bk)
\\ & \hphantom{ \frac{1}{2} \hbar \bigl\{ }
- a_{1}^{\dag\,-}(\bk) \circ a_{1}^{+}(\bk)
+ a_{2}^{\dag\,-}(\bk) \circ a_{2}^{+}(\bk)
\bigr\}
	\end{split}
\\			\label{7.13new}
\lindex[\mspace{-6mu}\tope{S}]{}{0}^3(\bk) \big|_{m=0}
=
\tope{S}^3(\bk) \big|_{m=0}
& =
- \frac{1}{2} \hbar \bigl\{
  a_{0}^{\dag\,+}(\bk) \circ a_{0}^{-}(\bk)
+ a_{0}^{\dag\,-}(\bk) \circ a_{0}^{+}(\bk)
\bigr\}
	\end{align}
in a frame in which $k^1=k^2=0$.
	From here, for $m\not=0$, follows that the spin projection, on the
third axis (of the chosen frame of reference) is $-\frac{1}{2}\hbar$ for
the particles corresponding to
  $a_{1}^{\dag\,+}(\bk)$, $a_{1}^{-}(\bk)$, $a_{1}^{\dag\,-}(\bk)$, and
  $a_{1}^{+}(\bk)$,
while it equals $+\frac{1}{2}\hbar$ for the ones corresponding to
  $a_{2}^{\dag\,+}(\bk)$, $a_{2}^{-}(\bk)$, $a_{2}^{\dag\,-}(\bk)$, and
  $a_{2}^{+}(\bk)$.
Similarly (see~\eref{7.13new}), for $m=0$, the spin
projection, on the third axis (of the chosen frame of reference) is
$-\frac{1}{2}\hbar$ for the particles corresponding to \emph{any}
creation/annihilation operator.%
\footnote{~%
After the normal ordering, this interpretation of $a_s^\pm(\bk)$ and
$a_{s}^{\dag\,\pm}(\bk)$ will be partially changed.  In that connection, the
reader may notice some contradiction of the interpretation of the creation
and annihilation operators, given in Sect.~\ref{Sect6}, and the
expressions~\eref{7.5}, \eref{7.8} and~\eref{7.11-3}. It will disappear after
normal ordering of products. For details, see Sect.~\ref{Sect10}.%
}
Further details regarding the spin angular momentum and its interpretation
can be found in~\cite{Bjorken&Drell-2,Roman-QFT,Itzykson&Zuber}.

	Let us express now the orbital angular momentum~\eref{7.12} in
terms of the operators $a_s^\pm(\bk)$ and $a_{s}^{\dag\,\pm}(\bk)$.
Substituting~\eref{6.20} and~\eref{7.5} into~\eref{7.12} and
applying~\eref{7.12-3}, \eref{5.21} and~\eref{5.22}, we obtain the time\ndash
independent part of the orbital angular momentum in Heisenberg picture as%
\footnote{~%
In momentum picture, the orbital angular momentum will be found in
Sect.~\ref{Sect8} --- see equations~\eref{8.8} and~\eref{8.11} below.%
}
	\begin{multline}	\label{7.20}
\lindex[\mspace{-6mu}{\tope{L}_{\mu\nu}}]{}{0}
=
\sum_{s}\int \Id^3\bk
 ( x_{0\,\mu}k_\nu - x_{0\,\nu}k_\mu ) |_{ k_0=\sqrt{m^2c^2+{\bs k}^2} }
\{
a_s^{\dag\,+}(\bk)\circ a_s^-(\bk)
-
a_s^{\dag\,-}(\bk)\circ a_s^+(\bk)
\}
\\
+ \frac{1}{2} \hbar \sum_{s,s'}\int \Id^3\bk
\bigl\{
l_{\mu\nu}^{ss',-}(\bk) a_s^{\dag\,+}(\bk) \circ a_{s'}^-(\bk)
+
l_{\mu\nu}^{ss',+}(\bk) a_s^{\dag\,-}(\bk) \circ a_{s'}^+(\bk)
\bigr\}
\displaybreak[2]\\
+ \frac{1}{2} \ih \sum_{s}\int \Id^3\bk
\Bigl\{
a_s^{\dag\,+}(\bk)
\Bigl( \xlrarrow{ k_\mu \frac{\pd}{\pd k^\nu} }
     - \xlrarrow{ k_\nu \frac{\pd}{\pd k^\mu} } \Bigr)
\circ a_s^-(\bk)
\\ +
a_s^{\dag\,-}(\bk)
\Bigl( \xlrarrow{ k_\mu \frac{\pd}{\pd k^\nu} }
     - \xlrarrow{ k_\nu \frac{\pd}{\pd k^\mu} } \Bigr)
\circ a_s^+(\bk)
\Bigr\} \Big|_{ k_0=\sqrt{m^2c^2+{\bs k}^2} } \ ,
	\end{multline}
where
	\begin{equation}	\label{7.21}
	\begin{split}
l_{\mu\nu}^{ss',\pm}(\bk)
: & =
\iu v^{\dag\,s,\mp}(\bk)
\Bigl( \xlrarrow{ k_\mu \frac{\pd}{\pd k^\nu} }
     - \xlrarrow{ k_\nu \frac{\pd}{\pd k^\mu} } \Bigr)
v^{s',\pm}(\bk)
= - l_{\nu\mu}^{ss',\pm}(\bk)
\\
& =
- 2\iu
  \Bigl( k_\mu \frac{\pd v^{\dag\,s,\mp}(\bk)}{\pd k^\nu}
        - k_\nu \frac{\pd v^{\dag\,s,\mp}(\bk)}{\pd k^\mu}
  \Bigr) v^{s',\pm}(\bk)
\\
& =
+2\iu v^{\dag\,s,\mp}(\bk) \Bigl(
  k_\mu \frac{\pd v^{s',\pm}(\bk)}{\pd k^\nu}
- k_\nu \frac{\pd v^{s',\pm}(\bk)}{\pd k^\mu} \Bigr) .
 	\end{split}
	\end{equation}
with the restriction $k_0=\sqrt{m^2c^2+{\bs k}^2}$ done after the
differentiation (so that the derivatives with respect to $k_0$ vanish).
The last two equalities in~\eref{7.21} are consequences of
(see~\eref{5.22})
	\begin{gather}	\label{7.22}
\frac{\pd v^{\dag\,s,\pm}(\bk)}{\pd k^\lambda} v^{s',\mp}(\bk)
+
v^{\dag\,s,\pm}(\bk) \frac{\pd v^{s',\mp}(\bk)}{\pd k^\lambda}
= 0 ,
\intertext{so that}		\label{7.23}
v^{\dag\,s,\pm}(\bk)
\xlrarrow{ k_\mu \frac{\pd}{\pd k^\nu} }
v^{s',\mp}(\bk)
=-2k_\mu \frac{\pd v^{\dag\,s,\pm}(\bk)}{\pd k^\nu} v^{s',\mp}(\bk)
= 2k_\mu v^{\dag\,s,\pm}(\bk) \frac{\pd v^{s',\mp}(\bk)}{\pd k^\nu} .
	\end{gather}
Notice, the equation~\eref{7.22} implies (see~\eref{5.21})
	\begin{equation}	\label{7.23new}
\bigl( l_{\mu\nu}^{ss',\pm}(\bk) \bigr)^\ast
=
l_{\mu\nu}^{s's,\pm}(\bk).
	\end{equation}
So, $l_{\mu\nu}^{ss,\pm}(\bk)$ are real and, by virtue of~\eref{6.21}, the
sums of the first/second terms in the last integrand in~\eref{7.20} are
Hermitian. Using~\eref{5.26}, one sees that the derivatives in the last
equality in~\eref{7.21} are
	\begin{equation}	\label{7.23new1}
	\begin{split}
\frac{\pd v^{s,\pm}(\bk)}{\pd k^0} &= 0
\\
\frac{\pd v^{s,\pm}(\bk)}{\pd k^a}
& =
\pm A_{\pm}(\bk)
\Bigl(- \gamma_a + \frac{k_a}{\sqrt{m^2c^2+\bk^2}} \gamma^0 \Bigr)
v_{0}^{s,\pm}
+
A_\pm^{-1}(\bk)
\frac{\pd A_\pm(\bk)}{\pd k^a}
v^{s,\pm}(\bk),
	\end{split}
	\end{equation}
where $a=1,2,3$, $A_{\pm}(\bk)$ are some normalization constants and
$v_{0}^{s,\pm}$ are given by~\eref{5.27}.

	The charge, spin and orbital angular momentum operators in
\emph{momentum picture} will be found in Sect.~\ref{Sect8}.

	Since the right-hand-sides of~\eref{7.11-3} and~\eref{7.20} are
constant (in spacetime) operators, we have
	\begin{equation}	\label{7.23-1}
\pd_\lambda \lindex[\mspace{-6mu}{\tope{S}_{\mu\nu}}]{}{0} = 0
\qquad
\pd_\lambda \lindex[\mspace{-6mu}{\tope{L}_{\mu\nu}}]{}{0} = 0 .
	\end{equation}
Besides, by virtue of~\eref{7.12-3},~\eref{7.12-5} and~\eref{2.11}, the total
angular momentum operator of the spinor field under consideration is
	\begin{equation}	\label{7.23-2}
\tope{M}_{\mu\nu}
=
\lindex[\mspace{-6mu}{\tope{S}_{\mu\nu}}]{}{0}
+
\lindex[\mspace{-6mu}{\tope{L}_{\mu\nu}}]{}{0}
	\end{equation}
and, as a result of~\eref{7.23-1}, satisfies the evident conservation
equation
	\begin{equation}	\label{7.23-3}
\pd_ \lambda \tope{M}_{\mu\nu} = 0 ,
	\end{equation}
which agrees with~\eref{2.16}. We shall call the conserved operators
 $\lindex[\mspace{-6mu}{\tope{S}_{\mu\nu}}]{}{0}$ and
 $\lindex[\mspace{-6mu}{\tope{L}_{\mu\nu}}]{}{0}$
the \emph{spin (or spin charge) operator} and the \emph{angular operator},
respectively, of the spinor field. In fact, these invariant characteristics,
not the non\ndash conserved spin and orbital angular momentum, of the field
are the ones which are used practically for the description of a free spinor
field; for instance, the vector components~\eref{7.13} and~\eref{7.13new}
are, actually, the only spin characteristics of a free spinor field examined
in the (text)books~\cite{Bogolyubov&Shirkov,Bjorken&Drell,Itzykson&Zuber}.

	Ending this section, we would like to make a comparison with the
expressions for the dynamical variables in terms of the creation/annihilation
operators $\tilde{a}_{s}^{\pm}(\bk)$ and $\tilde{a}_{s}^{\dag\,\pm}(\bk)$ in
(the momentum representation of) Heisenberg picture of
motion~\cite{Bogolyubov&Shirkov,Bjorken&Drell,Itzykson&Zuber,Roman-QFT}.
As a consequence of~\eref{5.16-1}, the analogues of the creation/annihilation
operators, defined in terms of spinors via~\eref{6.1} and~\eref{6.1new}, are
	\begin{equation}	\label{7.25}
	\begin{split}
\tope{\psi}_s^\pm(\bk)
& =
\e^{\pm\frac{1}{\ih} x_0^\mu k_\mu} \psi_s^\pm(\bk)
\quad
\tope{\opsi}_s^\pm(\bk)
=
\e^{\pm\frac{1}{\ih} x_0^\mu k_\mu} \opsi_s^\pm(\bk)
\quad
\bigl( k_0=\sqrt{m^2c^2+\bk^2} \bigr)
\\
\tope{\psi}^\pm(\bk)
& =
\e^{\pm\frac{1}{\ih} x_0^\mu k_\mu} \psi^\pm(\bk)
\quad
\tope{\opsi}^\pm(\bk)
=
\e^{\pm\frac{1}{\ih} x_0^\mu k_\mu} \opsi^\pm(\bk)
\quad
( k_0=\sqrt{m^2c^2+\bk^2} )
	\end{split}
	\end{equation}
in Heisenberg picture. Therefore, defining (cf.~\eref{6.20})
	\begin{equation}	\label{7.26}
\tope{\psi}_s^\pm(\bk)
=: (2\pi\hbar)^{-3/2} \tilde{a}_s^\pm(\bk) v^{s,\pm}(\bk)
\qquad
\tope{\opsi}_s^\pm(\bk)
=: (2\pi\hbar)^{-3/2} \tilde{a}_s^{\dag\,\pm}(\bk) \overline{v}^{s,\pm}(\bk) ,
	\end{equation}
we get the creation/annihilation operators in Heisenberg picture as
	\begin{gather}	\label{7.27}
\tilde{a}_s^\pm(\bk)
 = \e^{ \pm\frac{1}{\ih} x_0^\mu k_\mu } a_s^\pm(\bk)
\quad
\tilde{a}_s^{\dag\,\pm}(\bk)
 = e^{ \pm\frac{1}{\ih} x_0^\mu k_\mu } a_s^{\dag\,\pm}(\bk)
\quad
	\bigl( k_0=\sqrt{m^2c^2+\bk^2} \bigr) .
\intertext{Evidently, these operators satisfy the equations}
			\label{7.28}
\bigl( \tilde{a}_s^\pm(\bk) \bigr)^\dag = \tilde{a}_s^{\dag\,\mp}(\bk)
\quad
\bigl( \tilde{a}_s^{\dag\,\pm}(\bk) \bigr)^\dag = \tilde{a}_s^\mp(\bk) ,
	\end{gather}
due to~\eref{6.21}, and have all other properties of their momentum picture
counterparts described in Sect.~\ref{Sect6}.

	The connection~\eref{12.114} is not applicable to the
creation/annihilation operators, as well as to operators in momentum
representation (of momentum picture), \ie to ones depending on the momentum
variable $\bk$. In particular, the reader may verify, by using the results of
Sections~\ref{Sect5} and~\ref{Sect6}, the formulae
	\begin{equation}	\label{7.28-1}
	\begin{split}
a_s^\pm(\bk)
& =
\e^{\mp\iih x^\mu k_\mu}
\ope{U}(x,x_0)\circ \tilde{a}_s^\pm(k) \circ \ope{U}^{-1}(x,x_0)
\\
a_s^{\dag\,\pm}(\bk)
& =
\e^{\mp\iih x^\mu k_\mu}
\ope{U}(x,x_0)\circ \tilde{a}_s^{\dag\,\pm}(\bk) \circ \ope{U}^{-1}(x,x_0)
	\end{split}
\ \ \Bigg\} \quad
k_0=\sqrt{m^2c^2+{\bs k}^2} ,
	\end{equation}
from which equations~\eref{7.27} follow for $x=x_0$. (Notice, the right hand
sides of the equations~\eref{7.28-1} are independent of $x$, due to the
Heisenberg relations~\eref{2.28}.)

	From~\eref{7.4}--\eref{7.11},~\eref{7.11-3},
and~\eref{7.25}--\eref{7.27}, it is clear that all of the obtained
expressions for the momentum, charge and spin angular momentum operators in
terms of the (invariant) creation/annihilation operators remain unchanged in
Heisenberg picture; to obtain a Heisenberg version of these equations, one
has formally to add a tilde over the creation/annihilation operators in
momentum picture. However, this is not the case with the orbital
operator~\eref{7.20} because of the existence of derivatives in the integrands
in~\eref{7.12} and~\eref{7.20}. We leave to the reader to prove  as exercise
that, in terms of the operators~\eref{7.27}, in~\eref{7.12} the term
$x_{0\,\mu}\ope{P}_\nu - x_{0\,\nu}\ope{P}_\mu$ should be deleted and tildes
over the creation/annihilation operators must be added. Correspondingly,
equation~\eref{7.20} in Heisenberg picture will read
	\begin{multline}	\label{7.29}
\lindex[\mspace{-6mu}{\tope{L}_{\mu\nu}}]{}{0}
=
\frac{1}{2} \hbar \sum_{s,s'}\int \Id^3\bk
\bigl\{
l_{\mu\nu}^{ss',-}(\bk) \Tilde{a}_s^{\dag\,+}(\bk) \circ \Tilde{a}_{s'}^-(\bk)
+
l_{\mu\nu}^{ss',+}(\bk) \Tilde{a}_s^{\dag\,-}(\bk) \circ \Tilde{a}_{s'}^+(\bk)
\bigr\}
\displaybreak[2]\\
+ \frac{1}{2} \ih \sum_{s}\int \Id^3\bk
\Bigl\{
\Tilde{a}_s^{\dag\,+}(\bk)
\Bigl( \xlrarrow{ k_\mu \frac{\pd}{\pd k^\nu} }
     - \xlrarrow{ k_\nu \frac{\pd}{\pd k^\mu} } \Bigr)
\circ \Tilde{a}_s^-(\bk)
\\ +
\Tilde{a}_s^{\dag\,-}(\bk)
\Bigl( \xlrarrow{ k_\mu \frac{\pd}{\pd k^\nu} }
     - \xlrarrow{ k_\nu \frac{\pd}{\pd k^\mu} } \Bigr)
\circ \Tilde{a}_s^+(\bk)
\Bigr\} \Big|_{ k_0=\sqrt{m^2c^2+{\bs k}^2} } \ .
	\end{multline}


\section
[The field equations in terms of creation and annihilation operators]
{The field equations in terms of creation and\\ annihilation operators}
\label{Sect8}

	As we said at the beginning of Sect.~\ref{Sect5}, the
equalities~\eref{3.5}, \eref{2.6} and~\eref{3.17} form a closed
algebraic\ndash functional system of equations for determination of the
spinor field operators and, consequently, of the dynamical quantities
characterizing a free spinor field. Since, from sections~\ref{Sect6}
and~\ref{Sect7}, we know that the field operators and dynamical variables of
a free spinor field can be expressed uniquely via the creation and
annihilation operators $a_s^\pm(\bk)$ and $a_s^{\dag\,\pm}(\bk)$, which are
invariant (frame\ndash independent), in the present section we shall derive a
system of equations for these operators, which system is equivalent to the
one just described. In fact, this procedure will be equivalent to write the
field equations (in Heisenberg or momentum picture) in terms of creation and
annihilation operators.

	The problem, we want to analyze, is as follows. Given field operators
$\psi$ and $\opsi$ with decompositions (see~\eref{6.3}, \eref{6.4},
and~\eref{6.20})
	\begin{subequations}	\label{8.1}
	\begin{align}	\label{8.1a}
	\begin{split}
\psi
 =
\sum_{s} \int\Id\bk \bigl\{ \psi_s^+(\bk) + \psi_s^-(\bk) \bigr\}
 =
(2\pi\hbar)^{-3/2} \sum_{s} \int\Id\bk
\bigl\{ a_s^+(\bk) v^{s,+}(\bk) + a_s^-(\bk) v^{s,-}(\bk) \bigr\}
	\end{split}
\\			\label{8.1b}
	\begin{split}
\opsi
 =
\sum_{s} \int\Id\bk \bigl\{ \opsi_s^+(\bk) + \opsi_s^-(\bk) \bigr\}
 =
(2\pi\hbar)^{-3/2} \sum_{s} \int\Id\bk
\bigl\{ a_s^+(\bk) \overline{v}^{s,+}(\bk)
+
a_s^{\dag\,-}(\bk) \overline{v}^{s,-}(\bk) \bigr\} ,
	\end{split}
	\end{align}
	\end{subequations}
find (the explicit equations describing) $a_s^{\pm}(\bk)$ and
$a_s^{\dag\,\pm}(\bk)$ such that the equalities~\eref{6.5} hold.

	The results of sections~\ref{Sect5} and~\ref{Sect6} show
that~\eref{6.5} are the implicit equations of motion for $\psi$ and $\opsi$
as, under the definitions of the quantities in them, they are equivalent to
the initial Dirac equations~\eref{3.5}. The equations~\eref{6.5}
and~\eref{7.4} form a closed system of equations with respect to
$\psi_s^\pm(\bk)$ and $\opsi_s^\pm(\bk)$. The substitution of~\eref{7.4}
into~\eref{6.5} results into an explicit system of equations relative to
$\psi_s^\pm(\bk)$ and $\opsi_s^\pm(\bk)$; one can easily write it out in full.
However, it is rather complicated, which is due to the non\ndash invariant,
frame\ndash dependent, character of the operators $\psi_s^\pm(\bk)$ and
$\opsi_s^\pm(\bk)$. This dependence can be removed by specifying these
operators as in~\eref{6.20}. In this way one gets a closed system of
equations for the invariant, frame\ndash independent, operators
$a_s^\pm(\bk)$ and $a_s^{\dag\,\pm}(\bk)$. It can be derived in the
following way.

	Inserting~\eref{6.20} into~\eref{6.5} and taking into account that
 ${v}^{s,\pm}\not=0$ and $\overline{v}^{s,\pm}\not=0$, due to the linear
independence of these spinors (see also the normalization
conditions~\eref{5.22}), we see that~\eref{6.5} is tantamount to:
	\begin{subequations}	\label{8.2}
	\begin{align}
			\label{8.2a}
  [a_s^\pm(\bk),\ope{P}_\mu]_{\_}
& = \mp k_\mu a_s^\pm(\bk)
  = \mp\int q_\mu a_s^\pm(\bk) \delta^3(\bk-\bs q) \Id^3\bs q
\\			\label{8.2b}
  [a_s^{\dag\,\pm}(\bk),\ope{P}_\mu]_{\_}
& = \mp k_\mu a_s^{\dag\,\pm}(\bk)
  = \mp\int q_\mu a_s^{\dag\,\pm}(\bk) \delta^3(\bk-\bs q) \Id^3\bs q
\\			\label{8.2c}
k_0& =\sqrt{m^2c^2+{\bs k}^2}
\qquad
q_0=\sqrt{m^2c^2+{\bs q}^2} .
	\end{align}
	\end{subequations}
Substituting here~\eref{7.5}, with an integration variable $\bs q$ for $\bk$
and summation index $t$ for $s$, we get (do not sum over $s$!)
	\begin{subequations}	\label{8.3}
	\begin{align}
			\label{8.3a}
	\begin{split}
\sum_{t} \int q_\mu\big|_{q_0=\sqrt{m^2c^2+{\bs q}^2}}
\bigl\{
\bigl[ a_s^{\pm}(\bk)
& ,
a_t^{\dag\,+}(\bs q) \circ a_t^{-}(\bs q)
-
a_t^{\dag\,-}(\bs q) \circ a_t^{+}(\bs q)
\bigr]_{\_}
\\
& \pm a_s^{\pm}(\bk) \delta_{st} \delta^3(\bk-\bs q)
\bigr\} \Id^3\bs q = 0
	\end{split}
\\			\label{8.3b}
	\begin{split}
\sum_{t} \int q_\mu\big|_{q_0=\sqrt{m^2c^2+{\bs q}^2}}
\bigl\{
\bigl[ a_s^{\dag\,\pm}(\bk)
& ,
a_t^{\dag\,+}(\bs q) \circ a_t^{-}(\bs q)
-
a_t^{\dag\,-}(\bs q) \circ a_t^{+}(\bs q)
\bigr]_{\_}
\\
& \pm a_s^{\dag\,\pm}(\bk) \delta_{st} \delta^3(\bk-\bs q)
\bigr\} \Id^3\bs q = 0 .
	\end{split}
	\end{align}
	\end{subequations}
Consequently, the operators $a_s^{\pm}(\bk)$ and $a_s^{\dag\,\pm}(\bk)$ must
be solutions of
	\begin{subequations}	\label{8.4}
	\begin{align}
			\label{8.4a}
	\begin{split}
& \bigl[ a_s^{\pm}(\bk) ,
a_t^{\dag\,+}(\bs q) \circ a_t^{-}(\bs q)
 -
a_t^{\dag\,-}(\bs q) \circ a_t^{+}(\bs q)
\bigr]_{\_}
 \pm a_s^{\pm}(\bk) \delta_{st} \delta^3(\bk-\bs q)
= f_{st}^\pm(\bk,\bs q)
	\end{split}
\\			\label{8.4b}
	\begin{split}
& \bigl[ a_s^{\dag\,\pm}(\bk) ,
a_t^{\dag\,+}(\bs q) \circ a_t^{-}(\bs q)
 -
a_t^{\dag\,-}(\bs q) \circ a_t^{+}(\bs q)
\bigr]_{\_}
\pm  a_s^{\dag\,\pm}(\bk) \delta_{st} \delta^3(\bk-\bs q)
= f_{st}^{\dag\,\pm}(\bk,\bs q) ,
	\end{split}
	\end{align}
where $f_{st}^{\pm}(\bk,\bs q)$ and $f_{st}^{\dag\,\pm}(\bk,\bs q)$ are
(generalized) functions such that
	\begin{equation}	\label{8.4c}
	\begin{split}
\sum_{t} \int q_\mu\big|_{q_0=\sqrt{m^2c^2+{\bs q}^2}}
f_{st}^{\pm}(\bk,\bs q) \Id^3\bs{q}
= 0
\qquad
\sum_{t} \int q_\mu\big|_{q_0=\sqrt{m^2c^2+{\bs q}^2}}
f_{st}^{\dag\,\pm}(\bk,\bs q) \Id^3\bs{q}
& = 0 .
	\end{split}
	\end{equation}
	\end{subequations}

	Since any solution of the Dirac equations~\eref{3.5} can be written
in the form~\eref{8.1} with $a_s^{\pm}(\bk)$ and $a_s^{\dag\,\pm}(\bk)$ being
solutions of~\eref{8.4}, we can assert that the system~\eref{8.4} is
equivalent to the initial system~\eref{3.5} under the subsidiary
condition~\eref{2.28}. In this sense,~\eref{8.4} is the \emph{system of field
equations in terms of creation and annihilation operators in momentum
picture}. Comparing it with the system(s) of field equations in terms of
creation and annihilation operators for a free \emph{charged} scalar
field~\cite{bp-QFTinMP-scalars}, we see that the structure of these two types
of algebraic\ndash functional equations is quite similar; the only essential
difference being in the signs before the second terms in the commutators
in~\eref{8.4a} and~\eref{8.4b} as in a case of free charged scalar field in
this place stands $+a_s^{\dag\,-}(\bs q)\circ a_s^+(\bs q)$ instead of the
term $-a_s^{\dag\,-}(\bs q)\circ a_s^+(\bs q)$.%
\footnote{~%
This comparison is based on the `first' choice of a Lagrangian
and the `third' choice of an energy\ndash momentum operators
in~\cite{bp-QFTinMP-scalars}. Regarding some problems of a choice of
Lagrangian(s) and structure of conserved quantities --- see
Sect.~\ref{Sect12}.%
}

	As a first application of the field equations~\eref{8.4}, we shall
calculate the commutator
 $[\tope{S}_{\mu\nu},\ope{P}_\lambda]_{\_}$ between the spin angular momentum
operator in \emph{Heisenberg} picture, $\tope{S}_{\mu\nu}$, and the momentum
operator $\tope{P}_\lambda=\ope{P}_\lambda$. Applying the
equalities~\eref{7.11-3}, with $\bs q$ for the integration variable $\bk$,
and~\eref{7.5}, with summation variable $t$ for $s$, we get
	\begin{equation*}
[\lindex[\mspace{-6mu}{\tope{S}_{\mu\nu}}]{}{0} , \ope{P}_\lambda]_{\_}
=
\frac{1}{2} \hbar
\sum_{s,s',t} \int \Id^3\bs q\Id^3\bk
k_\lambda\big|_{k_0=\sqrt{m^2c^2+\bk^2}}
\{
  \sigma_{\mu\nu}^{ss',-}(\bs q) B_{ss't}^-(\bs q,\bk)
+ \sigma_{\mu\nu}^{ss',+}(\bs q) B_{ss't}^+(\bs q,\bk) \}
	\end{equation*}
where
	\begin{equation*}
B_{ss't}^\mp(\bs q,\bk)
:=
[
a_{s}^{\dag\,\pm}(\bs q) \circ a_{s'}^{\mp}(\bs q) ,
a_t^{\dag\,+}(\bk)\circ a_t^-(\bk)
-
a_t^{\dag\,-}(\bk)\circ a_t^+(\bk)
]_{\_} .
	\end{equation*}
Using the identity
$ [A\circ B,C]_{\_} = [A,C]_{\_}\circ B + A\circ [B,C]_{\_} $,~\eref{8.4a}
and~\eref{8.4b}, we see that (do not sum over $s$ and $s'$!)
	\begin{equation*}
B_{ss't}^\mp(\bs q,\bk)
=
(\mp \delta_{st} \pm \delta _{s't}) \delta^3(\bs q - \bk)
a_s^{\dag\,\pm}(\bs q)\circ a_{s'}^\mp(\bs q)
+ a_{s'}^{\mp}(\bs q) f_{st}^{\dag\,\pm}(\bs q,\bk)
+ a_{s}^{\dag\,\pm}(\bs q) f_{s't}^{\mp}(\bs q,\bk) .
	\end{equation*}
Substituting this result into the above expression for
$[\lindex[\mspace{-6mu}{\tope{S}_{\mu\nu}}]{}{0} , \ope{P}_\lambda]_{\_}$,
using~\eref{8.4c} (with $\bk$ for $\bs q$ and vice versa) and summing over
$t$, we obtain
	\begin{equation}	\label{8.5}
[\lindex[\mspace{-6mu}{\tope{S}_{\mu\nu}}]{}{0} , \ope{P}_\lambda]_{\_} = 0 .
	\end{equation}
So, for free spinor fields the \emph{spin operator in Heisenberg picture
commutes with the momentum operator in Heisenberg or momentum picture}
(see~\eref{2.3}).

	Similar manipulations with the time-dependent part~\eref{7.12-8} of
the spin angular momentum give the relation
	\begin{multline}	\label{8.5-1}
[\lindex[\mspace{-6mu}{\tope{S}_{\mu\nu}}]{}{1} (x^0) , \ope{P}_\lambda]_{\_}
=
\delta_{0\lambda} \hbar
\sum_{s,s'} \int \Id^3\bk \sqrt{m^2c^2+\bk^2}
\\\times
\bigl\{
- \tau_{\mu\nu}^{ss',+}(x^0,\bk) a_{s}^{\dag\,+}(\bk)\circ a_{s'}^{+}(-\bk)
+ \tau_{\mu\nu}^{ss',-}(x^0,\bk) a_{s}^{\dag\,-}(\bk)\circ a_{s'}^{-}(-\bk)
\bigr\},
	\end{multline}
which, by virtue of~\eref{8.5} and~\eref{7.12-3}, implies
	\begin{multline}	\label{8.5-2}
[\tope{S}_{\mu\nu},\ope{P}_\lambda]_{\_}
=
\delta_{0\lambda} \hbar
\sum_{s,s'} \int \Id^3\bk \sqrt{m^2c^2+\bk^2}
\\\times
\bigl\{
- \tau_{\mu\nu}^{ss',+}(x^0,\bk) a_{s}^{\dag\,+}(\bk)\circ a_{s'}^{+}(-\bk)
+ \tau_{\mu\nu}^{ss',-}(x^0,\bk) a_{s}^{\dag\,-}(\bk)\circ a_{s'}^{-}(-\bk)
\bigr\} .
	\end{multline}

	By virtue of~\eref{12.112} (or
footnote~\ref{CommutativityWithMomentum}) and~\eref{8.5}, we have
	\begin{equation}	\label{8.5new}
[\lindex[\mspace{-6mu}{\tope{S}_{\mu\nu}}]{}{0} , \ope{U}(x,x_0)]_{\_} = 0
	\end{equation}
where $\ope{U}(x,x_0)$ is the operator responsible for the transition from
Heisenberg to momentum picture. So,~\eref{12.114} implies
	\begin{equation}	\label{8.6}
\lindex[\mspace{-6mu}{\ope{S}_{\mu\nu}}]{}{0}
=
\lindex[\mspace{-6mu}{\tope{S}_{\mu\nu}}]{}{0}
	\end{equation}
\ie the \emph{spin operators is one and the same in Heisenberg and momentum
pictures}.%
\footnote{~%
From~\eref{8.6} does not follow the conservation of the spin operator;
it is a consequence from the considerations in Sect.~\ref{Sect7}.%
}
Combining the above results with~\eref{7.23-2}, \eref{7.12-5}, \eref{2.21},
\eref{2.23} and~\eref{2.25}, one can easily derive the following equalities:
	\begin{align}
 &			\label{8.7}
\tope{M}_{\mu\nu} =
\lindex[\mspace{-6mu}{\tope{L}_{\mu\nu}}]{}{0}
+
\lindex[\mspace{-6mu}{\tope{S}_{\mu\nu}}]{}{0}
\quad
\ope{M}_{\mu\nu} =
\lindex[\mspace{-6mu}{\ope{L}_{\mu\nu}}]{}{0}
+
\lindex[\mspace{-6mu}{\ope{S}_{\mu\nu}}]{}{0}
\\ &			\label{8.8}
\lindex[\mspace{-6mu}{\ope{L}_{\mu\nu}}]{}{0}
=
\lindex[\mspace{-6mu}{\tope{L}_{\mu\nu}}]{}{0}
+ (x_\mu-x_{0\,\mu}) \ope{P}_\nu - (x_\nu-x_{0\,\nu}) \ope{P}_\mu
\\ &			\label{8.9}
[\lindex[\mspace{-6mu}{\tope{L}_{\mu\nu}}]{}{0}, \ope{P}_\lambda]_{\_}=
 [\lindex[\mspace{-6mu}{\ope{L}_{\mu\nu}}]{}{0} , \ope{P}_\lambda]_{\_}
=
- \ih\{ \eta_{\lambda\mu}\tope{P}_\nu  - \eta_{\lambda\nu}\tope{P}_\mu \} .
\\ &			\label{8.10}
[\lindex[\mspace{-6mu}{\tope{L}_{\mu\nu}}]{}{0} , \ope{U}(x,x_0)]_{\_}
=
 [\lindex[\mspace{-6mu}{\ope{L}_{\mu\nu}}]{}{0} , \ope{U}(x,x_0)]_{\_}
\\ & \notag  \hphantom{ [\tope{L}_{\mu\nu}, \ope{U}(x,x_0)]_{\_} }
=
-  \{ (x_\mu-x_{0\,\mu}) \tope{P}_\nu
    - (x_\nu-x_{0\,\nu}) \tope{P}_\mu \} \circ \ope{U}(x,x_0) .
	\end{align}
As a result of~\eref{7.20} and~\eref{8.8}, the explicit form of the orbital
operator in momentum picture in terms of creation and annihilation operators
is
	\begin{multline}	\label{8.11}
\lindex[\mspace{-6mu}{\ope{L}_{\mu\nu}}]{}{0}
=
\sum_{s}\int \Id^3\bk
 ( x_{\mu}k_\nu - x_{\nu}k_\mu ) |_{ k_0=\sqrt{m^2c^2+{\bs k}^2} }
\{
a_s^{\dag\,+}(\bk)\circ a_s^-(\bk)
-
a_s^{\dag\,-}(\bk)\circ a_s^+(\bk)
\}
\\
+ \frac{1}{2} \hbar \sum_{s,s'}\int \Id^3\bk
\bigl\{
l_{\mu\nu}^{ss',-}(\bk) a_s^{\dag\,+}(\bk) \circ a_{s'}^-(\bk)
+
l_{\mu\nu}^{ss',+}(\bk) a_s^{\dag\,-}(\bk) \circ a_{s'}^+(\bk)
\bigr\}
\displaybreak[2]\\
+ \frac{1}{2} \ih \sum_{s}\int \Id^3\bk
\Bigl\{
a_s^{\dag\,+}(\bk)
\Bigl( \xlrarrow{ k_\mu \frac{\pd}{\pd k^\nu} }
     - \xlrarrow{ k_\nu \frac{\pd}{\pd k^\mu} } \Bigr)
\circ a_s^-(\bk)
\\ +
a_s^{\dag\,-}(\bk)
\Bigl( \xlrarrow{ k_\mu \frac{\pd}{\pd k^\nu} }
     - \xlrarrow{ k_\nu \frac{\pd}{\pd k^\mu} } \Bigr)
\circ a_s^+(\bk)
\Bigr\} \Big|_{ k_0=\sqrt{m^2c^2+{\bs k}^2} } \ .
	\end{multline}

	For the purposes of the present work, the equalities~\eref{8.6}
and~\eref{8.11} are very important as they, in view of~\eref{7.11-3}, give
explicit representations of the spin and orbital operators (and, hence, of the
total angular momentum operator) in momentum picture via the creation and
annihilation operators (in momentum picture).

	If we split $\lindex[\mspace{-6mu}{\ope{L}_{\mu\nu}}]{}{0}$ into a
sum of two \emph{conserved} operators as
	\begin{align}
			\label{8.11-1}
&
\lindex[\mspace{-6mu}{\ope{L}_{\mu\nu}}]{}{0}
=
\lindex[\mspace{-6mu}{\ope{L}_{\mu\nu}}]{}{2} +
\lindex[\mspace{-6mu}{\ope{L}_{\mu\nu}}]{}{3}
\\			\label{8.11-2}
&
\lindex[\mspace{-6mu}{\ope{L}_{\mu\nu}}]{}{2}
:=
+ \frac{1}{2} \hbar \sum_{s,s'}\int \Id^3\bk
\bigl\{
l_{\mu\nu}^{ss',-}(\bk) a_s^{\dag\,+}(\bk) \circ a_{s'}^-(\bk)
+
l_{\mu\nu}^{ss',+}(\bk) a_s^{\dag\,-}(\bk) \circ a_{s'}^+(\bk)
\bigr\}
	\end{align}
\vspace{-4ex}
	\begin{multline}
			\label{8.11-3}
\lindex[\mspace{-6mu}{\ope{L}_{\mu\nu}}]{}{3}
:=
\sum_{s}\int \Id^3\bk
 ( x_{\mu}k_\nu - x_{\nu}k_\mu ) |_{ k_0=\sqrt{m^2c^2+{\bs k}^2} }
\{
a_s^{\dag\,+}(\bk)\circ a_s^-(\bk)
-
a_s^{\dag\,-}(\bk)\circ a_s^+(\bk)
\}
\\
+ \frac{1}{2} \ih \sum_{s}\int \Id^3\bk
\Bigl\{
a_s^{\dag\,+}(\bk)
\Bigl( \xlrarrow{ k_\mu \frac{\pd}{\pd k^\nu} }
     - \xlrarrow{ k_\nu \frac{\pd}{\pd k^\mu} } \Bigr)
\circ a_s^-(\bk)
\\ +
a_s^{\dag\,-}(\bk)
\Bigl( \xlrarrow{ k_\mu \frac{\pd}{\pd k^\nu} }
     - \xlrarrow{ k_\nu \frac{\pd}{\pd k^\mu} } \Bigr)
\circ a_s^+(\bk)
\Bigr\} \Big|_{ k_0=\sqrt{m^2c^2+{\bs k}^2} } \ ,
	\end{multline}
then, from the proof of~\eref{8.5}, it immediately follows
	\begin{equation}	\label{8.11-4}
[\lindex[\mspace{-6mu}{\tope{L}_{\mu\nu}}]{}{2} , \ope{P}_\lambda]_{\_} = 0 ,
	\end{equation}
which implies
	\begin{gather}	\label{8.11-5}
[\lindex[\mspace{-6mu}{\tope{L}_{\mu\nu}}]{}{2} , \ope{U}(x,x_0)]_{\_} = 0
\\			\label{8.11-6}
\lindex[\mspace{-6mu}{\ope{L}_{\mu\nu}}]{}{2}
=
\lindex[\mspace{-6mu}{\tope{L}_{\mu\nu}}]{}{2}.
	\end{gather}
Combining the above results with~\eref{8.8}--\eref{8.10}, we obtain
	\begin{align}
			\label{8.11-7}
&
\lindex[\mspace{-6mu}{\ope{L}_{\mu\nu}}]{}{3}
=
\lindex[\mspace{-6mu}{\tope{L}_{\mu\nu}}]{}{3}
+ (x_\mu-x_{0\,\mu}) \ope{P}_\nu - (x_\nu-x_{0\,\nu}) \ope{P}_\mu
\\ &			\label{8.11-8}
[\lindex[\mspace{-6mu}{\tope{L}_{\mu\nu}}]{}{3}, \ope{P}_\lambda]_{\_}=
 [\lindex[\mspace{-6mu}{\ope{L}_{\mu\nu}}]{}{3} , \ope{P}_\lambda]_{\_}
=
- \ih\{ \eta_{\lambda\mu}\tope{P}_\nu  - \eta_{\lambda\nu}\tope{P}_\mu \} .
\\ &			\label{8.11-9}
[\lindex[\mspace{-6mu}{\tope{L}_{\mu\nu}}]{}{3} , \ope{U}(x,x_0)]_{\_}
=
 [\lindex[\mspace{-6mu}{\ope{L}_{\mu\nu}}]{}{3} , \ope{U}(x,x_0)]_{\_}
\\ & \notag  \hphantom{ [\tope{L}_{\mu\nu}, \ope{U}(x,x_0)]_{\_} }
=
-  \{ (x_\mu-x_{0\,\mu}) \tope{P}_\nu
    - (x_\nu-x_{0\,\nu}) \tope{P}_\mu \} \circ \ope{U}(x,x_0) .
	\end{align}
As a result of~\eref{8.11-7}, in Heisenberg picture and in terms of the
Heisenberg creation and annihilation operators~\eref{7.27}, the terms
proportional to the momentum operator in~\eref{8.11-3} disappear,
	\begin{multline} \label{8.11-9new}
\lindex[\mspace{-6mu}{\tope{L}_{\mu\nu}}]{}{3}
=
\frac{1}{2} \ih \sum_{s}\int \Id^3\bk
\Bigl\{
\tilde{a}_s^{\dag\,+}(\bk)
\Bigl( \xlrarrow{ k_\mu \frac{\pd}{\pd k^\nu} }
     - \xlrarrow{ k_\nu \frac{\pd}{\pd k^\mu} } \Bigr)
\circ \tilde{a}_s^-(\bk)
\\ +
\tilde{a}_s^{\dag\,-}(\bk)
\Bigl( \xlrarrow{ k_\mu \frac{\pd}{\pd k^\nu} }
     - \xlrarrow{ k_\nu \frac{\pd}{\pd k^\mu} } \Bigr)
\circ \tilde{a}_s^+(\bk)
\Bigr\} \Big|_{ k_0=\sqrt{m^2c^2+{\bs k}^2} } \ ,
	\end{multline}

	So, the conclusion can be made that the non\ndash commutativity
between the total angular momentum or orbital operator with the momentum
operator is entirely due to the operator~\eref{8.11-3} or, more precisely, to
the terms containing derivatives in~\eref{8.11-3}, \ie in the expression for
the orbital operator $\lindex[\mspace{-6mu}{\tope{L}_{\mu\nu}}]{}{0}$.

	It is important to be noted, the operator
$\lindex[\mspace{-6mu}{\tope{L}_{\mu\nu}}]{}{3}$ is a carrier of the pure
orbital angular momentum of a system while the sum
\(
\lindex[\mspace{-6mu}{\tope{L}_{\mu\nu}}]{}{2} +
\lindex[\mspace{-6mu}{\tope{S}_{\mu\nu}}]{}{0}
\)
represents its pure spin angular momentum.

	Using the explicit results~\eref{7.5},~\eref{7.8} and~\eref{7.20}, in
a way similar to the derivation of~\eref{8.5} (by essentially employing the
equation of motion~\eref{8.4}), one can derive by direct calculation the
relations:
	\begin{align}
			\label{8.12}
& [ \ope{P}_\mu ,\ope{P}_\nu ] = 0
\\			\label{8.13}
& [ \tope{Q},\ope{P}_\nu ] = 0
\\			\label{8.14}
& [ \lindex[\mspace{-6mu}{\ope{L}_{\mu\nu}}]{}{0} , \ope{P}_\lambda ]
 = - \ih ( \eta_{\lambda\mu}\ope{P}_\nu - \eta_{\lambda\nu}\ope{P}_\mu ) .
	\end{align}
We emphasize ones again, the relations~\eref{8.12}--\eref{8.14} are external
to the Lagrangian formalism and their validity depends on the Lagrangian one
employs. The equation~\eref{8.14} agrees with~\eref{8.9}(, obtained
from~\eref{8.6} and~\eref{2.21}) and, by virtue of~\eref{8.5}, \eref{8.6},
\eref{2.25new} and~\eref{2.11} implies
	\begin{align}
			\label{8.15}
[\tope{M}_{\mu\nu}, \tope{P}_\lambda]_{\_}
= 
- \ih\{ \eta_{\lambda\mu}\tope{P}_\nu  - \eta_{\lambda\nu}\tope{P}_\mu \} .
	\end{align}
which justifies the choice of the sign on the r.h.s.\ of~\eref{2.21}.
Notice,~\eref{8.11} agrees with~\eref{6.11} and~\eref{6.22c} (see also
Sect.~\ref{Sect11}) which will not be the case if on the r.h.s.\
of~\eref{2.21} stands $+\ih$ instead of $-\ih$.

	Since~\eref{8.13} implies (see
footnote~\ref{CommutativityWithMomentum} or~\eref{12.112})
	\begin{gather}	\label{8.16}
[\tope{Q},\ope{U}(x,x_0)]_{\_} = 0,
\\\intertext{from the general equation~\eref{12.114}, with $\ope{A}=\ope{Q}$,
we get}
			\label{8.17}
\ope{Q} = \tope{Q} .
	\end{gather}
So, the \emph{charge operator is one and the same in momentum and Heisenberg
pictures.}


\section {The anticommutation relations}
\label{Sect9}

	As we said in Sect.~\ref{Sect8}, the main difference
between~\eref{8.4} and a similar system of equations for a free charged
scalar field is in the sign in the second terms in the commutators in them.
There is also and a second difference, which is and the last one. We have in
mind  the presence of the polarization (spin) indices $s$ and $t$ (taking the
values~1 and~2 for $m\not=0$ and the value~0 for $m=0$) in~\eref{8.4}.
However, this second difference is (formally)
insignificant as we, for instance, can introduce new variables, say $\alpha$
and $\beta$, and put $\alpha=(s,\bk)$ and $\beta=(t,\bs q)$. So, if we set
$\delta(\alpha-\beta):=\delta_{st}\delta^3(\bk-\bs q)$ and
$\int\Id\beta:=\sum_{t}\int\Id^3\bs q$, the mentioned second difference
will disappear. Hence we can treat the polarization and momentum variables on
equal footing. These reasonings allow us to transfer the investigation of the
(anti)commutation relations in~\cite{bp-QFTinMP-scalars} for scalar fields to
spinor ones \emph{mutatis mutandis}, almost automatically.

	Applying the identity
	\begin{equation}	\label{9.1}
[A,B\circ C]_{\_}
= [A,B]_\varepsilon \circ C -\varepsilon B\circ [A,C]_\varepsilon ,
	\end{equation}
where $\varepsilon=\pm1$ and
$[A,B]_{\pm1}:=[A,B]_{\pm}:=A\circ B\pm B\circ A$ for any operators $A$ and
$B$ (with common domain), we rewrite~\eref{8.4} as%
\footnote{~%
Here and below, do not sum over $s$ in expressions like
$a_s^{\pm}(\bs k) \delta_{st}$.%
}
	\begin{subequations}	\label{9.2}
	\begin{gather}
				\label{9.2a}
	\begin{split}
[a_s^{\pm}(\bs k), a_t^{\dag\,+}(\bs q)]_{\varepsilon}
					\circ a_t^{-}(\bs q)
& - \varepsilon a_t^{\dag\,+}(\bs q) \circ
	[a_s^{\pm}(\bs k), a_t^{-}(\bs q) ]_{\varepsilon}
\\
- [a_s^{\pm}(\bs k), a_t^{\dag\,-}(\bs q)]_{\varepsilon}
					\circ a_t^{+}(\bs q)
& + \varepsilon a_t^{\dag\,-}(\bs q) \circ
	[a_s^{\pm}(\bs k), a_t^{+}(\bs q) ]_{\varepsilon}
\\
& \pm  a_s^{\pm}(\bs k) \delta_{st} \delta^3(\bs k-\bs q)
= f_{st}^{\pm}(\bs k,\bs q)
	\end{split}
\\
				\label{9.2b}
	\begin{split}
[a_s^{\dag\,\pm}(\bs k), a_t^{\dag\,+}(\bs q)]_{\varepsilon}
					\circ a_t^{-}(\bs q)
& - \varepsilon a_t^{\dag\,+}(\bs q) \circ
	[a_s^{\dag\,\pm}(\bs k), a_t^{-}(\bs q) ]_{\varepsilon}
\\
- [a_s^{\dag\,\pm}(\bs k), a_t^{\dag\,-}(\bs q)]_{\varepsilon}
					\circ a_t^{+}(\bs q)
& + \varepsilon a_t^{\dag\,-}(\bs q) \circ
	[a_s^{\dag\,\pm}(\bs k), a_t^{+}(\bs q) ]_{\varepsilon}
\\
& \pm  a_s^{\dag\,\pm}(\bs k) \delta_{st} \delta^3(\bs k-\bs q)
= f_{st}^{\dag\,\pm}(\bs k,\bs q) .
	\end{split}
	\end{gather}
	\end{subequations}
Writing these equalities explicitly for the upper, ``$+$'', and lower,
``$-$'', signs, we see that they can equivalently be written respectively as:
	\begin{subequations}	\label{9.3}
	\begin{gather}
				\label{9.3a}
	\begin{split}
\pm [a_s^{\pm}(\bs k), a_t^{\dag\,\pm}(\bs q)]_{\varepsilon}
					\circ a_t^{\mp}(\bs q)
& \pm \varepsilon a_t^{\dag\,\mp}(\bs q) \circ
	[a_s^{\pm}(\bs k), a_t^{\pm}(\bs q) ]_{\varepsilon}
\\
\mp [a_s^{\pm}(\bs k), a_t^{\dag\,\mp}(\bs q)]_{\varepsilon}
					\circ a_t^{\pm}(\bs q)
& \mp \varepsilon a_t^{\dag\,\pm}(\bs q) \circ
	[a_s^{\pm}(\bs k), a_t^{\mp}(\bs q) ]_{\varepsilon}
\\
& \pm  a_s^{\pm}(\bs k) \delta_{st} \delta^3(\bs k-\bs q)
= f_{st}^{\pm}(\bs k,\bs q)
	\end{split}
\\
				\label{9.3b}
	\begin{split}
\pm [a_s^{\dag\,\pm}(\bs k), a_t^{\dag\,\pm}(\bs q)]_{\varepsilon}
					\circ a_t^{\mp}(\bs q)
& \pm \varepsilon a_t^{\dag\,\mp}(\bs q) \circ
	[a_s^{\dag\,\pm}(\bs k), a_t^{\pm}(\bs q) ]_{\varepsilon}
\\
\mp [a_s^{\dag\,\pm}(\bs k), a_t^{\dag\,\mp}(\bs q)]_{\varepsilon}
					\circ a_t^{\pm}(\bs q)
& \mp \varepsilon a_t^{\dag\,\pm}(\bs q) \circ
	[a_s^{\dag\,\pm}(\bs k), a_t^{\mp}(\bs q) ]_{\varepsilon}
\\
& \pm  a_s^{\dag\,\pm}(\bs k) \delta_{st} \delta^3(\bs k-\bs q)
= f_{st}^{\dag\,\pm}(\bs k,\bs q) .
	\end{split}
	\end{gather}
	\end{subequations}

	Following the book~\cite[subsect.~10.1]{Bogolyubov&Shirkov}, we shall
assume, as an \emph{additional condition} that the (anti)commutators of
creation and annihilation operator are proportional to the identity
operator of system's Hilbert space $\Hil$ of states.%
\footnote{~%
The authors of~\cite{Bogolyubov&Shirkov} refer to~\cite[\S~21]{Dirac-PQM},
where the introduction of commutation relations in quantum mechanics is
discussed for systems which have a classical analogue. The introduction of
\emph{anti}commutation relations, which are adequate for description of spinor
fields, cannot be confirmed in this way as half\ndash integer spin
(Fermi\ndash Dirac) systems have not classical analogues.%
}
Namely, we \emph{assume} the following conditions:
	\begin{equation}	\label{9.4}
	\begin{split}
[a_s^{\pm}(\bs k), a_t^{\pm}(\bs q) ]_{\varepsilon}
	= g_{\varepsilon,st}^{\pm}(\bs k,\bs q) \id_\Hil
\qquad
[a_s^{\dag\,\pm}(\bs k), a_t^{\dag\,\pm}(\bs q) ]_{\varepsilon}
	= g_{\varepsilon,st}^{\dag\,\pm}(\bs k,\bs q) \id_\Hil
\\
[a_s^{\mp}(\bs k), a_t^{\pm}(\bs q) ]_{\varepsilon}
	= b_{\varepsilon,st}^{\pm}(\bs k,\bs q) \id_\Hil
\qquad
[a_s^{\dag\,\mp}(\bs k), a_t^{\dag\,\pm}(\bs q) ]_{\varepsilon}
	= b_{\varepsilon,st}^{\dag\,\pm}(\bs k,\bs q) \id_\Hil
\\
[a_s^{\pm}(\bs k), a_t^{\dag\,\pm}(\bs q) ]_{\varepsilon}
	= d_{\varepsilon,st}^{\pm}(\bs k,\bs q) \id_\Hil
\qquad
[a_s^{\dag\,\pm}(\bs k), a_t^{\pm}(\bs q) ]_{\varepsilon}
	= \varepsilon  d_{\varepsilon,st}^{\pm}(\bs q,\bs k) \id_\Hil
\\
[a_s^{\mp}(\bs k), a_t^{\dag\,\pm}(\bs q) ]_{\varepsilon}
	= e_{\varepsilon,st}^{\pm}(\bs k,\bs q) \id_\Hil
\qquad
[a_s^{\dag\,\mp}(\bs k), a_t^{\pm}(\bs q) ]_{\varepsilon}
	= \varepsilon  e_{\varepsilon,st}^{\mp}(\bs q,\bs k) \id_\Hil
	\end{split}
	\end{equation}
where $\varepsilon=\pm1$ and $g_{\varepsilon,st}^{\pm}$,
$g_{\varepsilon,st}^{\dag\,\pm}$, \dots, $e_{\varepsilon,st}^{\pm}$ are some
complex\ndash valued (generalized) functions, which we have to determine.
These last (generalized) functions are subjected to a number of restrictions
which were derived
in~\cite[sections~7 and~15]{bp-QFTinMP-scalars}.\typeout{Bozho, please check
the section numbers before publication!!!!!!!!} Their explicit form is:
	\begin{subequations}	\label{9.5}
	\begin{gather}	\label{9.5a}
(\bs k+\bs q) g_{\varepsilon,st}^{\pm} (\bs k,\bs q) = 0 \quad
(\bs k+\bs q) g_{\varepsilon,st}^{\dag\,\pm} (\bs k,\bs q) = 0 \quad
(\bs k+\bs q) d_{\varepsilon,st}^{\pm} (\bs k,\bs q) = 0
\\
			\label{9.5b}
(\bs k-\bs q) b_{\varepsilon,st}^{\pm} (\bs k,\bs q) = 0 \quad
(\bs k-\bs q) b_{\varepsilon,st}^{\dag\,\pm} (\bs k,\bs q) = 0 \quad
(\bs k-\bs q) e_{\varepsilon,st}^{\pm} (\bs k,\bs q) = 0
	\end{gather}
	\end{subequations}
\vspace{-4ex}
	\begin{subequations}	\label{9.6}
	\begin{gather}	\label{9.6a}
\bigl( \sqrt{m^2c^2+{\bs k}^2} + \sqrt{m^2c^2+{\bs q}^2} \bigr)
	\alpha(\bs k,\bs q) = 0
\qquad\text{for }
\alpha=g_{\varepsilon,st}^{\pm}, g_{\varepsilon,st}^{\dag\,\pm}, d_{\varepsilon,st}^{\pm}
\\
			\label{9.6b}
\bigl( \sqrt{m^2c^2+{\bs k}^2} - \sqrt{m^2c^2+{\bs q}^2} \bigr)
	\beta(\bs k,\bs q)  = 0
\qquad\text{for }
\beta=b_{\varepsilon,st}^{\pm}, b_{\varepsilon,st}^{\dag\,\pm}, e_{\varepsilon,st}^{\pm} .
	\end{gather}
	\end{subequations}

	Regarding $g_{\varepsilon,st}^{\pm}$, $g_{\varepsilon,st}^{\dag\,\pm}$,
\dots,$e_{\varepsilon,st}^{\pm}$ as distributions, from~\eref{9.5}, we
derive:
	\begin{subequations}	\label{9.7}
	\begin{gather}	\label{9.7a}
f(\bs q) \alpha  (\bs k,\bs q) =  f(-\bs k) \alpha  (\bs k,\bs q)
\qquad\text{for }
\alpha=g_{\varepsilon,st}^{\pm}, g_{\varepsilon,st}^{\dag\,\pm}, d_{\varepsilon,st}^{\pm}
\\
			\label{9.7b}
f(\bs q) \beta (\bs k,\bs q) =  f(+\bs k) \beta (\bs k,\bs q)
\qquad\text{for }
\beta=b_{\varepsilon,st}^{\pm}, b_{\varepsilon,st}^{\dag\,\pm}, e_{\varepsilon,st}^{\pm}
	\end{gather}
	\end{subequations}
for any function $f$ which is polynomial or convergent power series. In view
of~\eref{9.7}, the equalities~\eref{9.6b} are identically satisfied,
while~\eref{9.6a} are equivalent to the equations
	\begin{equation}	\label{9.8}
\sqrt{m^2c^2+{\bs k}^2} \alpha(\bs k,\bs q) = 0
\qquad\text{for }
\alpha= g_{\varepsilon,st}^{\pm}, g_{\varepsilon,st}^{\dag\,\pm}, d_{\varepsilon,st}^{\pm} .
	\end{equation}

	Substituting~\eref{9.4} into~\eref{9.3} and taking into account the
equalities~\eref{9.7}, we obtain, from~\eref{8.4c}, the following system of
equations for the unknown (generalized) functions $g_{\varepsilon,st}^{\pm}$,
$g_{\varepsilon,st}^{\dag\,\pm}$, \dots, $e_{\varepsilon,st}^{\pm}$:
	\begin{subequations}	\label{9.9}
	\begin{align}
	\begin{split}
				\label{9.9a}
& k_a \sum_{t} \int\Id^3\bs q \bigl\{
\mp a_t^{\mp}(\bs q) d_{\varepsilon,st}^{\pm}(\bs k,\bs q)
\mp \varepsilon a_t^{\dag\,\mp}(\bs q) g_{\varepsilon,st}^{\pm}(\bs k,\bs q)
\\
&\mp \varepsilon a_t^{\dag\,\pm}(\bs q) b_{\varepsilon,st}^{\mp}(\bs k,\bs q)
\mp a_t^{\pm}(\bs q)
	\bigl( e_{\varepsilon,st}^{\mp} (\bs k,\bs q)
      - \delta_{st} \delta^3(\bs k-\bs q) \bigr)
\bigr\}
= 0
	\end{split}
\\[1.25ex]				\label{9.9b}
	\begin{split}
 \sqrt{m^2c^2+{\bs k}^2} \sum_{t} \int\Id^3\bs q \bigl\{
\pm a_t^{\mp}(\bs q) d_{\varepsilon,st}^{\pm}(\bs k,\bs q)
\pm \varepsilon a_t^{\dag\,\mp}(\bs q) g_{\varepsilon,st}^{\pm}(\bs k,\bs q)
\\
 \mp \varepsilon a_t^{\dag\,\pm}(\bs q) b_{\varepsilon,st}^{\mp}(\bs k,\bs q)
\mp a_t^{\pm}(\bs q)
	\bigl( e_{\varepsilon,st}^{\mp} (\bs k,\bs q)
     -	\delta_{st} \delta^3(\bs k-\bs q) \bigr)
\bigr\}
= 0
	\end{split}
\displaybreak[1]\\[1.25ex]		\label{9.9c}
	\begin{split}
& k_a \sum_{t} \int\Id^3\bs q \bigl\{
\mp a_t^{\mp}(\bs q) g_{\varepsilon,st}^{\dag\,\pm}(\bs k,\bs q)
\mp a_t^{\dag\,\mp}(\bs q) d_{\varepsilon,st}^{\pm}(\bs q,\bs k)
\\
& \mp a_t^{\pm}(\bs q) b_{\varepsilon,st}^{\dag\,\mp}(\bs k,\bs q)
\mp a_t^{\dag\,\pm}(\bs q)
	\bigl( e_{\varepsilon,st}^{\pm} (\bs q,\bs k)
	- \delta_{st} \delta^3(\bs k-\bs q) \bigr)
\bigr\}
= 0
	\end{split}
\displaybreak[1]\\[1.25ex]	\label{9.9d}
	\begin{split}
& \sqrt{m^2c^2+{\bs k}^2} \sum_{t} \int\Id^3\bs q \bigl\{
\pm a_t^{\mp}(\bs q) g_{\varepsilon,st}^{\dag\,\pm}(\bs k,\bs q)
\pm a_t^{\dag\,\mp}(\bs q) d_{\varepsilon,st}^{\pm}(\bs q,\bs k)
\\
& \mp a_t^{\pm}(\bs q) b_{\varepsilon,st}^{\dag\,\mp}(\bs k,\bs q)
  \mp a_t^{\dag\,\pm}(\bs q)
	\bigl( e_{\varepsilon,st}^{\pm} (\bs q,\bs k)
	- \delta_{st} \delta^3(\bs k-\bs q) \bigr)
\bigr\}
= 0 ,
	\end{split}
	\end{align}
	\end{subequations}
where $\varepsilon=\pm1$ is a free parameter and $a=1,2,3$.

	Now we shall  impose as a \emph{second}, after~\eref{9.4},
\emph{additional condition} the requirement that~\eref{9.9} to be valid for
arbitrary $a_t^{\pm}(\bs q)$ and $a_t^{\dag\,\pm}(\bs q)$. In this case, for
$(\bs k,m)\not=(\bs0,0)$, the equations~\eref{9.9} have the following
unique solution with respect to the $g_{\varepsilon,st}^{\pm}$,
$g_{\varepsilon,st}^{\dag\,\pm}$, \dots, $e_{\varepsilon,st}^{\pm}$:

	\begin{subequations}\label{9.10}
	\begin{align}
				\label{9.10a}
  g_{\varepsilon,st}^{\pm} (\bs k,\bs q)
& = g_{\varepsilon,st}^{\dag\,\pm} (\bs k,\bs q)
= b_{\varepsilon,st}^{\pm} (\bs k,\bs q)
= b_{\varepsilon,st}^{\dag\,\pm} (\bs k,\bs q)
= d_{\varepsilon,st}^{\pm} (\bs k,\bs q)
= 0
\\				\label{9.10b}
e_{\varepsilon,st}^{\pm} (\bs k,\bs q)
& = \delta_{st} \delta^3(\bs k-\bs q) .
	\end{align}
	\end{subequations}
Evidently,~\eref{9.10a} convert~\eref{9.8} into identities, and, consequently,
under the hypotheses made,~\eref{9.10} is the general solution of our
problem (for $(\bs k,m)\not=(\bs0,0)$).

	Let us consider now the case $(\bs k,m)=(\bs0,0)$,
when~\eref{9.9} and~\eref{9.6} reduce to the identity $0=0$. In it, the
operators
	\begin{equation}	\label{9.11}
a_s^{\pm}(\bs k) \text{ and } a_s^{\dag\,\pm}(\bs k)
\qquad\text{for $m=0$ and $\bs k= \bs0$,}
	\end{equation}
must satisfy~\eref{9.4} in which the (generalized) functions
$g_{\varepsilon,st}^{\pm}(\bs0,\bs q)$,
$g_{\varepsilon,st}^{\dag\,\pm}(\bs0,\bs q)$, \dots,
$e_{\varepsilon,st}^{\pm}(\bs0,\bs q)$ remain as free parameters of the
theory. To insure a continuous limit $(\bs k,m)\to(\bs0,0)$, we shall assume,
\emph{by convention}, that these (generalized) functions are given
via~\eref{9.10} with $\bs k=\bs0$ (and $m=0$). Physically (see
Sect.~\ref{Sect6}), the operators~\eref{9.11} describe creation/annihilation
of massless particles (of massless spinor field) with vanishing 4\ndash
momentum and angular momentum, but carrying charge $\pm q$ and having
different spin operators. Thus, the \emph{theory admits existence of free,
charged, spin $\frac{1}{2}$, massless particles with vanishing spacetime
dynamical variables (4\ndash momentum and angular momentum).}%
\footnote{~%
It seems, until now such particles/fields have not been observed.%
}
This is the `quantized' version of the `exotic' solutions~\eref{5.1} (see the
paragraph containing equation~\eref{6.15} in Sect.~\ref{Sect6}). Meanwhile, we
notice that the solutions~\eref{5.1} are rejected by~\eref{9.10}
and~\eref{9.4} (see~\eref{6.15}--\eref{6.17}).

	The developed until now theory cannot make a distinction between the
cases $\varepsilon=-1$ and $\varepsilon=+1$ in~\eref{9.4}, corresponding to
quantization with commutators and anticommutators, respectively. For the
purpose, a \emph{new additional conditions} (hypotheses, postulate) is
required. As such additional condition can serve the spin\ndash statistics
theorem (or an equivalent to it assertion, like, e.g., the charge
symmetry)~\cite{Bogolyubov&Shirkov,Bjorken&Drell,Roman-QFT}. According to it,
the half\ndash integer spin fields should be quantized via
\emph{anti}commutators and, consequently, in our case the choice
	\begin{equation}	\label{9.12}
\varepsilon = +1
	\end{equation}
should be made. So, the equalities~\eref{9.4},~\eref{9.10} and~\eref{9.12}
imply the following anticommutation rules:
	\begin{align}	\notag
&[a_s^{\pm}(\bs k), a_t^{\pm}(\bs q) ]_{+}
	= 0
&&
[a_s^{\dag\,\pm}(\bs k), a_t^{\dag\,\pm}(\bs q) ]_{+}
	= 0
\\	\notag
&[a_s^{\mp}(\bs k), a_t^{\pm}(\bs q) ]_{+}
	=  0
&&
[a_s^{\dag\,\mp}(\bs k), a_t^{\dag\,\pm}(\bs q) ]_{+}
	= 0
\\	\notag
&[a_s^{\pm}(\bs k), a_t^{\dag\,\pm}(\bs q) ]_{+}
	= 0
&&
[a_s^{\dag\,\pm}(\bs k), a_t^{\pm}(\bs q) ]_{+}
	= 0
\\	\label{9.13}
&[a_s^{\mp}(\bs k), a_t^{\dag\,\pm}(\bs q) ]_{+}
	= \delta_{st} \delta^3(\bs k-\bs q) \id_\Hil
&&
[a_s^{\dag\,\mp}(\bs k), a_t^{\pm}(\bs q) ]_{+}
	= \delta_{st} \delta^3(\bs k-\bs q) \id_\Hil .
	\end{align}
These quantization rules will be accepted hereafter in the present work. In
Sect.~\ref{Sect12}, we shall show how they can be derived without invoking the
spin\ndash statistics theorem.

	As we have noted several times above, the concepts of a distribution
(generalized function) and operator-valued distribution appear during the
derivation of the commutation relations~\eref{9.13}. We first met them in the
tri\ndash linear relations~\eref{8.4}. In particular, the canonical
commutation relations~\eref{9.13} have a sense iff
the anticommutators in them are
operator-valued distributions (proportional to $\id_\Hil$), which is
\emph{not} the case if the field is described as an ordinary operator acting
on $\Hil$. These facts point to inherent contradiction of quantum field
theory if the field variables are considered as operators acting on a Hilbert
space. The rigorous mathematical setting requires the fields variables to be
regarded as operator\ndash valued distributions. However, such a setting is
out of the scope of the present work and the reader is referred to books
like~\cite{Streater&Wightman,Jost,
Bogolyubov&et_al.-AxQFT,Bogolyubov&et_al.-QFT}
for more details and realization of that program. In what follows, the
distribution character of the quantum fields will be encoded in the Dirac's
delta function, which will appear in relations like~\eref{8.4}
and~\eref{9.13}.

	As a first application of the anticommutation relations~\eref{9.13},
we shall calculate the commutator between the components~\eref{7.11-3} of the
spin operator. For the purpose, as well as for computing different
commutators between the dynamical variables, we shall use the following
commutation relations between quadratic combinations of creation and
annihilation operators:
	\begin{equation}	\label{9.14}
	\begin{split}
[ a_{s}^{\dag\,\pm}(\bk)  \circ a_{s'}^{\mp}(\bk)
&,
  a_{t}^{\dag\,\pm}(\bs p)\circ a_{t'}^{\mp} (\bs p) ]_{\_}
=
\{
- \delta_{st'} a_{t}^{\dag\,\pm}(\bs p)  \circ a_{s'}^{\mp}(\bk)
+ \delta_{s't} a_{s}^{\dag\,\pm}(\bs k)  \circ a_{t'}^{\mp}(\bs p)
\} \delta^3(\bk-\bs p)
\\
[ a_{s}^{\dag\,\pm}(\bk)  \circ a_{s'}^{\mp}(\bk)
&,
  a_{t}^{\mp}(\bs p)\circ a_{t'}^{\dag\,\pm} (\bs p) ]_{\_}
=
\{
- \delta_{st} a_{s'}^{\mp}(\bs k)  \circ a_{t'}^{\dag\,\pm}(\bs p)
+ \delta_{s't'} a_{t}^{\mp}(\bs p)  \circ a_{s}^{\dag\,\pm}(\bs k)
\} \delta^3(\bk-\bs p)
\\
[ a_{s}^{\pm}(\bk)  \circ a_{s'}^{\dag\,\mp}(\bk)
&,
  a_{t}^{\pm}(\bs p)\circ a_{t'}^{\dag\,\mp} (\bs p) ]_{\_}
=
\{
- \delta_{st'} a_{t}^{\pm}(\bs p) \circ a_{s'}^{\dag\,\mp}(\bk)
+ \delta_{s't} a_{s}^{\pm}(\bs k) \circ a_{t'}^{\dag\,\mp}(\bs p)
\} \delta^3(\bk-\bs p)
\\
[ a_{s}^{\dag\,\pm}(\bk)  \circ a_{s'}^{\mp}(\bk)
&,
  a_{t}^{\dag\,\mp}(\bs p)\circ a_{t'}^{\pm}  (\bs p) ]_{\_}
= 0
\\
[ a_{s}^{\dag\,\pm}(\bk)  \circ a_{s'}^{\mp}(\bk)
&,
  a_{t}^{\pm}(\bs p)\circ a_{t'}^{\dag\,\mp}  (\bs p) ]_{\_}
= 0
\\
[ a_{s}^{\pm}(\bk)  \circ a_{s'}^{\dag\,\mp}(\bk)
&,
  a_{t}^{\mp}(\bs p)\circ a_{t'}^{\dag\,\pm}  (\bs p) ]_{\_}
= 0  .
	\end{split}
	\end{equation}
These equalities are simple corollaries of the identities
\(
[A,B\circ C]_{\_} = [A,B]_{\_}\circ C + B\circ [A,C]_{\_}
\)
and
\(
[B\circ C,A]_{\_} = - [A,B]_{+}\circ C + B\circ [A,C]_{+}
\),
applied in this order to the left-hand-sides of~\eref{9.14}, and~\eref{9.13}.
Now, via a direct calculation by means of~\eref{7.11-3} and~\eref{9.14}, we
get
	\begin{multline}	\label{9.15}
[ \lindex[\mspace{-6mu}{\ope{S}_{\mu\nu}}]{}{0} ,
  \lindex[\mspace{-6mu}{\ope{S}_{\varkappa\lambda}}]{}{0} ]_{\_}
 =
\frac{1}{4}\hbar^2
\sum_{s,s',t} \int\Id^3\bk
\bigl\{
\bigl( \sigma_{\mu\nu}^{ss',-}(\bk) \sigma_{\varkappa\lambda}^{s't,-}(\bk)
     - \sigma_{\varkappa\lambda}^{ss',-}(\bk) \sigma_{\mu\nu}^{s't,-}(\bk)
\bigr) a_{s}^{\dag\,+}(\bk) \circ a_{t}^{-}(\bk)
\\ +
\bigl( \sigma_{\mu\nu}^{ss',+}(\bk) \sigma_{\varkappa\lambda}^{s't,+}(\bk)
     - \sigma_{\varkappa\lambda}^{ss',+}(\bk) \sigma_{\mu\nu}^{s't,+}(\bk)
\bigr) a_{s}^{\dag\,-}(\bk) \circ a_{t}^{+}(\bk)
\bigr\}  .
	\end{multline}
In particular, the equation
	\begin{equation}	\label{9.16new1}
[ \lindex[\mspace{-6mu}{\ope{S}_{\mu\nu}}]{}{0} ,
  \lindex[\mspace{-6mu}{\ope{S}_{\varkappa\lambda}}]{}{0} ]_{\_}
 = 0
\quad \text{for }
\delta^{0\mu} + \delta^{0\nu} + \delta^{0\varkappa} + \delta^{0\lambda}  \ge 1
	\end{equation}
is an evident consequence of~\eref{7.12-10} and~\eref{9.15}.
	Besides, in the massless case, when the polarization indices take the
single value 0, we obtain
	\begin{equation}	\label{9.16}
[ \lindex[\mspace{-6mu}{\ope{S}_{\mu\nu}}]{}{0} ,
  \lindex[\mspace{-6mu}{\ope{S}_{\varkappa\lambda}}]{}{0} ]_{\_} \big|_{m=0}
 = 0 .
	\end{equation}
For $m\not=0$, the r.h.s.\ of~\eref{9.15} is, generally, a non-zero operator,
which means that in the massive case not all of the spin operator components
are simultaneously measurable, contrary to the massless one. More details on
this problem will be given in Sect.~\ref{Sect11}
(see~\eref{11.9}--\eref{11.14}).

	For $m\not=0$, the summation over $s'$ in~\eref{9.15} can be
performed explicitly by means of~\eref{7.12-10}, \eref{6.23}, \eref{5.25},
\eref{5.20}, and
\(
[ k^\lambda \gamma_\lambda , \gamma^0 \sigma_{\mu\nu} ]_{+}
=
2k_0 \sigma_{\mu\nu} - 2\iu \gamma_0 ( k_\mu\gamma_\nu - k_\nu\gamma_\mu )
\)
(see~\eref{3.26}). The result reads:
	\begin{multline}	\label{9.16new}
\sum_{s'} \{
\hat{\sigma}_{\mu\nu}^{ss',\mp}(\bk)
			\hat{\sigma}_{\varkappa\lambda}^{s't,\mp}(\bk)
-
\hat{\sigma}_{\varkappa\lambda}^{ss',\mp}(\bk)
			\hat{\sigma}_{\mu\nu}^{s't,\mp}(\bk)
\}
\\
=
v^{\dag\, s, \pm}(\bk)
\Bigl\{
[\sigma_{\mu\nu},\sigma_{\varkappa\lambda}]_{\_}
- \iu\frac{\gamma^0}{k_0}
\bigl(
\sigma_{\mu\nu} (k_\varkappa\gamma_\lambda - k_\lambda\gamma_\varkappa )
-
\sigma_{\varkappa\lambda} (k_\mu\gamma_\nu -  k_\nu\gamma_\mu )
\bigr)
\Bigr\}
v^{\dag\, s, \mp}(\bk)
\\
=
- 2\iu \bigl\{
  \eta_{\mu\varkappa} \hat{\sigma}_{\nu\lambda}^{st,\mp}(\bk)
- \eta_{\nu\varkappa} \hat{\sigma}_{\mu\lambda}^{st,\mp}(\bk)
- \eta_{\mu\lambda} \hat{\sigma}_{\nu\varkappa}^{st,\mp}(\bk)
+ \eta_{\nu\lambda} \hat{\sigma}_{\mu\varkappa}^{st,\mp}(\bk)
\bigr\}
\\ -
\frac{\iu}{k_0}
v^{\dag\, s, \pm}(\bk)
\bigl\{
\gamma^0
\bigl(
\sigma_{\mu\nu} (k_\varkappa\gamma_\lambda - k_\lambda\gamma_\varkappa )
-
\sigma_{\varkappa\lambda} (k_\mu\gamma_\nu -  k_\nu\gamma_\mu )
\bigr)
\bigr\}
v^{\dag\, s, \mp}(\bk)
\qquad \text{ for } m\not=0 ,
	\end{multline}
where
\(
[\sigma_{\mu\nu},\sigma_{\varkappa\lambda}]_{\_}
=
- 2\iu \bigl\{
  \eta_{\mu\varkappa} {\sigma}_{\nu\lambda}
- \eta_{\nu\varkappa} {\sigma}_{\mu\lambda}
- \eta_{\mu\lambda} {\sigma}_{\nu\varkappa}
+ \eta_{\nu\lambda} {\sigma}_{\mu\varkappa}
\bigr\}
\)
(see~\eref{3.26}) was used and one should set
 $\mu,\nu,\varkappa,\lambda=1,2,3$
in the context of~\eref{9.15}, due to~\eref{7.12-10}.

	The commutativity between the spin and charge operators, \ie
	\begin{equation}	\label{9.17}
[ \lindex[\mspace{-6mu}{\ope{S}_{\mu\nu}}]{}{0} , \ope{Q} ]_{\_} = 0 ,
	\end{equation}
is an almost trivial corollary from~\eref{7.8},~\eref{7.11-3},
and~\eref{9.14}. This implies that the spin and charge are simultaneously
measurable quantities.

	Important corollaries from~\eref{9.13} are the commutation relations
between the field operators and the total angular momentum operator
$\ope{M}_{\mu\nu}$, \viz
	\begin{subequations}	\label{9.18}
	\begin{align}
			\label{9.18a}
& [\psi, \ope{M}_{\mu\nu}(x,x_0)]_{\_}
=
x_\mu [\psi ,\ope{P}_\nu]_{\_} - x_\nu [\psi ,\ope{P}_\mu]_{\_}
+ \frac{1}{2}\hbar \sigma_{\mu\nu}\psi
\\			\label{9.18b}
& [\opsi, \ope{M}_{\mu\nu}(x,x_0)]_{\_}
=
x_\mu [\opsi ,\ope{P}_\nu]_{\_} - x_\nu [\opsi ,\ope{P}_\mu]_{\_}
- \frac{1}{2}\hbar \opsi \sigma_{\mu\nu} ,
	\end{align}
	\end{subequations}
which in Heisenberg picture read
	\begin{subequations}	\label{9.19}
	\begin{align}
			\label{9.19a}
& [\tope{\psi}(x), \tope{M}_{\mu\nu}]_{\_}
=
\Bigl( x_\mu\frac{\pd}{\pd x^\nu} - x_\nu \frac{\pd}{\pd x^\mu}
\Bigr) \tope{\psi}(x)
+ \frac{1}{2}\hbar \sigma_{\mu\nu}\tope{\psi}(x)
\\			\label{9.19b}
& [\tope{\opsi}(x), \tope{M}_{\mu\nu}]_{\_}
=
\Bigl( x_\mu\frac{\pd}{\pd x^\nu} - x_\nu \frac{\pd}{\pd x^\mu}
\Bigr) \tope{\opsi}(x)
- \frac{1}{2}\hbar \tope{\opsi}(x) \sigma_{\mu\nu} ,
	\end{align}
	\end{subequations}
and, together with~\eref{2.28} for
$\tope{\varphi}_i(x)=\tope{\psi}(x),\tope{\opsi}(x)$, represent the
relativistic covariance of the theory
considered~\cite[\S~80]{Bjorken&Drell-2}.

	Since~\eref{9.18b} is a consequence of~\eref{9.18a},
$\opsi=\psi^\dag\gamma^0$, and~\eref{3.28new}, we shall prove only the
equation~\eref{9.18a}; besides, for brevity, only the massive case,
$m\not=0$, will be considered.%
\footnote{~%
One can prove independently~\eref{9.18b} in a similar way. For an alternative
proof of~\eref{9.19} --- see~\cite[\S~19.1]{Akhiezer&Berestetskii}.%
}
We shall prove that
	\begin{subequations}	\label{9.20}
	\begin{multline}	\label{9.20a}
[ \psi , \lindex[\mspace{-6mu}{\ope{S}_{\mu\nu}}]{}{0} ]_{\_}
\\
=
\begin{cases}
  \frac{1}{2}\hbar \sigma_{\mu\nu} \psi
+ \frac{1}{2}\ih \gamma^0
\times & \\ \times
\int\Id^3\bs p
\Bigl\{ \frac{1}{p_0} (p_\mu\gamma_\nu - p_\nu\gamma_\mu)
(\psi^+(\bs p) + \psi^-(\bs p))
\Bigr\} \Big|_{ p_0=\sqrt{m^2c^2+{\bs p}^2} }
	&	\text{ for } \mu,\nu=1,2,3
\\
0	&	\text{ for } \mu=0\text{ or } \nu=0
\end{cases}
	\end{multline}
\vspace{-4ex}
	\begin{multline}	\label{9.20b}
[ \psi , \lindex[\mspace{-6mu}{\ope{L}_{\mu\nu}}]{}{0} ]_{\_}
=
x_\mu [\psi ,\ope{P}_\nu]_{\_} - x_\nu [\psi ,\ope{P}_\mu]_{\_}
\\
-
\begin{cases}
\frac{1}{2}\ih \gamma^0
\times & \\ \times
\int\Id^3\bs p
\Bigl\{ \frac{1}{p_0} (p_\mu\gamma_\nu - p_\nu\gamma_\mu)
(\psi^+(\bs p) + \psi^-(\bs p))
\Bigr\} \Big|_{ p_0=\sqrt{m^2c^2+{\bs p}^2} }
	&	\text{ for } \mu,\nu=1,2,3
\\
- \frac{1}{2}\hbar \sigma_{\mu\nu} \psi
	&	\text{ for } \mu=0\text{ or } \nu=0
\end{cases}
	\end{multline}
	\end{subequations}
from where equation~\eref{9.18a} follows, due to~\eref{8.7}.

	For $\mu=0$ or $\nu=0$,~\eref{9.20a} is an evident corollary
of~\eref{7.12-10} and~\eref{7.11-3}.
	Let $\mu,\nu=1,2,3$. Substituting the equations~\eref{8.1a}
and~\eref{7.11-3} (see also~\eref{8.6}) in
\(
[ \psi , \lindex[\mspace{-6mu}{\ope{S}_{\mu\nu}}]{}{0} ]_{\_}
\)
and applying the identity
\(
[A,B\circ C]_{\_}
= [A,B]_{+}\circ C - B\circ [A,C]_{+} ,
\)
the anticommutation relations~\eref{9.13}, and then~\eref{7.12-10},
\eref{5.26} and~\eref{5.25}, we get ($\mu,\nu=1,2,3$)
\[
[ \psi , \lindex[\mspace{-6mu}{\ope{S}_{\mu\nu}}]{}{0} ]_{\_}
=
\frac{1}{2}\hbar \int\Id^3\bs p
\Bigl\{
\frac{p^\lambda\gamma_\lambda - mc}{2p_0}
	\gamma^0 \sigma_{\mu\nu} \psi^+(\bs p)
+
\frac{p^\lambda\gamma_\lambda + mc}{2p_0}
	\gamma^0 \sigma_{\mu\nu} \psi^- (\bs p)
\Bigr\} \Big|_{ p_0=\sqrt{m^2c^2+{\bs p}^2} } \ .
\]
Since
\(
[ p^\lambda \gamma_\lambda , \gamma^0 \sigma_{\mu\nu} ]_{+}
=
2p_0 \sigma_{\mu\nu} - 2\iu \gamma_0 ( p_\mu\gamma_\nu - p_\nu\gamma_\mu )
\)
(use
\(
[A,B\circ C]_{+}
= [A,B]_{+}\circ C - B\circ [A,C]_{\_} ,
\)
and~\eref{3.26}), the last equality implies~\eref{9.20a} for $\mu,\nu=1,2,3$,
as a result of~\eref{6.1new2}, \eref{6.1}, and~\eref{5.17} (or~\eref{6.20}
and~\eref{5.20}).

	Performing similar manipulations with the l.h.s.\ of~\eref{9.20b}, by
applying~\eref{7.20} for~\eref{7.11-3} and~\eref{7.21} for~\eref{7.12-10}, we
see that the last displayed equation should be replace with
	\begin{multline*}
[ \psi , \lindex[\mspace{-6mu}{\ope{L}_{\mu\nu}}]{}{0} ]_{\_}
=
x_\mu [\psi ,\ope{P}_\nu]_{\_} - x_\nu [\psi ,\ope{P}_\mu]_{\_}
+
\frac{1}{2}\ih (2\pi\hbar)^{-3/2} \sum_s \int\Id^3\bs p
\Bigl\{
p_\mu \Bigl( \frac{p^\lambda\gamma_\lambda - mc}{p_0} \gamma^0 - 2 \Bigr)
\\ \times
	\frac{\pd v^{s,+}(\bs p)}{\pd p^\nu} a_s^+(\bs p)
+
p_\mu \Bigl( \frac{p^\lambda\gamma_\lambda + mc}{p_0} \gamma^0 - 2 \Bigr)
	\frac{\pd v^{s,-}(\bs p)}{\pd p^\nu} a_s^-(\bs p)
-
(\mu \leftrightarrow \nu)
\Bigr\} \Big|_{ p_0=\sqrt{m^2c^2+{\bs p}^2} } \ ,
	\end{multline*}
where $-(\mu \leftrightarrow \nu)$ means that one should subtract the previous
terms in the braces with the change $\mu\leftrightarrow \nu$,
\eref{7.12-3new1} has been applied, and integration by parts of terms
proportional to
$\frac{\pd a^\pm_s(\bs p)}{\pd p^\nu}$
has been performed.
	Since $[p^\lambda\gamma_\lambda,\gamma^0]_{+}=2p_0\openone_4$
(see~\eref{3.15}) and $\frac{\pd v^{s,\pm}(\bs p)}{\pd p^0}\equiv 0$,
 the equation~\eref{9.20b} follows
from the last equality, due to~\eref{5.20}, \eref{6.20},
$\sigma_{0a}=\iu\gamma_0\gamma_a$ for $a=1,2,3$, (see~\eref{3.13-1}
and~\eref{3.15}, and the fact that for $\mu=0$ (resp.\ $\nu=0$) the
derivatives with respect to $\mu$ (resp.\ $\nu$) identically vanish, so that,
e.g., for $\mu=0$ and $\nu\not=0$ the terms denoted by
$-(\mu\leftrightarrow\nu)$ in the last equation identically
vanish.\QED

	It is easy to be seen, the terms containing commutators with the
momentum operator in~\eref{9.18} are due to the orbital operator
(see~\eref{8.11}) and, more precisely, the operator~\eref{8.11-3} is entirely
responsible for their appearance.

	We should also mentioned the equations
	\begin{equation}	\label{9.21}
[\psi, \ope{Q}]_{\_} =  q \psi
\quad
[\opsi, \ope{Q}]_{\_} = - q \opsi ,
	\end{equation}
which are trivial corollaries from~\eref{8.1},~\eref{8.7}, the identity
\(
[A,B\circ C]_{\_}
= [A,B]_{+}\circ C - B\circ [A,C]_{+} ,
\)
and~\eref{9.13}. Similarly one can prove that
	\begin{equation}	\label{9.22}
[a_s^\pm(\bk), \ope{Q}]_{\_} =  q a_s^\pm(\bk)
\quad
[a_s^{\dag\, \pm}(\bk), \ope{Q}]_{\_} = - q a_s^{\dag\, \pm}(\bk) .
	\end{equation}
Equations~\eref{9.21} and~\eref{9.22} entail that, if $\Gamma$ is any
$4\times4$ matrix and $\ope{G}(\bk,\bk')$ (resp.\ $\tope{G}(x)$) is any
matrix\ndash valued operator build from $\bk$, $\bk'$ (resp.\ $x$) and the
derivatives with respect to them, then
($\varepsilon,\varepsilon'=+.-$)
	\begin{gather}	\label{9.23}
	\begin{split}
&
[\opsi\Gamma\psi, \ope{Q}]_{\_} = 0
\quad
[\tope{\opsi}(x)\tope{G}(x)\tope{\psi}(x), \ope{Q}]_{\_} = 0
\\
&
[a_s^{\dag\, \varepsilon}(\bk) \ope{G}(\bk,\bk') \circ
	a_{s'}^{\varepsilon'}(\bk'), \ope{Q}]_{\_} = 0
\quad
[a_{s}^{\varepsilon}(\bk) \ope{G}(\bk,\bk') \circ
       a_{s'}^{\dag\, \varepsilon'}(\bk') , \ope{Q}]_{\_} = 0 .
	\end{split}
\intertext{In particular, we have}
			\label{9.24}
	\begin{split}
&
[\ope{P}_\mu, \ope{Q}]_{\_} = 0 \quad
[\ope{Q}, \ope{Q}]_{\_} = 0
\\
&
[\lindex[\mspace{-6mu}{\ope{S}_{\mu\nu}}]{}{0}  , \ope{Q}]_{\_}
=
[\lindex[\mspace{-6mu}{\ope{L}_{\mu\nu}}]{}{0}  , \ope{Q}]_{\_}
=
[\ope{M}_{\mu\nu}, \ope{Q}]_{\_} = 0 ,
	\end{split}
	\end{gather}
as a result of~\eref{7.5},~\eref{7.8},~\eref{7.11-3}, \eref{7.20},
\eref{7.23-2}, and~\eref{8.6}--\eref{8.8}.

	Ending this section, we would like to mention the relation
	\begin{equation}	\label{9.25}
[ \ope{M}_{\varkappa\lambda} , \ope{M}_{\mu\nu} ]_{\_}
=
- \ih \bigl\{
\eta_{\varkappa\mu}	\ope{M}_{\lambda\nu} -
\eta_{\lambda\mu}	\ope{M}_{\varkappa\nu} -
\eta_{\varkappa\nu}	\ope{M}_{\lambda\mu} +
\eta_{\lambda\nu}	\ope{M}_{\varkappa\mu}
\bigr\} ,
	\end{equation}
which can be proved via a long and tedious direct calculation based
on~\eref{8.7}, \eref{7.11-3}, \eref{8.8}, \eref{8.11} and~\eref{9.13}.%
\footnote{~%
The proof in Heisenberg picture is more simple, but also too long.%
}
It should be noted the opposite sign of the r.h.s.\ of~\eref{9.25} relative
to similar relations in the literature; see,
e.g.,~\cite[eq.~(9.1.15)]{Akhiezer&Berestetskii}
or~\cite[eq.~(2.84)]{Roman-QFT}. This sign, as well as the whole structure
of~\eref{9.25}, is in agreement with the first~4 terms, proportional to
$-2\iu$, in~\eref{9.16new} (see also~\eref{9.15} or similar result obtained
in~\cite{bp-QFTinMP-scalars} for a free scalar field).

	An alternative prove of~\eref{9.25} is based on~\eref{9.12}, which is
equivalent to ($k_0=\sqrt{m^2c^2+\bk^2}$)
	\begin{equation}	\label{9.26}
	\begin{split}
[ \tope{\psi}^\pm(\bk) , \tope{M}_{\mu\nu} ]_{\_}
& =
\ih \Bigl(
k_\mu \frac{\pd}{\pd k^\nu} -  k_\nu \frac{\pd}{\pd k^\mu}
\Bigr) \tope{\psi}^\pm(\bk)
+
\frac{1}{2} \ih \sigma_{\mu\nu} \tope{\psi}^\pm(\bk)
\\
[ \tope{\opsi}_\lambda^{\dag\,\pm}(\bk) , \tope{M}_{\mu\nu} ]_{\_}
& =
\ih \Bigl(
k_\mu \frac{\pd}{\pd k^\nu} -  k_\nu \frac{\pd}{\pd k^\mu}
\Bigr) \tope{\opsi}^{\pm}(\bk)
-
\frac{1}{2} \ih \tope{\opsi}^{\pm}(\bk) \sigma_{\mu\nu} ,
	\end{split}
	\end{equation}
where
\(
\tope{\psi}^\pm(\bk)
= \e^{\pm\frac{1}{\ih} x_0^\mu k_\mu} \ope{\psi}^\pm(\bk)
\)
and
\(
\tope{\opsi}^\pm(\bk)
= \e^{\pm\frac{1}{\ih} x_0^\mu k_\mu} \ope{\opsi}^\pm(\bk)
\)
are the Heisenberg analogues of $\ope{\psi}^\pm(\bk)$ and
$\ope{\opsi}^\pm(\bk)$, respectively, (see~\eref{7.25}).
Combining~\eref{9.26} with~\eref{7.9}--\eref{7.12-5} and applying the
identity $[A,B\circ C]_{\_} = [A,B]_{\_}\circ C + B\circ[A,C]_{\_}$, one can
easily verify~\eref{9.25} without invoking~\eref{9.13}. This quite more
simple derivation of~\eref{9.25} is remarkable with the fact that if one
imposes~\eref{9.19} (or~\eref{9.18} in momentum picture) as a subsidiary
restriction on the Lagrangian formalism, it immediately implies~\eref{9.25}
for a spin $\frac{1}{2}$ field regardless of the validity of the
anticommutation relations~\eref{9.13}. Similarly,~\eref{9.21}
entails~\eref{9.24} irrespectively of the validity of~\eref{9.13}.


\section {Vacuum and normal ordering}
\label{Sect10}

	The introduction of the vacuum (state) of a free scalar fields was
discussed in momentum picture in~\cite{bp-QFTinMP-scalars}.%
\footnote{~%
For similar problems in Heisenberg picture,
see~\cite{Bogolyubov&Shirkov,Bjorken&Drell,Roman-QFT}.%
}
Here is a brief \emph{mutatis mutandis} summary of the arguments leading to a
correct definition of the vacuum of free spinor field.

	The vacuum of a field is a particular its state which describes, in a
sense, the `absence' of the field itself. Since the field is considered as a
collection of particles, the vacuum should contain no particles of the field.
Therefore the conserved dynamical characteristics of the vacuum should vanish
as an `absent' particle has zero 4\ndash momentum, no charge, etc. Besides,
since the vacuum does not contains any particles, the action of an
annihilation operator on it should produce the zero state vector as one
cannot destroy something that does not exist. On the contrary, the action of
a creation operator on the vacuum should produce a non\ndash zero vector
describing a state with one particle in it and, consequently, the vacuum
cannot be represented by the zero vector. Applying these heuristic ideas to a
free spinor field, we may say that its vacuum should be represented by a
non\ndash vanishing state vector which has zero 4\ndash momentum, charge and
total angular momentum and the action of an annihilation operator on it
results in the zero vector. So, denoting by $\ope{X}_0$ the state vector
representing the vacuum, which is also called the vacuum, we should have
	\begin{equation}	\label{10.1}
	\begin{split}
& \ope{P}_\mu(\ope{X}_0) =0 \quad
  \ope{Q}(\ope{X}_0) =0 \quad
  \ope{M}_{\mu\nu}(\ope{X}_0) =0
\\ &
\ope{X}_0 \not= 0, \quad
a_s^{-}(\bk)(\ope{X}_0) = a_s^{\dag\,-}(\bk)(\ope{X}_0) =0.
	\end{split}
	\end{equation}
Besides, by virtue of $\ope{P}_\mu(\ope{X}_0) =0$,~\eref{12.112}
and~\eref{12.113}, one can expect the vacua in Heisenberg and momentum
picture to coincide, \ie
	\begin{equation}	\label{10.2}
\tope{\ope{X}}_0 =\ope{X}_0 .
	\end{equation}
However, one can easily see that the conditions~\eref{10.1} do not agree with
the expressions~\eref{7.5}, \eref{7.8}, \eref{7.11-3}, and~\eref{7.20} for
the conserved quantities of a spinor field. Indeed, substituting in them
(see~\eref{9.13})
\(
a_s^{\dag\,-}(\bk)\circ a_s^{+}(\bk)
=
- a_s^{+}(\bk) \circ a_s^{\dag\,-}(\bk)
+ \delta_{ss} \delta^3(\bk-\bk)
\)
and applying the so\ndash obtained operators on $\ope{X}_0$, we, in view
of~\eref{10.1}, get senseless combinations of infinities; e.g., for the
charge operator the result is
 $\ope{Q}(\ope{X}_0) \sim q \delta^3(\bs0) \infty^3 \times \ope{X}_0$
instead of the expected  $\ope{Q}(\ope{X}_0)=0$. The problem originates from
the terms $a_s^{\dag\,-}(\bk)\circ a_s^{+}(\bk)$ in the obtained expressions
for the dynamical variables. The accepted and well working procedure for its
removal is known as \emph{normal ordering of products (compositions)} of
creation and/or annihilation operators. It is described at length in the
literature~\cite{Bogolyubov&Shirkov,Bjorken&Drell,Roman-QFT,Wick} and
consists in the following, when applied to a free spin $\frac{1}{2}$ field.
The Lagrangian and dynamical variables should be written in terms of creation
and annihilation operators and, then, any composition (product) of such
operator must be replaced by its normally ordered form. By definition, the
\emph{normal form} of a composition of creation and/or annihilation operators
is called a composition of the same operators, in which all creation operators
stand to the \emph{left} relative to all annihilation operators, multiplied
by minus one or plus one, depending on is the permutation, bringing the
operators from the initial composition to the final one, odd or even,
respectively.%
\footnote{~%
The so-formulated definition holds only for half integer spin
fields/particles. The relative order of the creation/annihilation operators
with respect to each other is insignificant due to the anticommutation
relations~\eref{9.13}.%
}
The normal form of a composition of creation and/or annihilation operators is
known as their \emph{normal product (composition)} and the mapping assigning
to a product of such operators their normal product will be denoted by
$\ope{N}$ and it is called \emph{normal ordering (operator)}. The action of
$\ope{N}$ on polynomials or convergent power series  in creation and/or
annihilation operators is extended by linearity. The dynamical variables after
normal ordering are denoted by the same symbols as before this operation.

	Since, obviously,
	\begin{equation}	\label{10.3}
	\begin{split}
    \ope{N} \bigl( a_s^{-}(\bk)\circ a_t^{\dag\,+}(\bk) \bigr)
= - \ope{N} \bigl( a_t^{\dag\,+}(\bk)\circ a_s^{-}(\bk) \bigr)
= - a_t^{\dag\,+}(\bk)\circ a_s^{-}(\bk)
\\
    \ope{N} \bigl( a_s^{\dag\,-}(\bk)\circ a_t^{+}(\bk) \bigr)
= - \ope{N} \bigl( a_t^{+}(\bk)\circ a_s^{\dag\,-}(\bk) \bigr)
= - a_t^{+}(\bk)\circ a_s^{\dag\,-}(\bk) ,
	\end{split}
	\end{equation}
equations~\eref{7.5},~ \eref{7.8},~ \eref{7.11-3},~ \eref{7.13},
\eref{7.13new} and~\eref{8.11}, after normal ordering, take respectively the
forms:
	\begin{gather}	\label{10.4}
\ope{P}_\mu
=
\sum_{s}\int
  k_\mu |_{ k_0=\sqrt{m^2c^2+{\bs k}^2} }
\{
a_s^{\dag\,+}(\bk)\circ a_s^-(\bk)
+
a_s^+(\bk)\circ a_s^{\dag\,-}(\bk)
\}
\Id^3\bk
\displaybreak[1]\\			\label{10.5}
\ope{Q}
=
q \sum_{s}\int
\{
a_s^{\dag\,+}(\bk)\circ a_s^-(\bk) - a_s^+(\bk)\circ a_s^{\dag\,-}(\bk)
\} \Id^3\bk
\displaybreak[1]\\			\label{10.5new}
\lindex[\mspace{-6mu}{\ope{S}_{\mu\nu}}]{}{0}
=
  \frac{1}{2} \hbar
\sum_{s,s'} \!\int\!\!\Id^3\bk
\bigl\{
  \sigma_{\mu\nu}^{s s',-}(\bk) a_{s}^{\dag\,+}(\bk)\circ a_{s'}^{-}(\bk)
- \sigma_{\mu\nu}^{s s',+}(\bk) a_{s'}^{+}(\bk)\circ a_{s}^{\dag\,-}(\bk)
\bigr\}
\displaybreak[1]\\			\label{10.6}
	\begin{split}
\lindex[\mspace{-6mu}{\ope{S}^3}]{}{0} \big|_{m\not=0}
=
S^3(\bk) \big|_{m\not=0}
& =
\frac{1}{2} \hbar \bigl\{
- a_{1}^{\dag\,+}(\bk) \circ a_{1}^{-}(\bk)
+ a_{2}^{\dag\,+}(\bk) \circ a_{2}^{-}(\bk)
\\ & \hphantom{\frac{1}{2} \hbar \bigl\{}
+ a_{1}^{+}(\bk) \circ a_{1}^{\dag\,-}(\bk)
- a_{2}^{+}(\bk) \circ a_{2}^{\dag\,-}(\bk)
\bigr\}
	\end{split}
\displaybreak[1]\\			\label{10.6new}
\lindex[\mspace{-6mu}{\ope{S}^3}]{}{0}  \big|_{m=0}
=
S^3(\bk) \big|_{m=0}
 =
\frac{1}{2} \hbar \bigl\{
- a_{0}^{\dag\,+}(\bk) \circ a_{0}^{-}(\bk)
+ a_{0}^{+}(\bk) \circ  a_{0}^{\dag\,-}(\bk)
\bigr\}
	\end{gather}
\vspace{-3ex}
	\begin{multline}	\label{10.6new1}
\lindex[\mspace{-6mu}{\ope{L}_{\mu\nu}}]{}{0}
=
\sum_{s}\int \Id^3\bk
 ( x_{\mu}k_\nu - x_{\nu}k_\mu ) |_{ k_0=\sqrt{m^2c^2+{\bs k}^2} }
\{
a_s^{\dag\,+}(\bk)\circ a_s^-(\bk)
+
a_s^+(\bk) \circ a_s^{\dag\,-}(\bk)
\}
\\
+ \frac{1}{2} \hbar \sum_{s,s'}\int \Id^3\bk
\bigl\{
  l_{\mu\nu}^{ss',-}(\bk) a_s^{\dag\,+}(\bk) \circ a_{s'}^-(\bk)
- l_{\mu\nu}^{ss',+}(\bk) a_{s'}^+(\bk) \circ a_s^{\dag\,-}(\bk)
\bigr\}
\displaybreak[2]\\
+ \frac{1}{2} \ih \sum_{s}\int \Id^3\bk
\Bigl\{
a_s^{\dag\,+}(\bk)
\Bigl( \xlrarrow{ k_\mu \frac{\pd}{\pd k^\nu} }
     - \xlrarrow{ k_\nu \frac{\pd}{\pd k^\mu} } \Bigr)
\circ a_s^-(\bk)
\\ +
a_s^+(\bk)
\Bigl( \xlrarrow{ k_\mu \frac{\pd}{\pd k^\nu} }
     - \xlrarrow{ k_\nu \frac{\pd}{\pd k^\mu} } \Bigr)
\circ a_s^{\dag\,-}(\bk)
\Bigr\} \Big|_{ k_0=\sqrt{m^2c^2+{\bs k}^2} } \ .
	\end{multline}

	Accepting the above-described normal ordering procedure, we can
formalize the definition of the vacuum of a free spinor field as follows.

	\begin{Defn}	\label{Defn10.1}
The \emph{vacuum} of a free spinor field $\psi$ is its physical state that
contains no particles and has vanishing 4\ndash momentum, (total) charge and
(total) angular momentum. It is described by a state vector, denoted by
$\ope{X}_0$ (in momentum picture) and called also the vacuum (of the field),
such that:
	\begin{subequations}	\label{10.7}
	\begin{align}
				\label{10.7a}
& \ope{X}_0 \not= 0
\\				\label{10.7b}
& \ope{X}_0 = \tope{X}_0
\\				\label{10.7c}
& a_s^-(\bs k) (\ope{X}_0)
= a_s^{\dag\,-}(\bs k) (\ope{X}_0) = 0
\\				\label{10.7d}
& \langle\ope{X}_0|\ope{X}_0\rangle = 1
	\end{align}
	\end{subequations}
where $\langle\cdot | \cdot\rangle\colon\Hil\times\Hil\to\field[C]$ is the
(Hermitian) scalar product of system's (field's) Hilbert space of states.
	\end{Defn}

	As we noted in~\cite{bp-QFTinMP-scalars}, the condition~\eref{10.7d}
is of technical character and, usually, is imposed for pure computational
convenience.

	Taking into account the expressions of the dynamical variables after
normal ordering, we see that (cf.~\eref{10.1})
	\begin{equation}	\label{10.8}
\ope{P}_\mu(\ope{X}_0) =0 \quad
\ope{Q}(\ope{X}_0) =0 \quad
\ope{M}_{\mu\nu}(\ope{X}_0) =
\lindex[\mspace{-6mu}{\ope{L}_{\mu\nu}}]{}{0} (\ope{X}_0) =
\lindex[\mspace{-6mu}{\ope{S}_{\mu\nu}}]{}{0} (\ope{X}_0) = 0
	\end{equation}
which equalities solve the problem with the senseless eigenvalues of the
conserved quantities, corresponding to the vacuum before normal ordering.

	The normal form of the dynamical variables solves also two other
problems we premeditated did not mentioned earlier.

	The first problem is connected with the positivity of the energy
operator which is identified, up to a constant, with the zeroth component of
the momentum operator, \viz
	\begin{equation}	\label{10.9}
\ope{E} := c \ope{P}_0 .
	\end{equation}
Before a normal ordering, according to~\eref{7.5}, it is
	\begin{equation}	\label{10.10}
\ope{E}
=
\sum_{s}\int
\sqrt{m^2c^2+{\bs k}^2}
\{ a_s^{\dag\,+}(\bk)\circ a_s^-(\bk) - a_s^{\dag\,-}(\bk)\circ a_s^+(\bk)
\}
\Id^3\bk
	\end{equation}
which is not positive defined. After normal ordering, in view
of~\eref{10.4}, it takes the form
	\begin{equation}	\label{10.11}
\ope{E}
=
\sum_{s}\int
\sqrt{m^2c^2+{\bs k}^2}
\{ a_s^{\dag\,+}(\bk)\circ a_s^-(\bk) + a_s^+(\bk) \circ a_s^{\dag\,-}(\bk)
\}
\Id^3\bk
	\end{equation}
which is a Hermitian operator with positive eigenvalues.

	The second problem concerns the interpretation of the operators
 $a_s^{\pm}(\bk)$ and $a_s^{\dag\,\pm}(\bk)$
as creation/annihilation operators (see Sect.~\ref{Sect6}). Since the field
can be thought as a collection of particles, one can expect that the
dynamical variables should be expressible as  sums/integrals of the
corresponding individual characteristics of these particles; this is
rigorously expressed via results like~\eref{7.4}--\eref{7.12}
and~\eref{7.13}--\eref{7.20}. However, if we want to retain the
interpretation of
 $a_s^{\pm}(\bk)$ and $a_s^{\dag\,\pm}(\bk)$,
introduced in Sect.~\ref{Sect6}, the signs before the second terms in the
braces in~\eref{7.5} and~\eref{7.8} should be opposite. The reason being
quite simple: the 4\ndash momentum of a system of two particles, one created
by $a_s^{+}(\bk)$ and another one by $a_s^{\dag\,+}(\bk)$, should be a
sum of the 4\ndash momenta of these particles, while the charge of this
system should be the difference of the charges of the particles, measured in
the units $q$.%
\footnote{~%
Recall, $a_s^{+}(\bk)$ produces particles with charge $-q$, while
$a_s^{\dag\,+}(\bk)$ produces ones with charge $+q$. The 4\ndash momentum of
the both kind of particles is $(\sqrt{m^2c^2+{\bs k}^2},\bk)$.%
}
As we see from~\eref{10.4} and~\eref{10.5}, this problem automatically
vanishes after normal ordering.

	The equality~\eref{10.6} partially changes the interpretation of the
creation and annihilation operators with respect to the polarization (spin)
index $s$ (see~\eref{7.13} and the conclusions after it). So, in the frame in
which~\eref{10.6} is derived, the projection of the spin vector on the third
axis is $+\frac{1}{2}\hbar$ (resp.\ $-\frac{1}{2}\hbar$) for the particles
corresponding to
$a_{1}^{+}(\bk)$,     $a_{1}^{\dag\,-}(\bk)$,
$a_{2}^{-}(\bk)$, and $a_{2}^{\dag\,+}(\bk)$
(resp.\
$a_{1}^{-}(\bk)$,     $a_{1}^{\dag\,+}(\bk)$,
$a_{2}^{+}(\bk)$, and $a_{2}^{\dag\,-}(\bk)$).

	The general formulae~\eref{10.5new}, with $\mu\not=0$ or $\nu\not=0$,
and~\eref{10.6new1} for the spin and orbital operators agree
with~\eref{6.22c} and the interpretation of $a_{s}^{\pm}$ and
$a_{s}^{\dag\,\pm}$ given in Sect.~\ref{Sect6}. This problem will be
discussed and solved in Sect.~\ref{Sect12}; in particular, see
equation~\eref{11.2-2} below.

	The normal ordering changes not only the dynamical variables, but
also the field equations~\eref{8.4}. Since the quadratic combinations of
creation and annihilation operators in the commutators in~\eref{8.4}
originate from the momentum operator (see~\eref{8.2}), the field
equations~\eref{8.4}, after normal ordering, will read
	\begin{subequations}	\label{10.12}
	\begin{align}
			\label{10.12a}
	\begin{split}
\bigl[ a_s^{\pm}(\bk) ,
a_t^{\dag\,+}(\bs q) \circ a_t^{-}(\bs q)
 +
a_t^{+}(\bs q) \circ  a_t^{\dag\,-}(\bs q)
\bigr]_{\_}
	 \pm a_s^{\pm}(\bk) \delta_{ts} \delta^3(\bk-\bs q)
= f_{st}^\pm(\bk,\bs q)
	\end{split}
\\			\label{10.12b}
	\begin{split}
\bigl[ a_s^{\dag\,\pm}(\bk) ,
a_t^{\dag\,+}(\bs q) \circ a_t^{-}(\bs q)
 +
a_t^{+}(\bs q) \circ  a_t^{\dag\,-}(\bs q)
\bigr]_{\_}
	\pm  a_s^{\dag\,\pm}(\bk) \delta_{ts} \delta^3(\bk-\bs q)
= f_{st}^{\dag\,\pm}(\bk,\bs q)
	\end{split}
	\end{align}
\vspace{-3ex}
	\begin{equation}	\label{10.12c}
	\begin{split}
\sum_{t} \int q_\mu\big|_{q_0=\sqrt{m^2c^2+{\bs q}^2}}
f_{st}^{\pm}(\bk,\bs q) \Id^3\bs{q}
 = 0
\qquad
\sum_{t} \int q_\mu\big|_{q_0=\sqrt{m^2c^2+{\bs q}^2}}
f_{st}^{\dag\,\pm}(\bk,\bs q) \Id^3\bs{q}
 = 0 .
	\end{split}
	\end{equation}
	\end{subequations}
However, applying~\eref{9.1} with $\varepsilon=+1$, one can verify
that~\eref{10.12} is identically valid as a result of the anticommutation
relations~\eref{9.13}. This means that, under the hypotheses made when
deriving~\eref{9.13}, the anticommutation relations~\eref{9.13} play a role
of field equations with respect to the creation and annihilation operators,
considered as field operators (variables).

	The normal ordering influences the r.h.s.\ of~\eref{9.15} too. In
fact, applying~\eref{10.5new} and~\eref{9.14}, we, via a direct calculation,
see that
	\begin{multline}	\label{10.13}
[ \lindex[\mspace{-6mu}{\ope{S}_{\mu\nu}}]{}{0} ,
  \lindex[\mspace{-6mu}{\ope{S}_{\varkappa\lambda}}]{}{0} ]_{\_}
 =
\frac{1}{4}\hbar^2
\sum_{s,s',t} \int\Id^3\bk
\bigl\{
\bigl( \sigma_{\mu\nu}^{ss',-}(\bk) \sigma_{\varkappa\lambda}^{s't,-}(\bk)
     - \sigma_{\varkappa\lambda}^{ss',-}(\bk) \sigma_{\mu\nu}^{s't,-}(\bk)
\bigr) a_{s}^{\dag\,+}(\bk) \circ a_{t}^{-}(\bk)
\\ -
\bigl( \sigma_{\mu\nu}^{ss',+}(\bk) \sigma_{\varkappa\lambda}^{s't,+}(\bk)
     - \sigma_{\varkappa\lambda}^{ss',+}(\bk) \sigma_{\mu\nu}^{s't,+}(\bk)
\bigr) a_{t}^{+}(\bk) \circ  a_{s}^{\dag\,-}(\bk)
\bigr\}  ,
	\end{multline}
which agrees with~\eref{10.3}. Therefore, in the massless case, we have
	\begin{equation}	\label{10.14}
[ \lindex[\mspace{-6mu}{\ope{S}_{\mu\nu}}]{}{0} ,
  \lindex[\mspace{-6mu}{\ope{S}_{\varkappa\lambda}}]{}{0} ]_{\_} \big|_{m=0}
 = 0
	\end{equation}
as in it $s=s'=t=0$, \ie the relation~\eref{9.16} is preserved after normal
ordering of products. Notice, the equality
	\begin{equation}	\label{10.15}
[ \lindex[\mspace{-6mu}{\ope{S}_{\mu\nu}}]{}{0} ,
  \lindex[\mspace{-6mu}{\ope{S}_{\varkappa\lambda}}]{}{0} ]_{\_}
 = 0
\quad \text{for }
\delta^{0\mu} + \delta^{0\nu} + \delta^{0\varkappa} + \delta^{0\lambda}  \ge 1
	\end{equation}
is an evident corollary from~\eref{10.5new} and~\eref{7.12-10}.

	However, the commutativity between the charge and spin operators,
expressed by~\eref{9.17}, is not influenced by the normal ordering procedure
(see~\eref{10.5}, \eref{10.5new} and~\eref{9.14}). Similarly, the
relations~\eref{9.18}--\eref{9.21} remain valid after normal ordering.


\section {State vectors}
\label{Sect11}

	The description of the state vectors of a free spinor field is almost
identical with the one of state vectors of free charged (with non\ndash zero
charge) scalar field~\cite{bp-QFTinMP-scalars}. Formally, the only essential
difference is in the polarization index $s$ carried by the creation and
annihilation operators of a spinor field.

	In momentum picture, in accord with the general theory of
Sect.~\ref{Sect2}, the state vectors of a spinor field are spacetime\ndash
dependent, contrary to the field operators and dynamical variables
constructed from them. In view of~\eref{12.118}, the spacetime\ndash
dependence of a state vector $\ope{X}(x)$ is
	\begin{equation}	\label{11.1}
\ope{X}(x) = \ope{U}(x,x_0) (\ope{X}(x_0))
	\end{equation}
where $x_0$ is an arbitrarily fixed spacetime point and the \emph{evolution
operator} $\ope{U}(x,x_0)\colon\Hil\to\Hil$ is
	\begin{equation}	\label{11.2}
	\begin{split}
\ope{U}(x,x_0)
=
\exp
	\Bigl\{
\iih (x^\mu-x_0^\mu)
\! \sum_{s} \! \int \!
k_\mu |_{ k_0=\sqrt{m^2c^2+{\bs k}^2} }
\{
a_s^{\dag\,+}(\bs k)\circ a_s^{-}(\bs k)
 +
a_s^{+}(\bs k) \circ a_s^{\dag\,-}(\bs k)
\}
\Id^3\bs k
\Bigr\}  .
	\end{split}
	\end{equation}
due to~\eref{12.112} and~\eref{10.4} (see also~\eref{12.116}--\eref{12.118}).
The operator~\eref{11.2} plays also a role of an `$S$\ndash matrix'
determining the transition amplitudes between any initial and final states,
say $\ope{X}_i(x_i)$ and $\ope{X}_f(x_f)$ respectively. In fact, we have
	\begin{equation}	\label{11.3}
S_{fi}(x_f,x_i) := \langle\ope{X}_f(x_f) | \ope{X}_i(x_i)\rangle
=
\langle\ope{X}_f(x_f^{(0)}) | \ope{U}(x_i,x_f)
	(\ope{X}_i(x_i^{(0)}))\rangle .
	\end{equation}
For some purposes, the following expansion of $\ope{U}(x_i,x_f)$ into a power
series may turn to be useful:
	\begin{equation}	\label{11.4}
\ope{U}(x_i,x_f) = \id_\Hil + \sum_{n=1}^{\infty} \ope{U}^{(n)}(x_i,x_f)
	\end{equation}
\vspace{-2.4ex}
	\begin{multline}	\label{11.5}
\ope{U}^{(n)}(x_i,x_f)
:=
\frac{1}{n!} (x_i^{\mu_1}-x_f^{\mu_1}) \dots (x_i^{\mu_n}-x_f^{\mu_n})
\sum_{s_1,\dots,s_n}
	    \int \Id^3\bs k^{(1)}\dots\Id^3\bs k^{(n)}
k_{\mu_1}^{(1)}
\dotsb k_{\mu_n}^{(n)}
\\ \times
\bigl\{  a_{s_1}^{\dag\,+}(\bs k^{(1)})\circ a_{s_1}^-(\bs k^{(1)})
+
 a_{s_1}^+(\bs k^{(1)})\circ a_{s_1}^{\dag\,-}(\bs k^{(1)}) \bigr\}
\\
\circ\dotsb\circ
\bigl\{  a_{s_n}^{\dag\,+}(\bs k^{(n)})\circ a_{s_n}^-(\bs k^{(n)})
+
 a_{s_n}^+(\bs k^{(n)})\circ a_{s_n}^{\dag\,-}(\bs k^{(n)}) \bigr\}
	\end{multline}
where $k_0^{(a)}=\sqrt{m^2c^2+ ({\bs k}^{(a)})^2}$, $a=1,\dots,n$.

	According to~\eref{12.120} and the considerations in
Sect.~\ref{Sect6}, a state vector of a state containing $n'$
particles and $n^{\prime\prime}$ antiparticles,
$n^{\prime},n^{\prime\prime}\ge0$, such that
the $i^{\prime\,\text{th}}$ particle has 4\ndash momentum
$p'_{i'}$ and polarization $s'_{i'}$ and
the $i^{\prime\prime\,\text{th}}$ antiparticle has 4\ndash momentum
$p^{\prime\prime}_{i^{\prime\prime}}$ and polarization
$s^{\prime\prime}_{i^{\prime\prime}}$, where $i'=0,1,\dots,n'$ and
$i^{\prime\prime}=0,1,\dots,n^{\prime\prime}$, is given by the equality
	\begin{multline}	\label{11.6}
\ope{X}(
x;p'_1,s'_1;\ldots;p'_{n'}, s'_{n'};
  p^{\prime\prime}_1, s^{\prime\prime}_1;\ldots;
  p^{\prime\prime}_{n^{\prime\prime}}, s^{\prime\prime}_{n^{\prime\prime}}
)
\\
=
\frac{1}{\sqrt{n^{\prime}! n^{\prime\prime}!}}
\exp\Bigl\{
\iih (x^{\mu} - x_0^\mu) \sum_{i'=1}^{n'} (p'_{i'})_\mu
+
\iih (x^{\mu} - x_0^\mu) \sum_{i^{\prime\prime}=1}^{n^{\prime\prime}}
			 (p^{\prime\prime}_{i^{\prime\prime}})_\mu
\Bigr\}
\\ \times
\bigl(
 a_{s'_1}^+(\bs p'_1)\circ\dots\circ a_{s'_{n'}}^+(\bs p'_{n'})
\circ
 a_{s^{\prime\prime}_1}^{\dag\,+}(\bs p^{\prime\prime}_1)\circ\dots\circ
	a_{s^{\prime\prime}_{n^{\prime\prime}}}^{\dag\,+}
		(\bs p^{\prime\prime}_{n^{\prime\prime}})
\bigr)
(\ope{X}_0) ,
	\end{multline}
where, in view of the anticommutation relations~\eref{9.13}, the order  of
the creation operators is essential. Besides, the vector~\eref{11.6}
vanishes if two of the particles of the state it describes are identical,
\ie, \eg for a charged field, if either
$n^{\prime}\ge2$ and two of the pairs
\(
(p^{\prime}_{1},s^{\prime}_{1}), \dots,
(p^{\prime}_{n^{\prime}},s^{\prime}_{n^{\prime}})
\)
coincide or
$n^{\prime\prime}\ge2$ and two of the pairs
\(
(p^{\prime\prime}_{1},s^{\prime\prime}_{1}), \dots,
(p^{\prime\prime}_{n^{\prime\prime}},s^{\prime\prime}_{n^{\prime\prime}})
\)
coincide; to prove this, apply~\eref{9.13}.%
\footnote{~%
This is a demonstration of the so\ndash called Pauli principle: no more than
one particle can be in a given state of a system consisting of fermions.%
}
If $n'=0$ (resp.\ $n^{\prime\prime}=0$), the particle (resp.\ antiparticle)
creation operators and the first (resp.\ second) sum in the exponent should
be absent. In particular, the vacuum corresponds to~\eref{11.6} with
$n'=n^{\prime\prime}=0$. The state vector~\eref{11.6} is
	an eigenvector of the momentum operator~\eref{10.4}  with
eigenvalue (4\ndash momentum)
\(
\sum_{i'=1}^{n'} p'_{i'}
+
\sum_{i^{\prime\prime}=1}^{n^{\prime\prime}}
	p^{\prime\prime}_{i^{\prime\prime}}
\)
	and is also an eigenvector of the charge operator~\eref{10.5} with
eigenvalue  $(-q)(n'-n^{\prime\prime})$.%
\footnote{~%
Recall (see Sect.~\ref{Sect6}), the operator $ a_{s}^{+}(\bs k)$
creates a particle with 4\ndash momentum $k_\mu$ and charge $-q$, while
$ a_{s}^{\dag\,+}(\bs k)$ creates a particle with 4\ndash momentum $k_\mu$
and charge $+q$, where, in the both cases, $k_0=\sqrt{m^2c^2+{\bs k}^2}$. See
also equations~\eref{11.2-1}--\eref{11.2-3} below.%
}

	The reader may verify, using~\eref{9.13} and~\eref{6.21}, that
the transition amplitude between two states, like~\eref{11.6}, is:
	\begin{multline}	\label{11.7}
\langle
\ope{X}(
y;q'_1, t'_1;\ldots;q'_{n'}, t'_{n'};
  q^{\prime\prime}_1, t^{\prime\prime}_1;\ldots;
  q^{\prime\prime}_{n^{\prime\prime}}, t^{\prime\prime}_{n^{\prime\prime}}
)
\\
|
\ope{X}(
x;p'_1, s'_1;\ldots;p'_{m'},s'_{m'};
  p^{\prime\prime}_1, s^{\prime\prime}_1;\ldots;
  p^{\prime\prime}_{m^{\prime\prime}}, 	s^{\prime\prime}_{m^{\prime\prime}}
)
\rangle
\\
=
\frac{1}{n^{\prime}! n^{\prime\prime}!}
\delta_{m'n'} \delta_{m^{\prime\prime}n^{\prime\prime}}
\exp\Bigl\{
\iih (x^{\mu} - y^{\mu}) \sum_{i'=1}^{n'} (p'_{i'})_\mu
+
\iih (x^{\mu} - y^{\mu}) \sum_{i^{\prime\prime}=1}^{n^{\prime\prime}}
			 (p^{\prime\prime}_{i^{\prime\prime}})_\mu
\Bigr\}
\\ \times
\sum_{(i'_1,\dots,i'_{n'})}
		 \pi_{i_{1}^{\prime},\dots,i_{n^{\prime}}^{\prime}}
\delta_{s'_{n'} t'_{i'_1} }
	\delta^3(\bs p'_{n'} - \bs q'_{i'_1})
\delta_{s'_{n'-1}  t'_{i'_2} }
	\delta^3(\bs p'_{n'-1} - \bs q'_{i'_2})
\dots
\delta_{s'_{1}  t'_{i'_{n'}} }
	\delta^3(\bs p'_{1} - \bs q'_{i'_{n'}})
\\ \times
\!
\sum_{(i^{\prime\prime}_1,\dots,i_{n^{\prime\prime}}^{\prime\prime})}
	 \pi_{i_{1}^{\prime\prime},\dots,i_{n^{\prime\prime}}^{\prime\prime}}
\!
	\delta_{s^{\prime\prime}_{n^{\prime\prime}}
	  			t^{\prime\prime}_{i^{\prime\prime}_1} }
	\delta^3(\bs p^{\prime\prime}_{n^{\prime\prime}} -
		\bs q^{\prime\prime}_{i^{\prime\prime}_1})
	\delta_{s^{\prime\prime}_{n^{\prime\prime}-1}
				t^{\prime\prime}_{i^{\prime\prime}_2} }
	\delta^3(\bs p^{\prime\prime}_{n^{\prime\prime}-1} -
		\bs q^{\prime\prime}_{i^{\prime\prime}_2})
\dots
\delta_{s^{\prime\prime}_{1}
		 t^{\prime\prime}_{i^{\prime\prime}_{n^{\prime\prime}}} }
\delta^3(\bs p^{\prime\prime}_{1} -
\bs q^{\prime\prime}_{i^{\prime\prime}_{n^{\prime\prime}}})
	\end{multline}
where the summations are over all
permutations $(i'_1,\dots,i'_{n'})$ of $(1,\dots,n')$ and
$(i^{\prime\prime}_1,\dots,i_{n^{\prime\prime}}^{\prime\prime})$ of
$(1,\dots,n^{\prime\prime})$
and
$\pi_{i_{1}^{\prime},\dots,i_{n^{\prime}}^{\prime}}$
(resp.\
$\pi_{i_{1}^{\prime\prime},\dots,i_{n^{\prime\prime}}^{\prime\prime}}$)
equals to $+1$ or $-1$ depending on is the permutation
\(
(n^{\prime},\dots,1) \mapsto
( i_{1}^{\prime},\dots,i_{n^{\prime}}^{\prime} )
\)
(resp.\
\(
(n^{\prime\prime},\dots,1) \mapsto
( i_{1}^{\prime\prime},\dots,i_{n^{\prime\prime}}^{\prime\prime} )
\))
even or odd, respectively.
	The conclusions from this formula are similar to the ones concerning
free scalar fields~\cite{bp-QFTinMP-scalars}.
For instance, the only non\ndash forbidden transition from an
$n'$\ndash particle~+~$n^{\prime\prime}$\ndash antiparticle state is into
$n'$\ndash particle~+~$n^{\prime\prime}$\ndash antiparticle state; the both
states may differ only in the spacetime positions of the (anti)particles in
them. This result is quite natural as we are dealing with free
particles/fields.

	In particular, if $\ope{X}_n$ denotes any state containing $n$
particles and/or antiparticles, $n=0,1,\dots$, then~\eref{11.7} says that
	\begin{equation}	\label{11.8}
\langle \ope{X}_n|\ope{X}_0 \rangle = \delta_{n0} ,
	\end{equation}
which expresses the stability of the vacuum.

	We shall end the present section with a simple, but important,
example. Consider the one (anti)particle states
$a_{t}^{+}(\bs p)(\ope{X}_0)$ and
$a_{t}^{\dag\,+}(\bs p)(\ope{X}_0)$.
Applying~\eref{10.4}--\eref{10.6new1} and~\eref{9.13}, we find
($p_0:=\sqrt{m^2c^2+{\bs p}^2}$):%
\footnote{~%
The easiest way to derive~\eref{11.2-5} is by
applying~\eref{2.14},~\eref{7.2},~\eref{7.3} and~\eref{9.13}. Notice, in
Heisenberg picture and in terms of the Heisenberg creation/annihilation
operators~\eref{7.27}, equations~\eref{11.2-5} read
$\tope{L}_{\mu\nu}\bigl( a_{t}^{+}(\bs p)(\ope{X}_0) \bigr) = 0$ and
$\tope{L}_{\mu\nu}\bigl( a_{t}^{\dag\,+}(\bs p)(\ope{X}_0) \bigr) = 0$
which is quite understandable in view of the fact that $\tope{L}_{\mu\nu}$
is, in a sense, the average orbital momentum with respect to all spacetime
points, while $\ope{L}_{\mu\nu}(x,x_0)$ is the one relative to  $x$ and
$x_0$; the dependence on $x_0$ being hidden in $\ope{L}_{\mu\nu}$,
$a_{t}^{+}(\bs p)$ and $a_{t}^{\dag\,+}(\bs p)$.%
}
	\begin{gather}
			\label{11.2-1}
	\begin{split}
\ope{P}_\mu\bigl( a_{t}^{+}(\bs p)(\ope{X}_0) \bigr)
& =
p_\mu a_{t}^{+}(\bs p)(\ope{X}_0)
\quad \hphantom{+}
\ope{Q}\bigl( a_{t}^{+}(\bs p)(\ope{X}_0) \bigr)
=
 - q a_{t}^{+}(\bs p)(\ope{X}_0)
\\
\ope{P}_\mu\bigl( a_{t}^{\dag\,+}(\bs p)(\ope{X}_0) \bigr)
& =
p_\mu a_{t}^{\dag\,+}(\bs p)(\ope{X}_0)
\quad
\ope{Q}\bigl( a_{t}^{\dag\,+}(\bs p)(\ope{X}_0) \bigr)
=
 + q a_{t}^{\dag\,+}(\bs p)(\ope{X}_0)
	\end{split}
\displaybreak[1]\\		\label{11.2-2}
	\begin{split}
\lindex[\mspace{-6mu}{\ope{S}_{\mu\nu}}]{}{0}
	\bigl( a_{t}^{+}(\bs p)(\ope{X}_0) \bigr)
& =
- \frac{1}{2} \hbar \sum_{s}
\sigma_{\mu\nu}^{ts,+}(\bs p) a_{s}^{+}(\bs p)(\ope{X}_0)
\\
\lindex[\mspace{-6mu}{\ope{S}_{\mu\nu}}]{}{0}
	\bigl( a_{t}^{\dag\,+}(\bs p)(\ope{X}_0) \bigr)
& =
+ \frac{1}{2} \hbar \sum_{s}
\sigma_{\mu\nu}^{st,-}(\bs p) a_{s}^{\dag\,+}(\bs p)(\ope{X}_0)
	\end{split}
\displaybreak[1]\\		\label{11.2-3}
	\begin{split}
\ope{S}^3\big|_{m\not=0} \bigl( a_{t}^{+}(\bs p)(\ope{X}_0) \bigr)
& =
- (-1)^t \frac{1}{2}\hbar
\{ \delta_{t1} a_{1}^{+}(\bs p) + \delta_{t2} a_{2}^{+}(\bs p) \} (\ope{X}_0)
\\
\ope{S}^3\big|_{m\not=0} \bigl( a_{t}^{\dag\,+}(\bs p)(\ope{X}_0) \bigr)
& =
+ (-1)^t \frac{1}{2}\hbar
\{ \delta_{t1} a_{1}^{\dag\,+}(\bs p) + \delta_{t2} a_{2}^{\dag\,+}(\bs p) \}
   (\ope{X}_0)
	\end{split}
\displaybreak[1]\\		\label{11.2-4}
	\begin{split}
\ope{S}^3\big|_{m=0} \bigl( a_{0}^{+}(\bs p)(\ope{X}_0) \bigr)
& =
+ \frac{1}{2} \hbar  a_{0}^{+}(\bs p) (\ope{X}_0)
\\
\ope{S}^3\big|_{m=0} \bigl( a_{0}^{\dag\,+}(\bs p)(\ope{X}_0) \bigr)
& =
- \frac{1}{2} \hbar  a_{0}^{\dag\,+}(\bs p) (\ope{X}_0)
	\end{split}
\displaybreak[1]\\		\label{11.2-5}
	\begin{split}
\lindex[\mspace{-6mu}{\ope{L}_{\mu\nu}}]{}{0}
		\bigl( a_{t}^{+}(\bs p)(\ope{X}_0) \bigr)
& =
\Bigl\{
( x_\mu p_\nu - x_\nu p_\mu )
- \frac{1}{2} \hbar \sum_{s} l_{\mu\nu}^{ts,+}(\bs p)
-\ih \Big( p_\mu\frac{\pd}{\pd p^\nu} - p_\nu\frac{\pd}{\pd p^\mu} \Big)
\Bigr\}
\bigl( a_{t}^{+}(\bs p)(\ope{X}_0) \bigr)
\\
\lindex[\mspace{-6mu}{\ope{L}_{\mu\nu}}]{}{0}
		\bigl( a_{t}^{\dag\,+}(\bs p)(\ope{X}_0) \bigr)
&\! = \!
\Bigl\{
( x_\mu p_\nu - x_\nu p_\mu )
+ \frac{1}{2} \hbar \sum_{s} l_{\mu\nu}^{st,-}(\bs p)
-\ih \Big( p_\mu\frac{\pd}{\pd p^\nu} - p_\nu\frac{\pd}{\pd p^\mu} \Big)
\Bigr\}
\bigl( a_{t}^{\dag\,+}(\bs p)(\ope{X}_0) \bigr) ,
	\end{split}
	\end{gather}
where~\eref{11.2-3} and~\eref{11.2-4} are valid in a frame such that
$p^1=p^2=0$ and, consequently, in which~\eref{7.11-11} and~\eref{7.11-13} are
valid.
	Notice, the one\ndash(anti)par\-ticle massive states are
\emph{not}, generally, eigenstates of the spin (angular momentum) operator.
However, in a special frame in which $p^1=p^2=0$  they are eigenstates of
the third spin vector component. Conversely, the massless
one\ndash(anti)particle states are always eigenstates of the spin (angular
momentum) operator. It should be remarked the agreement
of~\eref{11.2-1}--\eref{11.2-5} with~\eref{6.22}.%
\footnote{~%
If the r.h.s.\ of~\eref{2.21} is with an opposite sign, this agreement
will be lost. Besides, due to~\eref{7.12-10}, the r.h.s.\ of~\eref{11.2-2}
vanishes for $\mu=0$ or $\nu=0$, which is not generally the case with the
spinor terms in~\eref{6.22c}.%
}

	The equations~\eref{11.2-1}--\eref{11.2-5} confirm the
interpretation, given in Sect.~\ref{Sect6}, of the operators
$a_{s}^{\pm}(\bk)$ and $a_{s}^{\dag\,\pm}(\bk)$ as ones describing
creation/annihilation of field's (anti)particles. For instance, the state
vector $a_{s}^{+}(\bk)(\ope{X}_0)$ can be interpreted as one representing a
particle with 4\ndash momentum $(\sqrt{m^2c^2+\bk^2},\bk)$, charge $(-q)$ and
polarization\ndash mixing matrices
$-\frac{1}{2}[\sigma_{\mu\nu}^{st,+}(\bk)]_{s,t=1}^{2-\delta_{0m}}$.

	Acting with~\eref{10.13} on the state vectors
$a_{t}^{+}(\bs p)(\ope{X}_0)$ and $a_{t}^{\dag\,+}(\bs p)(\ope{X}_0)$
and using~\eref{9.13}, we obtain
	\begin{subequations}	\label{11.9}
	\begin{equation} 	\label{11.9a}
[ \lindex[\mspace{-6mu}{\ope{S}_{\mu\nu}}]{}{0} ,
  \lindex[\mspace{-6mu}{\ope{S}_{\varkappa\lambda}}]{}{0} ]_{\_}
  \bigl( a_{t}^{+}(\bs p)(\ope{X}_0) \bigr)
 =
- \frac{1}{4}\hbar^2
\sum_{s',t'}
\bigl\{
  \sigma_{\mu\nu}^{ts',+}(\bs p) \sigma_{\varkappa\lambda}^{s't',+}(\bs p)
- \sigma_{\varkappa\lambda}^{ts',+}(\bs p) \sigma_{\mu\nu}^{s't',+}(\bs p)
\bigr\}  a_{t'}^{+}(\bs p)(\ope{X}_0)
	\end{equation}
\vspace{-4.3ex}
	\begin{equation} 	\label{11.9b}
[ \lindex[\mspace{-6mu}{\ope{S}_{\mu\nu}}]{}{0} ,
  \lindex[\mspace{-6mu}{\ope{S}_{\varkappa\lambda}}]{}{0} ]_{\_}
  \bigl( a_{t}^{\dag\,+}(\bs p)(\ope{X}_0) \bigr)
 =
+ \frac{1}{4}\hbar^2
\sum_{s',t'}
\bigl\{
  \sigma_{\mu\nu}^{t's',-}(\bs p) \sigma_{\varkappa\lambda}^{s't,-}(\bs p)
- \sigma_{\varkappa\lambda}^{t's',-}(\bs p) \sigma_{\mu\nu}^{s't,-}(\bs p)
\bigr\}  a_{t'}^{\dag\,+}(\bs p)(\ope{X}_0) .
	\end{equation}
	\end{subequations}
In particular, since for $m\not=0$ from~\eref{7.11-9} (or~\eref{9.16new} with
$\bk=\bs0$) follows (do not sum over $b$!)
	\begin{equation}	\label{11.10}
\sum_{s'}\bigl\{
\sigma_{ab}^{ss',\pm}(\bs 0) \sigma_{bc}^{s't,\pm}(\bs 0)
-
\sigma_{bc}^{ss',\pm}(\bs 0) \sigma_{ab}^{s't,\pm}(\bs 0)
\bigr\}
=
2\iu \sigma_{ca}^{st,\pm}(\bs 0) ,
	\end{equation}
where $(abc)$ is an even permutation of $(123)$, we see that~\eref{11.2-2}
and~\eref{11.9} imply
	\begin{equation}	\label{11.11}
[ \lindex[\mspace{-6mu}{\ope{S}_{ab}}]{}{0} ,
  \lindex[\mspace{-6mu}{\ope{S}_{bc}}]{}{0} ]_{\_}  \bigl( \ope{X} \bigr)
=
\ih
  \lindex[\mspace{-6mu}{\ope{S}_{ca}}]{}{0} \bigl( \ope{X} \bigr)
	\end{equation}
where
 $m\not=0$, $(abc)=(123),(231),(312)$, and
\(
\ope{X}=
a_{t}^{+}(\bs 0)(\ope{X}_0),
a_{t}^{\dag\,+}(\bs 0)(\ope{X}_0)
\).
The other non-vanishing components of~\eref{11.9} with $m\not=0$ and
$\bs p=\bs 0$ can be obtained from~\eref{11.11} by using the skewsymmetry of
$\lindex[\mspace{-6mu}{\ope{S}_{ab}}]{}{0}$ in the indices $a$ and $b$
(see~\eref{6.23}, \eref{7.12-10}, and~\eref{11.2-2}). Introducing the
conserved spin 3\ndash vector operator
	\begin{equation}	\label{11.12}
\lindex[\mspace{-6mu}{\ope{S}^{a}}]{}{0}
:=
\varepsilon^{abc} \lindex[\mspace{-6mu}{\ope{S}_{bc}}]{}{0},
	\end{equation}
we can rewrite~\eref{11.11} as
	\begin{equation}	\label{11.13}
[ \lindex[\mspace{-6mu}{\ope{S}^{a}}]{}{0} ,
  \lindex[\mspace{-6mu}{\ope{S}^{b}}]{}{0} ]_{\_} (\ope{X})
=
\ih \sum_{c=1}^{3}
  \varepsilon^{abc} \lindex[\mspace{-6mu}{\ope{S}^{c}}]{}{0} (\ope{X})
\qquad \text{ for } m\not=0
	\end{equation}
with $a,b=1,2,3$. Notice, in the massless case, we have
	\begin{equation}	\label{11.14}
[ \lindex[\mspace{-6mu}{\ope{S}^{a}}]{}{0} ,
  \lindex[\mspace{-6mu}{\ope{S}^{b}}]{}{0} ]_{\_} \big|_{m=0}
= 0 ,
	\end{equation}
due to~\eref{10.14}. Therefore, in the massless case all of the spin 3-vector
components are simultaneously measurable, contrary to the massive one, when
neither pair of them is simultaneously measurable (if the third one does not
vanish).


\section
[On the choice of Lagrangian and its consequences]
{On the choice of Lagrangian and its consequences}
\label{Sect12}

	The developed until now theory of a free spinor field (in momentum
picture) is based on the Lagrangian
	\begin{equation}	\label{12.1}
\tope{L}
=
\frac{1}{2}\ih c\{
  \tope{\opsi}(x)\gamma^\mu\circ(\pd_\mu\tope{\psi}(x))
- (\pd_\mu\tope{\opsi}(x))\gamma^\mu\circ \tope{\psi}(x)
\}
- mc^2 \tope{\opsi}(x)\circ \tope{\psi}(x) .
	\end{equation}
in Heisenberg picture, or
	\begin{equation}	\label{12.2}
\ope{L} = \tope{L}(\psi,\opsi,y_\mu,\overline{y}_\mu)
=
\frac{1}{2} c \{
\opsi\gamma^\mu\circ[\psi,\ope{P}_\mu]_{\_}
- [\opsi,\ope{P}_\mu]_{\_}\gamma^\mu\circ \psi
\}
-m c^2\opsi\circ\psi
	\end{equation}
in momentum picture. In this Lagrangian, the field $\psi$ and its Dirac
conjugate $\opsi$ are considered as independent variables. However, that
choice of a Lagrangian for description of a free spinor field is not the only
possible one. For instance, in~\cite[\S~78]{Bjorken&Drell-2} it is chosen as%
\footnote{~%
In~\cite{Bjorken&Drell-2} as an independent variable is chosen $\psi(x)$, not
$\psi(x)$ and $\opsi(x)$.%
}
	\begin{align}	\label{12.3}
\tope{L}_0
& =
\ih c
  \tope{\opsi}(x)\gamma^\mu\circ(\pd_\mu\tope{\psi}(x))
- mc^2 \tope{\opsi}(x)\circ \tope{\psi}(x)
\\			\label{12.4}
\ope{L}_0
& =
c
\opsi\gamma^\mu\circ[\psi,\ope{P}_\mu]_{\_}
-m c^2\opsi\circ\psi
	\end{align}
in respectively Heisenberg and momentum pictures. Since
	\begin{equation}	\label{12.6}
	\begin{split}
\frac{\pd\ope{L}_0}{\pd\psi} = \frac{\pd\tope{L}_0}{\pd\psi}
& =
 - m c^2\opsi
\qquad\qquad\qquad
\frac{\pd\ope{L}_0}{\pd y_\mu} = + \ih c \opsi \gamma^\mu
\\
\frac{\pd\ope{L}_0}{\pd\opsi} = \frac{\pd\tope{L}_0}{\pd\opsi}
& =
c \gamma^\mu [\psi,\ope{P}_\mu]_{\_} - m c^2\psi
\qquad
\frac{\pd\ope{L}_0}{\pd \overline{y}_\mu}
 = 0,
	\end{split}
	\end{equation}
the equations of motion~\eref{12.129} for the Lagrangian~\eref{12.4} coincide
with the system of Dirac equations~\eref{3.5} for the Lagrangian~\eref{12.2}.
As $\tope{L}$ and $\tope{L}_0$ differ by a full 4\ndash divergence,
\(
\tope{L}_0-\tope{L}
=
\pd_\mu\{ \frac{1}{2}\ih c \tope{\opsi}(x) \gamma^\mu \tope{\psi}(x) \} ,
\)
these two Lagrangians give rise (under some conditions at infinity) to
identical action integrals and, consequently, to identical theories.%
\footnote{~%
The expressions for the conserved currents (energy\ndash momentum, charge
current and angular momentum density) arising from the
Lagrangians~\eref{12.1} and~\eref{12.3} differ significantly. But the
conserved, time\ndash independent, quantities (momentum, charge and angular
momentum operators) for the both Lagrangians are identical.%
}
However, the Lagrangian~\eref{12.1} has two advantages: on one hand, it is
Hermitian contrary to~\eref{12.3} and, on the other hand, it vanishes when the
field variables satisfy the field equations, which simplifies some
calculations.

	Before going on, we emphasize that the calculation of the derivatives
in~\eref{12.6} and in~\eref{12.9} below according to the rules of the
classical analysis of commuting variables is not quite correct. This method
is harmless when the Euler-Lagrange equations are considered but it requires
some additional operator ordering rules for the expressions of the conserved
quantities; for details, see~\cite{bp-QFT-action-principle}. However, this
approach breaks down for the Lagrangian~\eref{12.19}
(or~\eref{12.16}) which we shall investigate below; the cause being that all
its derivatives vanish if they are calculated according to the rules
mentioned. For the correct treatment of that Lagrangian, the methods
developed in~\cite{bp-QFT-action-principle} are required.

	It is almost evident, the operator $\psi$ and its Dirac conjugate
$\opsi$ do not enter in~\eref{12.1}--\eref{12.4} on an equal footing, i.e.,
in some sense, $\opsi$ is `first' and $\psi$ is `second' in order (counting
from the left to the right) operator in these Lagrangians. One can revert the
situation by considering, for example, the Hermitian Lagrangian
	\begin{gather}	\label{12.7}
	\begin{split}
\tope{L}_1
=
 \frac{1}{2}\ih c\{
  \tope{\psi}^\top(x)\gamma^\mu\circ (\pd_\mu\tope{\opsi}^\top(x))
- (\pd_\nu\tope{\psi}^\top(x))\gamma^\mu\circ \tope{\opsi}^\top(x)
\}
- mc^2 \tope{\psi}^\top(x)\circ \tope{\opsi}^\top(x)
	\end{split}
\\			\label{12.8}
\ope{L}_1
=
\frac{1}{2} c \{
\psi^\top\gamma^\mu\circ[\opsi^\top,\ope{P}_\mu]_{\_}
- [\psi^\top,\ope{P}_\mu]_{\_}\gamma^\mu\circ \opsi^\top
\}
-m c^2\psi^\top\circ\opsi^\top
	\end{gather}
in Heisenberg and momentum picture, respectively.%
\footnote{~%
Here and below the matrix transposition sign $\top$ is required to be ensured
a proper matrix multiplication -- see the conventions at the beginning of
Sect.~\ref{Sect3}.%
}
A straightforward calculation gives:
	\begin{equation}	\label{12.9}
	\begin{split}
\frac{\pd\ope{L}_1}{\pd\psi^\top} = \frac{\pd\tope{L}_1}{\pd\psi^\top}
& =
+\frac{1}{2}c \gamma^\mu [\opsi^\top,\ope{P}_\mu]_{\_}  - m c^2\opsi^\top
\qquad
\frac{\pd\ope{L}}{\pd y_\mu^\top} = -\frac{1}{2} \ih c \gamma^\mu \opsi^\top
\\
\frac{\pd\ope{L}_1}{\pd\opsi^\top} = \frac{\pd\tope{L}_1}{\pd\opsi^\top}
& =
-\frac{1}{2}c [\psi^\top,\ope{P}_\mu]_{\_} \gamma^\mu - m c^2\psi^\top
\qquad
\frac{\pd\ope{L}_1}{\pd \overline{y}_\mu^\top}
=
+ \frac{1}{2} \ih c \psi^\top \gamma^\mu .
	\end{split}
	\end{equation}
Therefore, in view of~\eref{12.129}, the field equations for the
Lagrangian~\eref{12.8} are
	\begin{equation}	\label{12.10}
[\psi^\top,\ope{P}_\mu]_{\_} \gamma^\mu + m c\psi^\top = 0
\qquad
\gamma^\mu [\opsi^\top,\ope{P}_\mu]_{\_} - m c\opsi^\top = 0 .
	\end{equation}
So, as one can expect, we get the system of Dirac equations~\eref{3.5} with
interchanged places of $\psi$ and $\opsi$.%
\footnote{~%
Recall, the matrix transposition sign in~\eref{12.10} serves to provide a
proper matrix multiplication as we consider $\psi$ as vector\ndash column and
$\opsi$ as vector\ndash row. If we write~\eref{12.10} and~\eref{3.5} in
components, it will be obvious that these systems of~8 equations are
identical up to notation.%
}

	In the Lagrangian~\eref{12.7}, the fields $\psi$ and $\opsi$ differ by
their positions too; now $\psi$ is `first' and $\opsi$ is `second'. However,
since we are dealing with a free field, we, due to the charge
symmetry~\cite{Bjorken&Drell,Bogolyubov&Shirkov}, can try to start from a
Lagrangian which describes in a symmetric way the field's particles and
antiparticles. Since $\psi$ is connected with field's particles and $\opsi$
with its antiparticles, a hypothesis can be made that this may be achieved
via a Lagrangian in which $\psi$ and $\opsi$ enter symmetrically. For
example, as a possible candidate, the half sum of~\eref{12.2} and~\eref{12.7}
can be taken, \viz
	\begin{multline}	\label{12.11}
\ope{L}_2
=
\frac{1}{4} c
\{
 \opsi\gamma^\mu\circ[\psi,\ope{P}_\mu]_{\_}
- [\opsi,\ope{P}_\mu]_{\_}\gamma^\mu\circ \psi
+
 \tope{\psi}^\top(x)\gamma^\mu\circ (\pd_\mu\tope{\opsi}^\top(x))
- (\pd_\nu\tope{\psi}^\top(x))\gamma^\mu\circ \tope{\opsi}^\top(x)
\}
\\
 - \frac{1}{2}m c^2 \{
  \opsi\circ\psi
+ \tope{\psi}^\top(x)\circ \tope{\opsi}^\top(x) \} .
	\end{multline}

	Obviously, this Lagrangian has all required properties, including the
symmetry $\psi_\mu\leftrightarrow\opsi_\mu$, which in matrix notation reads
$\psi\leftrightarrow\opsi^\top$ and $\opsi\leftrightarrow\psi^\top$. However,
it turns that the field described by the Lagrangian~\eref{12.11} has nothing
to do with a spinor field. Indeed, one may verify that the field equations
for it are
	\begin{gather*}
\sum_{\beta}	[\psi_\beta,\ope{P}_\mu]_{\_}
		\overline{\gamma}_{\alpha\beta}^{\mu}
- mc\psi_\alpha
= 0
\qquad
\sum_{\beta}	[\opsi_\beta,\ope{P}_\mu]_{\_}
		\overline{\gamma}_{\alpha\beta}^{\mu}
- mc\opsi_\alpha
= 0
\intertext{or, in a matrix form,}
\overline{\gamma}^{\mu} [\psi,\ope{P}_\mu]_{\_} - mc\psi = 0
\qquad
[\opsi,\ope{P}_\mu]_{\_} \overline{\gamma}^{\mu} + mc\opsi = 0 ,
	\end{gather*}
where $\overline{\gamma}^{\mu}:=\frac{1}{2}(\gamma^\mu-(\gamma^\mu)^\top)$ is
the antisymmetric part of $\gamma^\mu$. Besides, now $\psi$ and $\opsi$ do
not satisfy the Klein\ndash Gordon equations~\eref{5.11} and the
Lagrangian~\eref{12.11} happens to be singular.%
\footnote{~%
For instance, in the representation ~\eref{3.1-1} of the $\gamma$\ndash
matrices, in which $\gamma^0$ is diagonal, we have in Heisenberg picture
\(
\frac{\pd\tope{L}_2}{\pd(\pd_0\psi)}
=
\frac{\pd\tope{L}_2}{\pd(\pd_0\opsi)}
= 0.
\)%
}
The last fact means that~\eref{12.11} describes a system with constraints that
requires a different treatment (see, e.g.,~\cite{Dirac-LQM}).

	Similar will be the consequences of any Lagrangian symmetric in
$\psi$ and $\opsi$: it will not describe the field we are investigating here.
One of the formal reasons for such a conclusion is that the field equations
for $\psi$ and $\opsi$, implied by a Lagrangian symmetric in them, will be
identical, contrary to our expectation that they should be (equivalent
to)~\eref{3.5}. The physical reason for this situation is that the change
particle$\leftrightarrow$antiparticle is describe by
 $a_s^\pm(\bk)\leftrightarrow a_s^{\dag\,\pm}(\bk)$, \emph{not} by
 $\psi\leftrightarrow\opsi$,
which is due to the half integer spin of Dirac spinor field (and its
transformation properties under Lorentz transformations).%
\footnote{~%
One should compare this situation with the one for free scalar field
$\varphi$ where the change $\varphi\leftrightarrow\varphi^\dag$ is equivalent
to the transformation particle$\leftrightarrow$antiparticle;
see~\cite{bp-QFTinMP-scalars} for considerations in momentum picture and,
e.g.,~\cite{Bogolyubov&Shirkov} for treatment in Heisenberg picture.%
}

	It is known (see, for instance,~\cite[\S~13.4]{Bogolyubov&Shirkov},
\cite[\S~99]{Bjorken&Drell}, and~\cite[p.~114]{Roman-QFT}), in terms of
$\psi$ and $\opsi$, the change particle$\leftrightarrow$antiparticle is
describe by the replacement
	\begin{gather}	\label{12.12}
\psi\mapsto \bpsi := C\opsi^\top = (\opsi C^\top)^\top
\qquad
\opsi\mapsto \bopsi := \psi^\top C^{-1\,\top} = (C^{-1}\psi)^\top ,
\intertext{called \emph{charge conjugation}, where the matrix $C$ satisfies
the conditions}
			\label{12.13}
C^{-1} \gamma^\mu C = -\gamma^{\mu\,\top}:=-(\gamma^\mu)^\top
\qquad
C^\top = - C
	\end{gather}
and, in a representation, like~\eref{3.1-1}, in which
$\gamma^\mu=(-1)^\mu\gamma^{\mu\,\top}$, $\mu=0,1,2,3$, can be chosen as
$C=b\gamma^0\gamma^2$ with $b\in\field[C]\backslash\{0\}$. The
Lagrangian~\eref{12.1}, under the conditions~\eref{9.13} and after (resp.\
before) normal ordering, is invariant (resp.\ changes sign) under the
change~\eref{12.12}.

	So, if we want, from the very beginning, to have a suitable
description of particle\ndash antiparticle properties of a spinor field, we
have to describe it via the variables $\psi$ and $\bpsi$, not via $\psi$ and
$\opsi$. In these new variables, due to $\opsi=-\bpsi^\top C^{-1}$
(see~\eref{12.12} and~\eref{12.13}), the Lagrangian~\eref{12.1} reads
	\begin{equation}	\label{12.14}
\tope{L}^{\prime}
=
- \frac{1}{2}\ih c\{
  \tope{\bpsi}^\top(x) C^{-1}\gamma^\mu\circ(\pd_\mu\tope{\psi}(x))
- (\pd_\mu\tope{\bpsi}^\top(x)) C^{-1}\gamma^\mu\circ \tope{\psi}(x)
\}
+ mc^2 \tope{\bpsi}^\top(x) C^{-1}\circ \tope{\psi}(x) .
	\end{equation}
We would like to emphasize on the change of the signs and the appearance
of the matrix $C$ in~\eref{12.14} with respect to~\eref{12.1}. An alternative
to this Lagrangian  is a one with changed positions of $\psi$ and $\bpsi$,
\viz
	\begin{equation}	\label{12.15}
\tope{L}^{\prime\prime}
= \!
- \frac{1}{2}\ih c\{
  \tope{\psi}^\top(x) C^{-1}\gamma^\mu\circ(\pd_\mu\tope{\bpsi}(x))
- (\pd_\mu\tope{\psi}^\top(x)) C^{-1}\gamma^\mu\circ \tope{\bpsi}(x)
\}
+ mc^2 \tope{\psi}^\top(x) C^{-1}\circ \tope{\bpsi}(x) .
	\end{equation}
Notice, the last Lagrangian is completely different from~\eref{12.7} and
cannot be obtained from it by putting
$\tope{\opsi}^\top(x)=C^{-1}\tope{\bpsi}(x)$. Evidently, the variables $\psi$
and $\bpsi$ do not enter in~\eref{12.14} and~\eref{12.15} on equal footing.
We shall try to `symmetrize' the situation by considering a Lagrangian which
is the half sum of the last two ones, \ie
%
%
%
	\begin{multline}	\label{12.16}
\tope{L}^{\prime\prime\prime}
=
  \frac{1}{4}\ih c\{
- \tope{\bpsi}^\top(x) C^{-1}\gamma^\mu\circ(\pd_\mu\tope{\psi}(x))
+ (\pd_\mu\tope{\bpsi}^\top(x)) C^{-1}\gamma^\mu\circ \tope{\psi}(x)
\\
-  \tope{\psi}^\top(x) C^{-1}\gamma^\mu\circ(\pd_\mu\tope{\bpsi}(x))
+ (\pd_\mu\tope{\psi}^\top(x)) C^{-1}\gamma^\mu\circ \tope{\bpsi}(x)
\}
\\
+ \frac{1}{2} mc^2
\{	\tope{\bpsi}^\top(x) C^{-1}\circ \tope{\psi}(x)
      +	\tope{\psi}^\top(x) C^{-1}\circ \tope{\bpsi}(x)
\} .
	\end{multline}

	According to~\eref{2.5}, the Lagrangians~\eref{12.14}--\eref{12.16} in
momentum picture respectively are:
	\begin{gather}	\label{12.17}
\ope{L}^{\prime}
=
- \frac{1}{2} c \{
\bpsi^\top C^{-1}\gamma^\mu\circ [\psi,\ope{P}_\mu]_{\_}
- [\bpsi^\top,\ope{P}_\mu]_{\_} C^{-1}\gamma^\mu\circ \psi
\}
+ m c^2\bpsi^\top C^{-1}\circ\psi
\\			\label{12.18}
\ope{L}^{\prime\prime}
=
- \frac{1}{2} c \{
\psi^\top C^{-1}\gamma^\mu\circ [\bpsi,\ope{P}_\mu]_{\_}
- [\psi^\top,\ope{P}_\mu]_{\_} C^{-1}\gamma^\mu\circ \bpsi
\}
+ m c^2\psi^\top C^{-1}\circ\bpsi
	\end{gather}
\vspace{-4.5ex}
	\begin{multline}	\label{12.19}
\ope{L}^{\prime\prime\prime}
=
  \frac{1}{4} c \{
- \bpsi^\top C^{-1}\gamma^\mu\circ [\psi,\ope{P}_\mu]_{\_}
+ [\bpsi^\top,\ope{P}_\mu]_{\_} C^{-1}\gamma^\mu\circ \psi
\\
- \psi^\top C^{-1}\gamma^\mu\circ [\bpsi,\ope{P}_\mu]_{\_}
+ [\psi^\top,\ope{P}_\mu]_{\_} C^{-1}\gamma^\mu\circ \bpsi
\}
+ \frac{1}{2} m c^2
 \{ \bpsi^\top C^{-1}\circ\psi + \psi^\top C^{-1}\circ\bpsi \} .
	\end{multline}

	The consequences of these Lagrangians will be described and compared
below. But, before proceeding with this task, we
\emph{must make an important remark regarding the Lagrangian}~\eref{12.16}
(or~\eref{12.19} in momentum picture). If one computes its derivatives with
respect to $\psi$ and $\bpsi$ and their partial derivatives, according to the
silently accepted rules valid for classical fields, one will see that all of
them vanish.%
\footnote{~%
It is quite possible, the first reaction to such a result will be the
rejection of the Lagrangian~\eref{12.16} as it leads to identities, like
$0=0$, instead to field equations and to identically vanishing dynamical
variables. However, this will be a hasty conclusion --- \emph{vide infra}.%
}
In a sense, this will mean that the Lagrangian~\eref{12.16} is constant as a
functional of the field variables, which, obviously, is not the case. This
`paradox' is explained in the paper~\cite{bp-QFT-action-principle} in which it
is proved that it is due to an incorrect transferring of the differentiation
rules from the classical analysis of commuting variables and functions of
them to the `quantum' one of \emph{non\ndash commuting} variables, such as
the operators on  Hilbert spaces, and functions of them. In this work, it is
shown that the derivatives of Lagrangians like~\eref{12.16} should be
computed in a way different from the one of the classical analysis (and, in
fact, are operators on the operator space over $\Hil$); in particular, the
Euler\ndash Lagrange equations for the Lagrangian~\eref{12.16} are
identities, like $0=0$, but it implies field equations according to a
rigorous modified action principle. Without going into details, the new
procedure, when applied to the Lagrangian~\eref{12.16}, is equivalent to the
standard one with \emph{new definitions} of the derivatives of~\eref{12.16}
with respect to its operator arguments. Here is the list of the derivatives
of the Lagrangians~\eref{12.17}--\eref{12.19}:
	\begin{equation}	\label{12.20}
	\begin{split}
\frac{\pd\ope{L}^{\prime}}{\pd\psi^\top} =
- \frac{\pd\ope{L}^{\prime\prime}}{\pd\psi^\top} =
  \frac{\pd\ope{L}^{\prime\prime\prime}}{\pd\psi^\top}
& :=
+ \frac{1}{2}c C^{-1}\gamma^\mu [\bpsi,\ope{P}_\mu]_{\_} - m c^2 C^{-1}\bpsi
\\
\frac{\pd\ope{L}^{\prime}}{\pd\bpsi^\top} =
- \frac{\pd\ope{L}^{\prime\prime}}{\pd\bpsi^\top} =
  \frac{\pd\ope{L}^{\prime\prime\prime}}{\pd\bpsi^\top}
& :=
- \frac{1}{2}c C^{-1}\gamma^\mu [\psi,\ope{P}_\mu]_{\_} + m c^2 C^{-1}\psi
\\
\pi^\mu
= \frac{\pd\ope{L}^{\prime}}{\pd y_\mu^\top}
=  - \frac{\pd\ope{L}^{\prime\prime}}{\pd y_\mu^\top}
=    \frac{\pd\ope{L}^{\prime\prime\prime}}{\pd y_\mu^\top}
& :=
- \frac{1}{2} \ih c C^{-1}\gamma^\mu \bpsi
\\
\Breve{\pi}^\mu
= \frac{\pd\ope{L}^{\prime}}{\pd \Breve{y}_\mu^\top}
= - \frac{\pd\ope{L}^{\prime\prime}}{\pd \Breve{y}_\mu^\top}
=   \frac{\pd\ope{L}^{\prime\prime\prime}}{\pd \Breve{y}_\mu^\top}
& :=
+ \frac{1}{2} \ih c C^{-1}\gamma^\mu \psi  .
	\end{split}
	\end{equation}
Notice, in this list the expressions for the derivatives of
$\ope{L}^{\prime\prime\prime}$ are definitions, while the derivatives of
$\ope{L}^{\prime}$ and $\ope{L}^{\prime\prime}$ are derived, using the
equality
	\begin{equation}	\label{12.19new}
\bigl( C^{-1}\gamma^\mu \bigr)^\top = C^{-1}\gamma^\mu
	\end{equation}
implied by~\eref{12.13}, according to the rules of the analysis of classical
fields. As an exercise, the reader may wish to convert~\eref{12.20} into the
`more usual' Heisenberg picture of motion.
	As a result of~\eref{12.20}, the field equations~\eref{12.129} for
all of the Lagrangians~\eref{12.17}--\eref{12.19} now read:
	\begin{equation}	\label{12.21}
\gamma^\mu [\psi,\ope{P}_\mu]_{\_} - m c\psi = 0
\qquad
\gamma^\mu [\bpsi,\ope{P}_\mu]_{\_} - m c\bpsi = 0 .
	\end{equation}
Consequently, as one may expect, the spinor field and its \emph{charge
conjugate} $\bpsi:=C\opsi^\top=C\gamma^{0\,\top}\psi^{\dag\,\top}$ are
solutions of one and the same Dirac equation.%
\footnote{~%
It is a simple algebra to be proved, the substitution $\bpsi:=C\opsi^\top$
transforms the second equation in~\eref{3.5} into the second equation
in~\eref{12.21}.%
}

	Regardless of the identical equations of motion obtained from the
Lagrangians~\eref{12.17}--\eref{12.19}, these Lagrangians entail completely
different conserved quantities (dynamical variables).
In~\cite{bp-QFT-action-principle} is shown that to any operator Lagrangian in
quantum field theory there corresponds a unique set of conserved
operators.%
\footnote{~%
The meaning of the last theorem is that there is a criterion for selecting
the order of all operators in compositions (products) of operators
appearing in the conserved quantities obtained from the Schwinger's variational
principle~\cite{Roman-QFT} and/or from the first Noether
theorem~\cite{Bogolyubov&Shirkov}. For rigorous derivation of
equations~\eref{12.22}--\eref{12.24} below, apply the results
of~\cite[sec.~4]{bp-QFT-action-principle} to the Lagrangians
under consideration here.%
}
The corresponding to the Lagrangians~\eref{12.14}--\eref{12.16}
energy-momentum, current and spin angular momentum density operators are:
 \newcommand{\bpi}{\Breve{\pi}}
	\begin{subequations}	\label{12.22}
	\begin{align}
			\label{12.22a}
	\begin{split}
& \tope{T}_{\mu\nu}^{\prime}
 =
  \tope{\pi}_\mu^\top \circ (\pd_\nu\tope{\psi})
+ (\pd_\nu\tope{\bpsi}^\top) \circ \tope{\bpi}_\mu
\\ & \hphantom{\tope{T}_{\mu\nu}^{\prime}}
=
\frac{1}{2} \ih c \{
 - \tope{\bpsi}^\top C^{-1}\gamma_\mu \circ (\pd_\nu\tope{\psi})
 + (\pd_\nu\tope{\bpsi}^\top)  C^{-1}\gamma_\mu \circ \tope{\psi}
\}
	\end{split}
\\			\label{12.22b}
& \tope{J}_{\mu}^{\prime}
 = \frac{q}{\ih} \{
  \tope{\pi}_\mu^\top \circ \tope{\psi}
- \tope{\bpsi}^\top \circ \tope{\bpi}_\mu
\}
=
- q c \tope{\bpsi}^\top C^{-1}\gamma_\mu \circ \tope{\psi}
\\			\label{12.22c}
	\begin{split}
& \tope{S}_{\mu\nu}^{\prime\lambda}
 = - \{
  \tope{\pi}^{\lambda\,\top} \circ (I_{\psi\mu\nu} \tope{\psi})
+ (I_{\bpsi\mu\nu} \tope{\bpsi})^\top \circ \tope{\bpi}^\lambda
\}
=
\frac{1}{4} \hbar c
  \tope{\bpsi}^\top C^{-1}
\{ \gamma^\lambda \sigma_{\mu\nu} + \sigma_{\mu\nu} \gamma^\lambda \}
   \circ \tope{\psi}
	\end{split}
	\end{align}
	\end{subequations}
\vspace{-2ex}
	\begin{subequations}	\label{12.23}
	\begin{align}
			\label{12.23a}
	\begin{split}
\tope{T}_{\mu\nu}^{\prime\prime}
& =
- \tope{\bpi}_\mu^\top \circ (\pd_\nu\tope{\bpsi})
- (\pd_\nu\tope{\psi}^\top) \circ \tope{\pi}_\mu
\\
& =
\frac{1}{2} \ih c \{
 - \tope{\psi}^\top C^{-1}\gamma_\mu \circ (\pd_\nu\tope{\bpsi})
 + (\pd_\nu\tope{\psi}^\top)  C^{-1}\gamma_\mu \circ \tope{\bpsi}
\}
	\end{split}
\\			\label{12.23b}
\tope{J}_{\mu}^{\prime\prime}
& = \frac{q}{\ih} \{
  \tope{\bpi}_\mu^\top \circ \tope{\bpsi}
- \tope{\psi}^\top \circ \tope{\pi}_\mu
\}
=
+ q c \tope{\psi}^\top C^{-1}\gamma_\mu \circ \tope{\bpsi}
\\			\label{12.23c}
	\begin{split}
\tope{S}_{\mu\nu}^{\prime\prime\lambda}
& = + \{
  \tope{\bpi}^{\lambda\,\top} \circ (I_{\bpsi\mu\nu} \tope{\bpsi})
+ (I_{\psi\mu\nu} \tope{\psi})^\top \circ \tope{\pi}^\lambda
\}
\frac{1}{4} \hbar c
  \tope{\psi}^\top C^{-1}
\{ \gamma^\lambda \sigma_{\mu\nu} + \sigma_{\mu\nu} \gamma^\lambda \}
   \circ \tope{\bpsi}
	\end{split}
	\end{align}
	\end{subequations}
\vspace{-2ex}
	\begin{subequations}	\label{12.24}
	\begin{align}
			\label{12.24a}
	\begin{split}
\tope{T}_{\mu\nu}^{\prime\prime\prime}
& = \frac{1}{2} \{
  \tope{\pi}_\mu^\top \circ (\pd_\nu\tope{\psi})
- (\pd_\nu\tope{\psi}^\top) \circ \tope{\pi}_\mu
- \tope{\bpi}_\mu^\top \circ (\pd_\nu\tope{\bpsi})
+ (\pd_\nu\tope{\bpsi}^\top) \circ \tope{\bpi}_\mu
\}
\\
& =
\frac{1}{4} \ih c \{
 - \tope{\bpsi}^\top C^{-1}\gamma_\mu \circ (\pd_\nu\tope{\psi})
 + (\pd_\nu\tope{\bpsi}^\top)  C^{-1}\gamma_\mu \circ \tope{\psi}
\\
& \hphantom{= \frac{1}{4} \ih c \{ }
 - \tope{\psi}^\top C^{-1}\gamma_\mu \circ (\pd_\nu\tope{\bpsi})
  + (\pd_\nu\tope{\psi}^\top)  C^{-1}\gamma_\mu \circ \tope{\bpsi}
\}
	\end{split}
\\			\label{12.24b}
	\begin{split}
\tope{J}_{\mu}^{\prime\prime\prime}
& = \frac{1}{2}\frac{q}{\ih} \{
  \tope{\pi}_\mu^\top \circ \tope{\psi}
+ \tope{\psi}^\top \circ \tope{\pi}_\mu
+  \tope{\bpi}_\mu^\top \circ \tope{\bpsi}
+ \tope{\bpsi}^\top \circ \tope{\bpi}_\mu
\}
\\
& = - \frac{1}{2}  q c \{
  \tope{\bpsi}^\top C^{-1}\gamma_\mu \circ \tope{\psi}
- \tope{\psi}^\top C^{-1}\gamma_\mu \circ \tope{\bpsi}
\}
	\end{split}
\\			\label{12.24c}
	\begin{split}
\tope{S}_{\mu\nu}^{\prime\prime\prime\lambda}
& = - \frac{1}{2} \{
  \tope{\pi}^{\lambda\,\top} \circ (I_{\psi\mu\nu} \tope{\psi})
- (I_{\psi\mu\nu} \tope{\psi})^\top \circ \tope{\pi}^\lambda
\\
& \hphantom{= - \frac{1}{2} \{ }
-  \tope{\bpi}^{\lambda\,\top} \circ (I_{\bpsi\mu\nu} \tope{\bpsi})
+ (I_{\bpsi\mu\nu} \tope{\bpsi})^\top \circ \tope{\bpi}^\lambda
\}
\\
& =
\frac{1}{8} \hbar c \bigl\{
  \tope{\bpsi}^\top C^{-1}
( \gamma^\lambda \sigma_{\mu\nu} + \sigma_{\mu\nu} \gamma^\lambda )
  \circ \tope{\psi}
+
  \tope{\psi}^\top C^{-1}
( \gamma^\lambda \sigma_{\mu\nu} + \sigma_{\mu\nu} \gamma^\lambda )
   \circ \tope{\bpsi}
\bigr\},
	\end{split}
	\end{align}
	\end{subequations}
where~\eref{12.20}, $\varepsilon(\psi)=-\varepsilon(\bpsi)=+1$,
 $I_{\psi\mu\nu}  = I_{\bpsi\mu\nu} = -\frac{1}{2}\iu\sigma_{\mu\nu}$
(see~\eref{3.13new}),
 $C\sigma_{\mu\nu}^{\top}C^{-1} = -\sigma_{\mu\nu}$ (see~\eref{3.13-1}
and~\eref{12.13}), and~\eref{12.19new} were used.
	The corresponding expressions for the orbital angular momentum are
obtained according to the general formula~\eref{2.12a}.
	The reader may easily write~\eref{12.22}--\eref{12.24} in momentum
picture; formally, this can be done by replacing
$\pd_\nu\tope{\psi}$ and $\pd_\nu\tope{\bpsi}$ by respectively
$y_\nu:=\iih[\psi,\ope{P}_\nu]_{\_}$
and
$\Breve{y}_\nu:=\iih[\bpsi,\ope{P}_\nu]_{\_}$
and by removing the tildes (waves) from the remaining symbols.

	Looking over~\eref{12.16} (or~\eref{12.19})
and~\eref{12.22}--\eref{12.24}, one can notice that the Lagrangian,
energy\ndash momentum and spin angular momentum density operators are
\emph{symmetric} while the charge current operator is \emph{antisymmetric}
under the charge conjugation, \ie under the change
$\psi\leftrightarrow\bpsi$, as it should
be~\cite{Bjorken&Drell,Bogolyubov&Shirkov,Roman-QFT}. In this way, we shall
see that the symmetry particle\ndash antiparticle for a free spinor field
should be encoded into a Lagrangian which is symmetric relative to the spinor
operators describing the corresponding fields/particles.

	Since the field $\psi$ and its charge conjugate $\bpsi:=C\opsi^\top$
satisfy identical Dirac equations for all of the
Lagrangians~\eref{12.17}--\eref{12.19} (see~\eref{12.21}), the field $\psi$
and its Dirac conjugate $\opsi:=\psi^\dag\gamma^0=-\bpsi^\top C^{-1}$ satisfy
the system of Dirac equations~\eref{3.5} for these Lagrangians. Thus  for
$\psi$ and $\opsi$ are valid all considerations and results in
sections~\ref{Sect5} and~\ref{Sect6}.%
\footnote{~%
The only exception being the concrete forms of the expression in the r.h.s.\
of~\eref{5.4} and~\eref{5.5} for the `degenerate' solutions~\eref{5.1}.%
}
As the Dirac conjugate field $\opsi$ enters linearly
everywhere in sections~\ref{Sect5} and~\ref{Sect6}, all of the material in
these sections is valid \emph{mutatis mutandis} for the charge conjugate
field $\bpsi=C\opsi^\top=(\opsi C^\top)^\top=-(\opsi C)^\top$
whose components are linear combinations of those of $\opsi$; the only thing
one should do, to rewrite this material in terms of $\bpsi$, is to replace
$\opsi$ with $\bpsi$ or, more formally, to write the breve accent sign
``$\Breve{\mspace{9mu}}$'' for the line
``\hspace{0.3ex}$\overline{\vphantom{m}\mspace{9mu}}$\hspace{0.3ex}'' over a
symbol's sign.  For example, the decomposition~\eref{5.19b} implies
	\begin{equation}	\label{12.25}
\bpsi
= \sum_{s} \int\Id^3\bk \bigl\{
  \Breve{f}_{s,+}(\bk) \bpsi_{s,(+)}(\bk)
+ \Breve{f}_{s,-}(\bk) \bpsi_{s,(-)}(\bk) \bigr\}
	\end{equation}
where $\Breve{f}_{s,\pm}(\bk)$ are some (, possibly, generalized) functions
and $\bpsi_{s,(\pm)}(\bk)$ are spinor operators. It is essential to be
mentioned the connection
	\begin{equation}	\label{12.26}
 \bpsi_{s,(\pm)}(\bk)
=  C \opsi_{s,(\pm)}^\top(\bk)
= - \bigl( \opsi_{s,(\pm)}(\bk) C \bigr)^\top
	\end{equation}
which is a consequence of~\eref{5.19b} and~\eref{12.12}. Further, the
creation/annihi\-lation operators for $\bpsi$ should be introduced via the
second formula in~\eref{6.1} with $\bpsi$ for $\opsi$ and $\Breve{f}$ for
$\overline{f}$ and, by virtue of~\eref{12.26}, are
	\begin{equation}	\label{12.27}
 \bpsi_{s}^{\pm}(\bk)
=  C \bigl( \opsi_{s}^{\pm}(\bk) \bigr)^\top
= - \bigl( \opsi_{s}^{\pm}(\bk) C \bigr)^\top .
	\end{equation}
All of the remaining material of Sect.~\ref{Sect6} concerning
$\opsi_{s}^{\pm}(\bk)$ is then automatically valid for
$\bpsi_{s}^{\pm}(\bk)$, provided the replacement $\opsi\mapsto\bpsi$ is made.
In particular, the interpretation of $\bpsi_{s}^{\pm}(\bk)$ as
creation/annihilation operators of (anti)particles with 4\ndash momentum
$(\sqrt{m^2c^2 + \bk^2},\bk)$, charge $q$, orbital angular momentum
$(x_\mu k_\nu- x_\nu k_\mu)\bigr|_{k_0=\sqrt{m^2c^2 + \bk^2}}\openone_4$, and
spin mixing angular momentum matrices
$-\frac{1}{2}\hbar\sigma_{\mu\nu}$ is valid. The invariant,
frame\ndash independent, creation/annihilation operators
$a_{s}^{\dag\,\pm}(\bk)$ for $\bpsi_{s}^{\pm}(\bk)$ are the same as for
$\opsi_{s}^{\pm}(\bk)$ and, as a consequence of~\eref{12.27} and~\eref{6.20},
are introduce via the equation
	\begin{equation}	\label{12.28}
\bpsi_s^\pm(\bk)
=
(2\pi\hbar)^{-3/2} a_s^{\dag\,\pm}(\bk) C ( \overline{v}^{s,\pm}(\bk) )^\top
=
- (2\pi\hbar)^{-3/2} a_s^{\dag\,\pm}(\bk)
\bigl(  \overline{v}^{s,\pm}(\bk) C \bigr)^\top
	\end{equation}
where ${v}^{s,\pm}(\bk)$ are defined by~\eref{5.20}--\eref{5.22}. So, the
equalities~\eref{6.20} and the following from them interpretation of
$a_{s}^{\pm}(\bk)$ and $a_{s}^{\dag\,\pm}(\bk)$ hold without changes.

	All of the above said, concerning $\bpsi$ and its (frequency)
decompositions, is valid for any one of the
Lagrangians~\eref{12.17}--\eref{12.19} as they entail identical field
equations. However, form~\eref{12.22}--\eref{12.24}, it is evident that the
energy\ndash momentum, charge and angular momenta densities for these
Lagrangians are completely different. The corresponding to them 4\ndash
momentum, charge and angular momenta can easily be calculated on the base
of the results of Sect.~\ref{Sect7} and the decomposition~\eref{12.28} in the
following way.

	At first, notice that the Lagrangians~\eref{12.17}--\eref{12.19}
vanish if $\psi$ and $\bpsi$ satisfy the field equations~\eref{12.21}, \ie
	\begin{equation}	\label{12.29}
\ope{L}^{\prime}
= \ope{L}^{\prime\prime}
= \ope{L}^{\prime\prime\prime}
= 0
\qquad \text{if}\quad
\gamma^\mu[\chi,\ope{P}_\mu]_{\_} - mc\chi = 0,
\quad
\chi=\psi,\bpsi,
	\end{equation}
and $\ope{L}=\ope{L}^{\prime}$ as $\bpsi=C\opsi^\top$ (see~\eref{12.2}).
Moreover, $\ope{L}^{\prime\prime}$  (resp.\ $\ope{L}^{\prime\prime\prime}$)
and the results following from it can be obtained by interchanging the
positions of $\psi$ and $\bpsi$, possibly with some indices and arguments,
(resp.\ by forming the half sum of $\ope{L}^{\prime}$ and
$\ope{L}^{\prime\prime}$ and the results corresponding to them).
Consequently, the results~\eref{7.2}, \eref{7.4}, \eref{7.6},  \eref{7.7},
\eref{7.9}--\eref{7.12}, all with
$\opsi_s^{\pm}= -\bigl(\bpsi_s^{\pm})^\top C^{-1}$, are valid for
$\ope{L}=\ope{L}^{\prime}$, but for $\ope{L}^{\prime\prime}$ and
$\ope{L}^{\prime\prime\prime}$ the just describe changes must be made.

	Applying~\eref{12.28},~\eref{6.20} and~\eref{12.13}, we derive:
	\begin{subequations}	\label{12.30}
	\begin{align}	\label{12.30a}
( \opsi_s^\varepsilon(\bk) ) \Lambda \circ \psi_{s'}^{\varepsilon'}(\bk')
& =
- (\bpsi_s^\varepsilon(\bk))^\top C^{-1} \Lambda \circ
				\psi_{s'}^{\varepsilon'}(\bk')
\\ \notag &
=
+ (2\pi\hbar)^{-3}
a_{s}^{\dag\,\varepsilon}(\bk)
\bigl\{ \overline{v}^{s,\varepsilon}(\bk) \Lambda {v}^{s',\varepsilon'}(\bk')
\bigr\}
\circ a_{s'}^{\varepsilon'}(\bk')
\\			\label{12.30b}
- (\psi_s^\varepsilon(\bk))^\top C^{-1}\Lambda \circ
			\bpsi_{s'}^{\varepsilon'}(\bk')
& =
- (2\pi\hbar)^{-3}
a_{s}^{\varepsilon}(\bk)
\bigl\{ \overline{v}^{s',\varepsilon'}(\bk')
\bigl( C^{-1}\Lambda C \bigr)^\top
{v}^{s,\varepsilon}(\bk)
\bigr\}
\circ a_{s'}^{\dag\,\varepsilon'}(\bk')
\\			\label{12.30c}
- (\psi_s^\varepsilon(\bk))^\top C^{-1}\gamma^\mu \circ
			\bpsi_{s'}^{\varepsilon'}(\bk')
& =
+ (2\pi\hbar)^{-3}
a_{s}^{\varepsilon}(\bk) \circ a_{s'}^{\dag\,\varepsilon'}(\bk')
\bigl\{\overline{v}^{s',\varepsilon'}(\bk')
\gamma^\mu {v}^{s,\varepsilon}(\bk)
\bigr\}
\\			\label{12.30d}
- (\psi_s^\varepsilon(\bk))^\top C^{-1}\gamma^0 \sigma_{ab} \circ
			\bpsi_{s'}^{\varepsilon'}(\bk')
& =
- (2\pi\hbar)^{-3}
a_{s}^{\varepsilon}(\bk) \circ a_{s'}^{\dag\,\varepsilon'}(\bk')
\bigl\{ \overline{v}^{s',\varepsilon'}(\bk') \gamma^0 \sigma_{ab}
					{v}^{s,\varepsilon}(\bk)
\bigr\}
	\end{align}
	\end{subequations}
where $\varepsilon,\varepsilon'=+,-$, $\Lambda$ is a constant $4\times4$
matrix (e.g., $\Lambda=\gamma^\mu,\gamma^0\sigma_{ab}$), or an operator
(e.g., $\Lambda=k_\mu\frac{\pd}{\pd k^\nu}$) or matrix operator
(e.g., $\Lambda=\gamma^0 k_\mu\frac{\pd}{\pd k^\nu}$),%
\footnote{~%
If $\Lambda$ is an operator, one should take care of the direction to which
it acts, as well as on what arguments it acts; e.g., if
$\Lambda=\frac{\pd}{\pd {k'}^\nu}$, then the r.h.s.\ of~\eref{12.30b} is
equal to a sum of two terms, due to the left action of $\Lambda$ on
$\overline{v}^{s',\varepsilon'}(\bk')$ and the right action of $\Lambda$ on
$a_{s'}^{\dag\,\varepsilon'}(\bk')$.%
}
$s,s'=1,2$ for $m\not=0$, and $s,s'=0$ for $m=0$.
The first of these equalities, together with the
derivation of~\eref{7.5}, \eref{7.8}, \eref{7.11-3}, and~\eref{7.20},
implies the validity of~\eref{7.5}, \eref{7.8} and~\eref{7.11-3} for
$\ope{L}^{\prime}$. The equations~\eref{12.30c} and~\eref{12.30d}, in view of
the derivation of~\eref{7.5}, \eref{7.8}, \eref{7.11-3}, and~\eref{7.20},
mean that the change $\psi\leftrightarrow\bpsi$ is equivalent to any one of
the replacements
$\psi_s^\pm(\bk)\leftrightarrow\bpsi_s^\pm(\bk)$ or
$a_s^\pm(\bk)\leftrightarrow a_s^{\dag\,\pm}(\bk)$,
both combined with the change of the sign of the charge operator and the
changes
$\sigma_{\mu\nu}^{s s',\pm}(\bk)\mapsto \sigma_{\mu\nu}^{s's,\mp}(\bk)$
and
$l_{\mu\nu}^{s s',\pm}(\bk)\mapsto l_{\mu\nu}^{s's,\mp}(\bk)$. The
equations~\eref{12.30}, combined with the above\ndash made
conclusions,~\eref{12.22}--\eref{12.24}, \eref{7.5}, \eref{7.8},
\eref{7.11-3} and~\eref{8.11}, imply that the momentum, charge, spin and
orbital (angular momentum) operators for the
Lagrangians~\eref{12.17}--\eref{12.19} respectively are:
	\begin{subequations}	\label{12.31}
	\begin{align}	\label{12.31a}
\ope{P}_\mu^{\prime}
& =
\sum_{s}\int
  k_\mu |_{ k_0=\sqrt{m^2c^2+{\bs k}^2} }
\{
a_s^{\dag\,+}(\bk)\circ a_s^-(\bk)
-
a_s^{\dag\,-}(\bk)\circ a_s^+(\bk)
\}
\Id^3\bk
\\			\label{12.31b}
\ope{Q}^{\prime}
& =
+ q \sum_{s}\int
\{
a_s^{\dag\,+}(\bk)\circ a_s^-(\bk) + a_s^{\dag\,-}(\bk)\circ a_s^+(\bk)
\} \Id^3\bk
\\			\label{12.31c}
\lindex[\mspace{-6mu}{\ope{S}_{\mu\nu}^{\prime}}]{}{0}
& =
+ \frac{1}{2} \hbar
\sum_{s,s'} \!\int\!\!\Id^3\bk
\bigl\{
\sigma_{\mu\nu}^{s s',-}(\bk) a_{s}^{\dag\,+}(\bk)\circ a_{s'}^{-}(\bk)
+
\sigma_{\mu\nu}^{s s',+}(\bk) a_{s}^{\dag\,-}(\bk)\circ a_{s'}^{+}(\bk)
\bigr\}
	\end{align}
\vspace{-3ex}
	\begin{multline}	\label{12.31d}
\lindex[\mspace{-6mu}{\ope{L}_{\mu\nu}^{\prime}}]{}{0}
=
\sum_{s}\int \Id^3\bk
 ( x_{\mu}k_\nu - x_{\nu}k_\mu ) |_{ k_0=\sqrt{m^2c^2+{\bs k}^2} }
\{
a_s^{\dag\,+}(\bk)\circ a_s^-(\bk)
-
a_s^{\dag\,-}(\bk)\circ a_s^+(\bk)
\}
\\
+ \frac{1}{2} \hbar \sum_{s,s'}\int \Id^3\bk
\bigl\{
l_{\mu\nu}^{ss',-}(\bk) a_s^{\dag\,+}(\bk) \circ a_{s'}^-(\bk)
+
l_{\mu\nu}^{ss',+}(\bk) a_s^{\dag\,-}(\bk) \circ a_{s'}^+(\bk)
\bigr\}
\displaybreak[2]\\
+ \frac{1}{2} \ih \sum_{s}\int \Id^3\bk
\Bigl\{
a_s^{\dag\,+}(\bk)
\Bigl( \xlrarrow{ k_\mu \frac{\pd}{\pd k^\nu} }
     - \xlrarrow{ k_\nu \frac{\pd}{\pd k^\mu} } \Bigr)
\circ a_s^-(\bk)
\\ +
a_s^{\dag\,-}(\bk)
\Bigl( \xlrarrow{ k_\mu \frac{\pd}{\pd k^\nu} }
     - \xlrarrow{ k_\nu \frac{\pd}{\pd k^\mu} } \Bigr)
\circ a_s^+(\bk)
\Bigr\} \Big|_{ k_0=\sqrt{m^2c^2+{\bs k}^2} } ,
	\end{multline}
	\end{subequations}
	\begin{subequations}	\label{12.32}
	\begin{align}	\label{12.32a}
\ope{P}_\mu^{\prime\prime}
& =
\sum_{s}\int
  k_\mu |_{ k_0=\sqrt{m^2c^2+{\bs k}^2} }
\{
a_s^{+}(\bk)\circ a_s^{\dag\,-}(\bk)
-
a_s^{-}(\bk)\circ a_s^{\dag\,+}(\bk)
\}
\Id^3\bk
\\			\label{12.32b}
\ope{Q}^{\prime\prime}
& =
- q \sum_{s}\int
\{
a_s^{+}(\bk)\circ a_s^{\dag\,-}(\bk) + a_s^{-}(\bk)\circ a_s^{\dag\,+}(\bk)
\}
\Id^3\bk
\\			\label{12.32c}
\lindex[\mspace{-6mu}{\ope{S}_{\mu\nu}^{\prime\prime}}]{}{0}
& =
- \frac{1}{2} \hbar
\sum_{s,s'} \!\int\!\!\Id^3\bk
\bigl\{
\sigma_{\mu\nu}^{s s',+}(\bk) a_{s'}^{+}(\bk)\circ a_{s}^{\dag\,-}(\bk)
+
\sigma_{\mu\nu}^{s s',-}(\bk) a_{s'}^{-}(\bk)\circ a_{s}^{\dag\,+}(\bk)
\bigr\}
	\end{align}
\vspace{-3ex}
	\begin{multline}	\label{12.32d}
\lindex[\mspace{-6mu}{\ope{L}_{\mu\nu}^{\prime\prime}}]{}{0}
=
\sum_{s}\int \Id^3\bk
 ( x_{\mu}k_\nu - x_{\nu}k_\mu ) |_{ k_0=\sqrt{m^2c^2+{\bs k}^2} }
\{
a_s^{+}(\bk)\circ a_s^{\dag\,-}(\bk)
-
a_s^{-}(\bk)\circ a_s^{\dag\,+}(\bk)
\}
\\
- \frac{1}{2} \hbar \sum_{s,s'}\int \Id^3\bk
\bigl\{
l_{\mu\nu}^{ss',+}(\bk) a_{s'}^{+}(\bk) \circ a_{s}^{\dag\,-}(\bk)
+
l_{\mu\nu}^{ss',-}(\bk) a_{s'}^{-}(\bk) \circ a_{s}^{\dag\,+}(\bk)
\bigr\}
\displaybreak[2]\\
+ \frac{1}{2} \ih \sum_{s}\int \Id^3\bk
\Bigl\{
a_s^{+}(\bk)
\Bigl( \xlrarrow{ k_\mu \frac{\pd}{\pd k^\nu} }
     - \xlrarrow{ k_\nu \frac{\pd}{\pd k^\mu} } \Bigr)
\circ a_s^{\dag\,-}(\bk)
\\ +
a_s^{-}(\bk)
\Bigl( \xlrarrow{ k_\mu \frac{\pd}{\pd k^\nu} }
     - \xlrarrow{ k_\nu \frac{\pd}{\pd k^\mu} } \Bigr)
\circ a_s^{\dag\,+}(\bk)
\Bigr\} \Big|_{ k_0=\sqrt{m^2c^2+{\bs k}^2} } ,
	\end{multline}
	\end{subequations}
	\begin{subequations}	\label{12.33}
	\begin{align}	\label{12.33a}
	\begin{split}
\ope{P}_\mu^{\prime\prime\prime}
& =
\frac{1}{2} \sum_{s}\int
  k_\mu |_{ k_0=\sqrt{m^2c^2+{\bs k}^2} }
\{
a_s^{\dag\,+}(\bk)\circ a_s^-(\bk)
-
a_s^{\dag\,-}(\bk)\circ a_s^+(\bk)
\\
& \hphantom{ \sum_{s}\int  k_\mu |_{ k_0=\sqrt{m^2c^2+{\bs k}^2} } \{ }
+ a_s^{+}(\bk)\circ a_s^{\dag\,-}(\bk)
-
a_s^{-}(\bk)\circ a_s^{\dag\,+}(\bk)
\}
\Id^3\bk
	\end{split}
\displaybreak[1]\\			\label{12.33b}
	\begin{split}
\ope{Q}^{\prime\prime\prime}
& =
+ \frac{1}{2} q \sum_{s}\int
\{
a_s^{\dag\,+}(\bk)\circ a_s^-(\bk) + a_s^{\dag\,-}(\bk)\circ a_s^+(\bk)
\\
& \hphantom{+ \frac{1}{2} q \sum_{s}\int}
- a_s^{+}(\bk)\circ a_s^{\dag\,-}(\bk) - a_s^{-}(\bk)\circ a_s^{\dag\,+}(\bk)
\} \Id^3\bk .
	\end{split}
\displaybreak[1]\\			\label{12.33c}
	\begin{split}
\lindex[\mspace{-6mu}{\ope{S}_{\mu\nu}^{\prime\prime\prime}}]{}{0}
& =
+ \frac{1}{4} \hbar \sum_{s,s'} \int\!\!\Id^3\bk
\bigl\{
\sigma_{\mu\nu}^{s s',-}(\bk)
[ a_{s}^{\dag\,+}(\bk) ,  a_{s'}^{-}(\bk) ]_{\_}
+
\sigma_{\mu\nu}^{s s',+}(\bk)
[ a_{s}^{\dag\,-}(\bk) , a_{s'}^{+}(\bk) ]_{\_}
\bigr\}
	\end{split}
\displaybreak[1]\\			\label{12.33d}
\lindex[\mspace{-6mu}{\ope{L}_{\mu\nu}^{\prime\prime\prime}}]{}{0}
& = \frac{1}{2} \bigl(
\lindex[\mspace{-6mu}{\ope{L}_{\mu\nu}^{\prime}}]{}{0}
+
\lindex[\mspace{-6mu}{\ope{L}_{\mu\nu}^{\prime\prime}}]{}{0}
\bigr) .
	\end{align}
	\end{subequations}
(To save some space, we do not write the evident, but rather long, explicit
formula for
$\lindex[\mspace{-6mu}{\ope{L}_{\mu\nu}^{\prime\prime\prime}}]{}{0}$;
the reader can easily write it down with some patience. For the same reason we
do not write at all the total angular momentum operators; they are simply sums
of the corresponding spin and orbital (angular momentum) operators. Besides,
we have omit the evident expressions for the time\ndash dependent part of the
spin angular momentum (see~\eref{7.12-8}, which is insignificant here and
below.)
	Notice, in the above formulae the operators $a_s^{\pm}$ and
$a_s^{\dag\,\pm}$ do not depend on the Lagrangians~\eref{12.17}--\eref{12.19}
we started off.
	As one can expect, the r.h.s. of~\eref{12.33a},~\eref{12.33c}
and~\eref{12.33d} (resp.~\eref{12.33b}) is symmetric (resp.\ antisymmetric)
with respect to the change particle$\leftrightarrow$antiparticle.

	Since the obtained expressions for the momentum operators
corresponding to the Lagrangians~\eref{12.17}--\eref{12.19} are different,
the field equations for them in terms of the creation and annihilation
operators $a_s^{\pm}(\bk)$ and $a_s^{\dag\,\pm}(\bk)$ will be different,
regardless of their coincidence in terms of $\psi$ and $\bpsi$
(see~\eref{12.21}). These equations can be derive in the same way as we did
in Sect.~\ref{Sect8} for the Lagrangian~\eref{12.2}, but the results at our
disposal give us the possibility to write them without any calculations. As
the expressions~\eref{12.31a} and~\eref{7.5} for the momentum operators
of~\eref{12.17} and~\eref{12.2}, respectively, coincide and the Lagrangians
$\ope{L}$ and $\ope{L}'$ are equal up to a change of the independent
variables in them, from the decompositions~\eref{8.1}, a similar one for
$\bpsi$ (see~\eref{12.25} and~\eref{12.28}), and the derivation of~\eref{8.4}
follows that the field equations for the Lagrangian $\ope{L}'$, given
by~\eref{12.17}, in terms of  $a_s^{\pm}$ and $a_s^{\dag\,\pm}$ coincide with
the ones for $\ope{L}$, given by~\eref{12.2}. Therefore, the equations of
motion for  $\ope{L}'$ in terms of creation and annihilation operators
are~\eref{8.4}. Since~\eref{12.31a} and~\eref{12.32a} differ by the change
$a_s^{\pm}\leftrightarrow a_s^{\dag\,\pm}$, the field equations for
$\ope{L}^{\prime\prime}$, given via~\eref{12.18}, can be obtained
form~\eref{8.4} by making the replacement
$a_s^{\pm}(\bs q)\leftrightarrow a_s^{\dag\,\pm}(\bs q)$,
\ie they are
	\begin{subequations}	\label{12.33-1}
	\begin{align}
			\label{12.33-1a}
	\begin{split}
\bigl[ a_s^{\pm}(\bk) ,
a_t^{+}(\bs q) \circ a_t^{\dag\,-}(\bs q)
& -
a_t^{-}(\bs q) \circ a_t^{\dag\,+}(\bs q)
\bigr]_{\_}
\pm a_s^{\pm}(\bk) \delta_{ts} \delta^3(\bk-\bs q)
= {\lindex[\mspace{-3mu}f]{}{\prime\prime}}_{st}^{\pm} (\bs k,\bs q)
	\end{split}
\\			\label{12.33-1b}
	\begin{split}
\bigl[ a_s^{\dag\,\pm}(\bk) ,
a_t^{+}(\bs q) \circ a_t^{\dag\,-}(\bs q)
& -
a_t^{-}(\bs q) \circ a_t^{\dag\,+}(\bs q)
\bigr]_{\_}
\pm a_s^{\dag\,\pm}(\bk) \delta_{ts} \delta^3(\bk-\bs q)
= {\lindex[\mspace{-3mu}f]{}{\prime\prime}}_{st}^{\dag\,\pm} (\bs k,\bs q) ,
	\end{split}
	\end{align}
where
${\lindex[\mspace{-3mu}f]{}{\prime\prime}}_{st}^{\pm}(\bk,\bs q)$ and
${\lindex[\mspace{-3mu}f]{}{\prime\prime}}_{st}^{\dag\,\pm}(\bk,\bs q)$ are
(generalized) functions such that
	\begin{equation}	\label{12.33-1c}
	\begin{split}
\sum_{t} \int q_\mu\big|_{q_0=\sqrt{m^2c^2+{\bs q}^2}}
{\lindex[\mspace{-3mu}f]{}{\prime\prime}}_{st}^{\pm} (\bs k,\bs q)
\Id^3\bs{q}
 = 0
\qquad
\sum_{t} \int q_\mu\big|_{q_0=\sqrt{m^2c^2+{\bs q}^2}}
{\lindex[\mspace{-3mu}f]{}{\prime\prime}}_{st}^{\dag\,\pm} (\bs k,\bs q)
\Id^3\bs{q}
 = 0.
	\end{split}
	\end{equation}
	\end{subequations}
At last, as~\eref{12.33a} is the half sum of~\eref{12.31a} and~\eref{12.32a},
the field equations for $\ope{L}^{\prime\prime\prime}$, given
via~\eref{12.19}, in terms of creation and annihilation operators can be
obtained from~\eref{8.4} by replacing the first terms in it by the half sum
of the first terms in left-hand-sides of~\eref{8.4} and~\eref{12.33-1}, \ie
these equations are:
	\begin{subequations}	\label{12.34}
	\begin{align}
			\label{12.34a}
	\begin{split}
\bigl[ a_s^{\pm}(\bk) ,
a_t^{\dag\,+}(\bs q) \circ a_t^{-}(\bs q)
& -
a_t^{\dag\,-}(\bs q) \circ a_t^{+}(\bs q)
\bigr]_{\_}
\\
+ \bigl[ a_s^{\pm}(\bk) ,
 a_t^{+}(\bs q) \circ a_t^{\dag\,-}(\bs q)
& -
 a_t^{-}(\bs q) \circ a_t^{\dag\,+}(\bs q)
\bigr]_{\_}
\\
& \pm 2 a_s^{\pm}(\bk) \delta_{ts} \delta^3(\bk-\bs q)
= {\lindex[\mspace{-3mu}f]{}{\prime\prime\prime}}_{st}^{\pm} (\bs k,\bs q)
	\end{split}
\\			\label{12.34b}
	\begin{split}
\bigl[ a_s^{\dag\,\pm}(\bk) ,
a_t^{\dag\,+}(\bs q) \circ a_t^{-}(\bs q)
& -
a_t^{\dag\,-}(\bs q) \circ a_t^{+}(\bs q)
\bigr]_{\_}
\\
+
\bigl[ a_s^{\dag\,\pm}(\bk) ,
a_t^{+}(\bs q) \circ a_t^{\dag\,-}(\bs q)
& -
a_t^{-}(\bs q) \circ a_t^{\dag\,+}(\bs q)
\bigr]_{\_}
\\&
\pm 2 a_s^{\dag\,\pm}(\bk) \delta_{ts} \delta^3(\bk-\bs q)
= {\lindex[\mspace{-3mu}f]{}{\prime\prime\prime}}_{st}^{\dag\,\pm}
							(\bs k,\bs q) ,
	\end{split}
	\end{align}
where
${\lindex[\mspace{-3mu}f]{}{\prime\prime\prime}}_{st}^{\pm}(\bk,\bs q)$ and
${\lindex[\mspace{-3mu}f]{}{\prime\prime\prime}}_{st}^{\dag\,\pm}(\bk,\bs q)$ are
(generalized) functions such that
	\begin{equation}	\label{12.34c}
	\begin{split}
\sum_{t} \int\! q_\mu\big|_{q_0=\sqrt{m^2c^2+{\bs q}^2}}
{\lindex[\mspace{-3mu}f]{}{\prime\prime\prime}}_{st}^{\pm} (\bs k,\bs q)
\Id^3\bs{q}
 = 0
\qquad
\sum_{t} \int\! q_\mu\big|_{q_0=\sqrt{m^2c^2+{\bs q}^2}}
{\lindex[\mspace{-3mu}f]{}{\prime\prime\prime}}_{st}^{\dag\,\pm}
					(\bs k,\bs q)
\Id^3\bs{q}
 = 0.
	\end{split}
	\end{equation}
	\end{subequations}

	Consider now the problem regarding the possible (anti)commutation
relations for the Lagrangians~\eref{12.17}--\eref{12.19}. As we saw in
Sect.~\ref{Sect9}, the field equations~\eref{8.4a}--\eref{8.4b} are
equivalent to~\eref{9.2} (with $\varepsilon=\pm1$). Similarly,
applying~\eref{9.1}, we can rewrite
equivalently~\eref{12.33-1a}--\eref{12.33-1b}
and~\eref{12.34a}--\eref{12.34b} respectively as
	\begin{subequations}	\label{12.35}
	\begin{gather}
				\label{12.35a}
	\begin{split}
[a_s^{\pm}(\bs k), a_t^{+}(\bs q)]_{\varepsilon}
					\circ a_t^{\dag\,-}(\bs q)
& - \varepsilon a_t^{+}(\bs q) \circ
	[a_s^{\pm}(\bs k), a_t^{\dag\,-}(\bs q) ]_{\varepsilon}
\\
- [a_s^{\pm}(\bs k), a_t^{-}(\bs q)]_{\varepsilon}
					\circ a_t^{\dag\,+}(\bs q)
& + \varepsilon a_t^{-}(\bs q) \circ
	[a_s^{\pm}(\bs k), a_t^{\dag\,+}(\bs q) ]_{\varepsilon}
\\
& \pm  a_s^{\pm}(\bs k) \delta_{st} \delta^3(\bs k-\bs q)
= {\lindex[\mspace{-3mu}f]{}{\prime\prime}}_{st}^{\pm}(\bs k,\bs q)
	\end{split}
\\
				\label{12.35b}
	\begin{split}
[a_s^{\dag\,\pm}(\bs k), a_t^{+}(\bs q)]_{\varepsilon}
					\circ a_t^{\dag\,-}(\bs q)
& - \varepsilon a_t^{+}(\bs q) \circ
	[a_s^{\dag\,\pm}(\bs k), a_t^{\dag\,-}(\bs q) ]_{\varepsilon}
\\
- [a_s^{\dag\,\pm}(\bs k), a_t^{-}(\bs q)]_{\varepsilon}
					\circ a_t^{\dag\,+}(\bs q)
& + \varepsilon a_t^{-}(\bs q) \circ
	[a_s^{\dag\,\pm}(\bs k), a_t^{\dag\,+}(\bs q) ]_{\varepsilon}
\\
& \pm  a_s^{\dag\,\pm}(\bs k) \delta_{st} \delta^3(\bs k-\bs q)
= {\lindex[\mspace{-3mu}f]{}{\prime\prime}}_{st}^{\dag\,\pm}(\bs k,\bs q)
	\end{split}
	\end{gather}
	\end{subequations}
	\begin{subequations}	\label{12.36}
	\begin{gather}
				\label{12.36a}
	\begin{split}
[a_s^{\pm}(\bs k), a_t^{\dag\,+}(\bs q)]_{\varepsilon}
					\circ a_t^{-}(\bs q)
& - \varepsilon a_t^{\dag\,+}(\bs q) \circ
	[a_s^{\pm}(\bs k), a_t^{-}(\bs q) ]_{\varepsilon}
\\
- [a_s^{\pm}(\bs k), a_t^{\dag\,-}(\bs q)]_{\varepsilon}
					\circ a_t^{+}(\bs q)
& + \varepsilon a_t^{\dag\,-}(\bs q) \circ
	[a_s^{\pm}(\bs k), a_t^{+}(\bs q) ]_{\varepsilon}
\\
+ [a_s^{\pm}(\bs k), a_t^{+}(\bs q)]_{\varepsilon}
					\circ a_t^{\dag\,-}(\bs q)
& - \varepsilon a_t^{+}(\bs q) \circ
	[a_s^{\pm}(\bs k), a_t^{\dag\,-}(\bs q) ]_{\varepsilon}
\\
- [a_s^{\pm}(\bs k), a_t^{-}(\bs q)]_{\varepsilon}
					\circ a_t^{\dag\,+}(\bs q)
& + \varepsilon a_t^{-}(\bs q) \circ
	[a_s^{\pm}(\bs k), a_t^{\dag\,+}(\bs q) ]_{\varepsilon}
\\
& \pm 2 a_s^{\pm}(\bs k) \delta_{st} \delta^3(\bs k-\bs q)
= {\lindex[\mspace{-3mu}f]{}{\prime\prime\prime}}_{st}^{\pm}(\bs k,\bs q)
	\end{split}
\\
				\label{12.36b}
	\begin{split}
[a_s^{\dag\,\pm}(\bs k), a_t^{\dag\,+}(\bs q)]_{\varepsilon}
					\circ a_t^{-}(\bs q)
& - \varepsilon a_t^{\dag\,+}(\bs q) \circ
	[a_s^{\dag\,\pm}(\bs k), a_t^{-}(\bs q) ]_{\varepsilon}
\\
- [a_s^{\dag\,\pm}(\bs k), a_t^{\dag\,-}(\bs q)]_{\varepsilon}
					\circ a_t^{+}(\bs q)
& + \varepsilon a_t^{\dag\,-}(\bs q) \circ
	[a_s^{\dag\,\pm}(\bs k), a_t^{+}(\bs q) ]_{\varepsilon}
\\
[a_s^{\dag\,\pm}(\bs k), a_t^{+}(\bs q)]_{\varepsilon}
					\circ a_t^{\dag\,-}(\bs q)
& - \varepsilon a_t^{+}(\bs q) \circ
	[a_s^{\dag\,\pm}(\bs k), a_t^{\dag\,-}(\bs q) ]_{\varepsilon}
\\
- [a_s^{\dag\,\pm}(\bs k), a_t^{-}(\bs q)]_{\varepsilon}
					\circ a_t^{\dag\,+}(\bs q)
& + \varepsilon a_t^{-}(\bs q) \circ
	[a_s^{\dag\,\pm}(\bs k), a_t^{\dag\,+}(\bs q) ]_{\varepsilon}
\\
& \pm 2  a_s^{\dag\,\pm}(\bs k) \delta_{st} \delta^3(\bs k-\bs q)
= {\lindex[\mspace{-3mu}f]{}{\prime\prime\prime}}_{st}^{\dag\,\pm}(\bs k,\bs q) .
	\end{split}
	\end{gather}
	\end{subequations}
The analysis of~\eref{12.35} and~\eref{12.36} is practically identical with
the one of~\eref{9.2} in Sect.~\ref{Sect9} with one very important new
consequence from~\eref{12.36}. Imposing the additional conditions~\eref{9.4},
we see that equations~\eref{12.36} reduce to
	\begin{subequations}	\label{12.37}
	\begin{gather}
				\label{12.37a}
	\begin{split}
(1 + \varepsilon) \bigl\{
[a_s^{\pm}(\bs k), a_t^{\dag\,+}(\bs q)]_{\varepsilon}
					\circ a_t^{-}(\bs q)
& - a_t^{\dag\,+}(\bs q) \circ
	[a_s^{\pm}(\bs k), a_t^{-}(\bs q) ]_{\varepsilon}
\\
- [a_s^{\pm}(\bs k), a_t^{\dag\,-}(\bs q)]_{\varepsilon}
					\circ a_t^{+}(\bs q)
& + a_t^{\dag\,-}(\bs q) \circ
	[a_s^{\pm}(\bs k), a_t^{+}(\bs q) ]_{\varepsilon}
\bigr\}
\\
& \pm  2 a_s^{\pm}(\bs k) \delta_{st} \delta^3(\bs k-\bs q)
= {\lindex[\mspace{-3mu}f]{}{\prime\prime\prime}}_{st}^{\pm}(\bs k,\bs q)
	\end{split}
\\
				\label{12.37b}
	\begin{split}
(1 + \varepsilon) \bigl\{
[a_s^{\dag\,\pm}(\bs k), a_t^{\dag\,+}(\bs q)]_{\varepsilon}
					\circ a_t^{-}(\bs q)
& - a_t^{\dag\,+}(\bs q) \circ
	[a_s^{\dag\,\pm}(\bs k), a_t^{-}(\bs q) ]_{\varepsilon}
\\
- [a_s^{\dag\,\pm}(\bs k), a_t^{\dag\,-}(\bs q)]_{\varepsilon}
					\circ a_t^{+}(\bs q)
& + a_t^{\dag\,-}(\bs q) \circ
	[a_s^{\dag\,\pm}(\bs k), a_t^{+}(\bs q) ]_{\varepsilon}
\bigr\}
\\
& \pm 2  a_s^{\dag\,\pm}(\bs k) \delta_{st} \delta^3(\bs k-\bs q)
= {\lindex[\mspace{-3mu}f]{}{\prime\prime\prime}}_{st}^{\dag\,\pm}
							(\bs k,\bs q) .
	\end{split}
	\end{gather}
	\end{subequations}

	Let us see what entails~\eref{12.37} for $\varepsilon=-1$, which
corresponds to quantization of a spinor field via \emph{commutators}, not via
\emph{anti}commutators. Inserting~\eref{12.37} with $\varepsilon=-1$
into~\eref{12.34c}, we get
	\begin{gather}	\label{12.38}
k_\mu|_{k_0=\sqrt{m^2c^2+{\bs k}^2}} a_s^\pm(\bs k) = 0
\quad
k_\mu|_{k_0=\sqrt{m^2c^2+{\bs k}^2}} a_s^{\dag\,\pm}(\bs k) = 0
\qquad
\text{for }\varepsilon =-1,
\intertext{which, by virtue of~\eref{12.33a}, imply}
			\label{12.39}
\ope{P}_\mu^{\prime\prime\prime} = 0
\qquad
\text{for }\varepsilon =-1
\\\intertext{which, in its turn, reduce the Dirac equations~\eref{12.21} to}
			\label{12.40}
mc\psi = 0 \quad mc\bpsi = 0
\qquad
\text{for }\varepsilon =-1.
	\end{gather}

	Thus, we see that, when one starts from the Lagrangian~\eref{12.19},
the only free spinor fields that are possible to be quantized via commutators
are~\eref{5.1}, in the massless case $m=0$, and the `missing' field
$\psi=\bpsi=0$ in the massive case $m\not=0$. The former solutions of the
field equations were investigated at length in the previous sections and the
latter solution, $\psi=\bpsi=0$, of the field equations is completely
unphysical as it cannot lead to any physically observable consequences.
But, moreover, these solutions are rejected by the properties~\eref{9.5}
and~\eref{9.6} of the (generalized) functions in the right-hand-sides of the
equations in~\eref{9.4}. Therefore, the
\emph{Lagrangian~\eref{12.19} does not admit quantization by commutators},
contrary to the Lagrangians~\eref{12.17} and~\eref{12.18}, and, consequently,
for it we must put $\varepsilon=+1$ in~\eref{9.4}.

	The further analysis of~\eref{12.35} with $\varepsilon=\pm1$
and~\eref{12.36} with $\varepsilon=+1$ is practically identical to the one
of~\eref{9.2} and leads to the anticommutation relations~\eref{9.13}.%
\footnote{~%
One can derive~\eref{9.13} without any calculations by taking into account
that the l.h.s.\  of~\eref{12.35} is equal to the l.h.s.\ of~\eref{9.2}
combined with the change
$a_s^\pm(\bs q)\leftrightarrow a_s^{\dag\,\pm}(\bs q)$ and that the l.h.s.\
of~\eref{12.36} is equal to the sum of the left-hand-sides of~\eref{9.2} and
~\eref{12.35}. This implies similar changes in~\eref{9.13} under which it is
invariant.%
}

	Consequently, the Lagrangians~\eref{12.17}--\eref{12.19} lead to
identical anticommutation relations, viz.~\eref{9.13}, and in this direction
the only difference between them is that~\eref{12.19} does not require as an
additional condition the quantization via anticommutators, or an equivalent
to it hypothesis, like the spin\ndash statistics theorem or charge symmetry,
as this condition is encoded in it from the very beginning.

	Since the anticommutation relations for the
Lagrangians~\eref{12.17}--\eref{12.19} are identical, the vacuum and normal
ordering for them should be introduced in an identical way, \viz as it was
described in Sect.~\ref{Sect10}. As a result of~\eref{12.31}--\eref{12.33},
after normal ordering, the operators of the dynamical variables for them
become identical and are given by~\eref{10.4}--\eref{10.6new1}.

	Let us summarize. The Lagrangians~\eref{12.17}--\eref{12.19}
(or~\eref{12.14}--\eref{12.16} in Heisenberg picture) are equivalent in a
sense that they entail identical final quantum field theories. The principle
difference between them is that in~\eref{12.19} is encoded additionally the
charge symmetry of (or, equivalently, the spin\ndash statistics theorem for)
a free spinor field and there is not a need to impose it as additional
condition on a later stage of theory's development.


\section {Conclusion}
\label{Conclusion}

	In this paper, we made a more or less comprehensive investigation of
free spin~$\frac{1}{2}$ quantum fields in momentum picture of motion in
Lagrangian quantum field theory. The methods we used and the results
obtained are near (similar) to the ones in~\cite{bp-QFTinMP-scalars},
concerning free charged scalar fields. The main difference from the scalar
case comes from the multi\ndash component character of a spinor field, which
is due to its non\ndash zero spin.

	A feature of the momentum picture is the introduction of creation
and annihilation operators without invoking explicitly the Fourier transform
(as in Heisenberg picture). We have written the field equations in terms of
these operators. In this form they turn to be trilinear algebraic(\ndash
functional) equations relative to the creation and annihilation operators. On
this base, the standard anticommutation relations are derived under some
explicitly presented additional conditions. Under these conditions, the
anticommutation relations are tantamount to the initial field equations for
free spinor fields. We have also briefly studied the state vectors, vacuum
and normal ordering procedure for these fields.

	An analysis of the initial Lagrangian, from which the quantum theory
of free spinor field is derived, is presented. The consequences of several
Lagrangians are described and the `best' one of them is single out. It is the
one which is charge symmetric, \ie it is the one which is invariant under the
change particle$\leftrightarrow$antiparticle described in suitable variables.
This is the Lagrangian~\eref{12.16} in Heisenberg picture or~\eref{12.19} in
momentum picture. It entails, before quantization and normal ordering, the
field equations~\eref{12.34} in terms of creation and annihilation operators.
Evidently, these equations can be rewritten as
	\begin{subequations}	\label{c.1}
	\begin{align}
			\label{c.1a}
	\begin{split}
\bigl[ [a_t^{\dag\,+}(\bs q) , a_t^{-}(\bs q)]_{\_} ,
	a_s^{\pm}(\bk) \bigr]_{\_}
& +
\bigl[ [a_t^{+}(\bs q) , a_t^{\dag\,-}(\bs q)]_{\_} ,
	a_s^{\pm}(\bk) \bigr]_{\_}
\\
& =
\pm 2 a_s^{\pm}(\bk) \delta_{st} \delta^3(\bk-\bs q)
-
{\lindex[\mspace{-3mu}f]{}{\prime\prime\prime}}_{st}^{\pm} (\bs k,\bs q)
	\end{split}
\\			\label{c.1b}
	\begin{split}
\bigl[ [a_t^{\dag\,+}(\bs q) , a_t^{-}(\bs q)]_{\_} ,
	a_s^{\dag\,\pm}(\bk) \bigr]_{\_}
& +
\bigl[ [a_t^{+}(\bs q) , a_t^{\dag\,-}(\bs q)]_{\_} ,
	a_s^{\dag\,\pm}(\bk) \bigr]_{\_}
\\
& =
\pm 2 a_s^{\dag\,\pm}(\bk) \delta_{st} \delta^3(\bk-\bs q)
-
{\lindex[\mspace{-3mu}f]{}{\prime\prime\prime}}_{st}^{\dag\,\pm}
							     (\bs k,\bs q).
	\end{split}
	\end{align}
	\end{subequations}
Trilinear relations, like equations~\eref{c.1}, are typical for the so\ndash
called para\-statistics and parafield
theory~\cite{Green-1953,Volkov-1959,Volkov-1960,Greenberg&Messiah-1964,
Greenberg&Messiah-1965}, where they play a role of (para)com\-mu\-tation
relations. Elsewhere we shall show how from~\eref{c.1} the parafermi
commutation relations for a free spinor field can be derived.

	We shall end with the evident remark that the particles and
antiparticles of a free spinor field (, described via some of the Lagrangians
considered in this paper) are always different, due to the fact that the
charge of field's particles and antiparticles is $-q$ and $+q$, respectively,
and their spin and orbital (angular momentum) operators are always non\ndash
zero and different (see, e.g,~\eref{11.2-1}, \eref{11.2-2}, \eref{7.11-3}
and~\eref{7.11-13}).




\addcontentsline{toc}{section}{References}
\bibliography{bozhopub,bozhoref}
\bibliographystyle{unsrt}
\addcontentsline{toc}{subsubsection}{This article ends at page}

\end{document}